\documentclass[traditabstract]{aa} 

\usepackage{graphicx,dblfloatfix}
\usepackage{txfonts}
\usepackage{natbib}
\usepackage[colorlinks=true,urlcolor=dblue,citecolor=dblue,
            linkcolor=dblue]{hyperref}

  \def\st{{\circ\hspace{-0.86ex}-}}
  \def\pst{{p^\st}}
  \def\pvap{p^{\rm vap}}
  \def\G{{\Delta G^{\hspace{0.1ex}\st}\!}}
  \def\GG{{G^{\hspace{0.1ex}\st}}}
  \def\GNIST{{G_{\rm NIST}^{\hspace{0.1ex}\st}}}
  \def\GSU{{G_{\rm SU}^{\hspace{0.1ex}\st}}}
  \def\dG{{\Delta G_{\rm\hspace{-0.2ex}f}^{\hspace{0.1ex}\st}}}
  \def\dGSU{{\Delta G_{\rm\hspace{-0.2ex}f,\rm SU}^{\hspace{0.1ex}\st}}}
  \def\dGNIST{{\Delta G_{\rm\hspace{-0.2ex}f,\rm NIST}^{\hspace{0.1ex}\st}}}
  
  \def\Tref{{T_{\rm ref}}}
  \def\mol{{{\rm A}_a{\rm B}_b{\rm C}_c}}
  \def\H{{\rm\langle H\rangle}}
  \def\nH{n_\H}
  \def\Sj{S_{\hspace{-0.45ex}j}}
  \def\ek{\epsilon_{\hspace{0.25ex}k}}
  \def\cj{c_{\hspace{-0.1ex}j}}
  \def\pj{p_{\hspace{-0.05ex}j}}

\begin{document}

\definecolor{GreenYellow}  {cmyk}{0.15,0,0.69,0}
\definecolor{Yellow}{cmyk}{0,0,1,0}
\definecolor{Goldenrod}{cmyk}{0,0.10,0.84,0}
\definecolor{Dandelion}{cmyk}{0,0.29,0.84,0}
\definecolor{Apricot}  {cmyk}{0,0.32,0.52,0}
\definecolor{Peach}    {cmyk}{0,0.50,0.70,0}
\definecolor{Melon}    {cmyk}{0,0.46,0.50,0}
\definecolor{YellowOrange}  {cmyk}{0,0.42,1,0}
\definecolor{Orange}   {cmyk}{0,0.61,0.87,0}
\definecolor{BurntOrange}   {cmyk}{0,0.51,1,0}
\definecolor{Bittersweet}   {cmyk}{0,0.75,1,0.24}
\definecolor{RedOrange}{cmyk}{0,0.77,0.87,0}
\definecolor{Mahogany} {cmyk}{0,0.85,0.87,0.35}
\definecolor{Maroon}   {cmyk}{0,0.87,0.68,0.32}
\definecolor{BrickRed} {cmyk}{0,0.89,0.94,0.28}
\definecolor{Red} {cmyk}{0,1,1,0}
\definecolor{OrangeRed}{cmyk}{0,1,0.50,0}
\definecolor{RubineRed}{cmyk}{0,1,0.13,0}
\definecolor{WildStrawberry}{cmyk}{0,0.96,0.39,0}
\definecolor{Salmon}   {cmyk}{0,0.53,0.38,0}
\definecolor{CarnationPink} {cmyk}{0,0.63,0,0}
\definecolor{Magenta}  {cmyk}{0,1,0,0}
\definecolor{VioletRed}{cmyk}{0,0.81,0,0}
\definecolor{Rhodamine}{cmyk}{0,0.82,0,0}
\definecolor{Mulberry} {cmyk}{0.34,0.90,0,0.02}
\definecolor{RedViolet}{cmyk}{0.07,0.90,0,0.34}
\definecolor{Fuchsia}{cmyk}{0.47,0.91,0,0.08}
\definecolor{Lavender} {cmyk}{0,0.48,0,0}
\definecolor{Thistle}{cmyk}{0.12,0.59,0,0}
\definecolor{Orchid}{cmyk}{0.32,0.64,0,0}
\definecolor{DarkOrchid}{cmyk}{0.40,0.80,0.20,0}
\definecolor{Purple}{cmyk}{0.45,0.86,0,0}
\definecolor{Plum}{cmyk}{0.50,1,0,0}
\definecolor{Violet} {cmyk}{0.79,0.88,0,0}
\definecolor{RoyalPurple} {cmyk}{0.75,0.90,0,0}
\definecolor{BlueViolet}{cmyk}{0.86,0.91,0,0.04}
\definecolor{Periwinkle}{cmyk}{0.57,0.55,0,0}
\definecolor{CadetBlue}{cmyk}{0.62,0.57,0.23,0}
\definecolor{CornflowerBlue}{cmyk}{0.65,0.13,0,0}
\definecolor{MidnightBlue}{cmyk}{0.98,0.13,0,0.43}
\definecolor{NavyBlue} {cmyk}{0.94,0.54,0,0}
\definecolor{RoyalBlue}{cmyk}{1,0.50,0,0}
\definecolor{Blue}{cmyk}{1,1,0,0}
\definecolor{Cerulean} {cmyk}{0.94,0.11,0,0}
\definecolor{Cyan}{cmyk}{1,0,0,0}
\definecolor{ProcessBlue} {cmyk}{0.96,0,0,0}
\definecolor{SkyBlue}{cmyk}{0.62,0,0.12,0}
\definecolor{Turquoise}{cmyk}{0.85,0,0.20,0}
\definecolor{TealBlue} {cmyk}{0.86,0,0.34,0.02}
\definecolor{Aquamarine}{cmyk}{0.82,0,0.30,0}
\definecolor{BlueGreen}{cmyk}{0.85,0,0.33,0}
\definecolor{Emerald}{cmyk}{1,0,0.50,0}
\definecolor{JungleGreen} {cmyk}{0.99,0,0.52,0}
\definecolor{SeaGreen} {cmyk}{0.69,0,0.50,0}
\definecolor{Green}{cmyk}{1,0,1,0}
\definecolor{ForestGreen} {cmyk}{0.91,0,0.88,0.12}
\definecolor{PineGreen}{cmyk}{0.92,0,0.59,0.25}
\definecolor{LimeGreen}{cmyk}{0.50,0,1,0}
\definecolor{YellowGreen} {cmyk}{0.44,0,0.74,0}
\definecolor{SpringGreen} {cmyk}{0.26,0,0.76,0}
\definecolor{OliveGreen}{cmyk}{0.64,0,0.95,0.40}
\definecolor{RawSienna}{cmyk}{0,0.72,1,0.45}
\definecolor{Sepia}{cmyk}{0,0.83,1,0.70}
\definecolor{Brown}{cmyk}{0,0.81,1,0.60}
\definecolor{Tan} {cmyk}{0.14,0.42,0.56,0}
\definecolor{Gray}{cmyk}{0,0,0,0.50}
\definecolor{Black}{cmyk}{0,0,0,1}
\definecolor{White}{cmyk}{0,0,0,0}

\definecolor{StaubBraun}{cmyk}{0,0.61,0.87,0.20}
\definecolor{lyellow}{cmyk}{0,0,0.2,0}
\definecolor{lred}{cmyk}{0,0.4,0.4,0}
\definecolor{lgray}{cmyk}{0,0,0,0.15}
\definecolor{lgreen}{cmyk}{0.3,0,0.8,0}
\definecolor{lblue}{cmyk}{0.4,0,0,0}
\definecolor{dgreen}{rgb}{0.05,0.5,0.1}
\definecolor{dred}{rgb}{0.85,0.0,0.0}
\definecolor{dblue}{rgb}{0.0,0.0,0.6}
\definecolor{brown}{rgb}{0.6,0.1,0.1}

\title{Equilibrium chemistry down to 100\,K}
\subtitle{Impact of silicates and phyllosilicates on carbon/oxygen ratio}   

   \author{P.~Woitke\inst{1,2}, 
           Ch.~Helling\inst{1,2,5},
           G.~H.~Hunter\inst{1,2},
           J.~D.~Millard\inst{1,2},
           G.~E.~Turner\inst{1,2},
           M.~Worters\inst{1,2},
           J.~Blecic\inst{3},
           J.~W.~Stock\inst{4}
   }
   
   \authorrunning{P.~Woitke et al.}

   \institute{ 
             SUPA School of Physics \& Astronomy, University of St Andrews,
             North Haugh, KY16\,9SS, St Andrews, UK
             \and  
             Centre for Exoplanet Science, University of St Andrews, 
             St Andrews, UK
             \and 
             New York University Abu Dhabi, Abu Dhabi, United Arab Emirates
             \and 
             Department of Chemistry and Environmental Science, Medgar
             Evers College-City University of New York, 1650 Bedford
             Avenue, Brooklyn, NY 11235, US
             \and 
             Anton Pannekoek Institute for Astronomy, University of 
             Amsterdam, Science Park 904, 1098 XH Amsterdam, NL
             }

   \date{Received\ \ Oct.\,27$^{\rm th}$, 2017;\ \ accepted
     Dec.\,2$^{\rm nd}$, 2017}

   \abstract{We introduce a fast and versatile computer code, {\sc
       GGchem}, to determine the chemical composition of gases in
     thermo-chemical equilibrium down to 100\,K, with or without
     equilibrium condensation. We review the data for molecular
     equilibrium constants, $k_p(T)$, from several sources and discuss
     which functional fits are most suitable for low temperatures.  We
     benchmark our results against another chemical equilibrium code.
     We collect Gibbs free energies, $\dG$, for about 200 solid and
     liquid species from the NIST-JANAF database and the geophysical
     database SUPCRTBL. We discuss the condensation sequence of the
     elements with solar abundances in phase equilibrium down to
     100\,K. Once the major magnesium silicates $\rm Mg_2SiO_4$[s] and
     $\rm MgSiO_3$[s] have formed, the dust/gas mass ratio jumps to a
     value of about 0.0045 which is significantly lower than the often
     assumed value of 0.01. Silicate condensation is found to
     increase the carbon/oxygen ratio (C/O) in the gas from its solar
     value of $\sim$\,0.55 up to $\sim$\,0.71, and, by the additional
     intake of water and hydroxyl into the solid matrix, the formation
     of phyllosilicates at temperatures below $\sim$\,400\,K increases
     the gaseous C/O further to about 0.83. Metallic tungsten (W) is
     the first condensate found to become thermodynamically stable
     around $1600-2200$\,K (depending on pressure), several hundreds
     of Kelvin before subsequent materials like zirconium dioxide
     ($\rm ZrO_2$) or corundum ($\rm Al_2O_3$) can condense. We
     briefly discuss whether tungsten, despite its low abundance of
     $\sim\!2\times10^{-7}$ times the silicon abundance, could provide
     the first seed particles for astrophysical dust formation. The
     {\sc GGchem} code is publicly available
     at\;\,\url{https://github.com/pw31/GGchem}.}

   \maketitle

\section{Introduction}

Deriving the chemical composition of gases in the atmospheres of
stars, brown dwarfs and planets is one of the most fundamental
modelling tasks in astronomy.  All hydrodynamical,
evolutionary and radiative processes depend on the thermodynamics
of the gas. In particular, the interpretation of spectral observations
requires detailed knowledge about the gas composition.  Only then, the
inverse problem of infering the internal physico-chemical structure of
an object in space from spectroscopic observations can be successfully solved.

Instrument development has progressed tremendously over the past
decades, allowing us to use a large range of wavelengths to analyse
e.g.\ the chemical composition of brown dwarfs (see
\citealt{Allard1997, hc2014} and references in these reviews) and
exoplanets \citep[e.g.][]{Desert2008, Grillmair2008, Kok2013,
  Kreidberg2014, Fraine2014, Birkby2017}. We can also revisit the
long-standing question of what drives the mass loss of oxygen-rich
Asymptotic Giant Branch (AGB) stars \citep{Gail1998, Woitke2006,
  Hofner2007, Golriz2014, Gail2016, Decin2017}, including the problem
of seed particle formation and the search for the first condensate in
space \citep[see e.g.][]{Patzer1995, Patzer2007}. High-resolution
spectroscopy of molecular lines from CO and H$_2$O are also used, for
example, to derive winds properties of extrasolar planets
\citep{Brogi2016}.
Infrared instruments on board of the {\sc James-Webb Space Telescope}
are expected to revolutionise our understanding of cool objects like
exoplanets, brown dwarfs and cool stars
\citep[e.g.][]{Marley2009,Beichmann2014}. High-resolution ground-based
spectroscopy with e.g.\ {\sc Carmenes} \citep[e.g.][]{Sanches2017} aim
at a better understanding of late-type main-sequence stars and their
low-mass planets.  In the more distant future, {\sc Plato}
\citep{Rauer2016} will combine astroseismology for solar-like stars
with the search for habitable, Earth-like planets. Exploring such
high-end quality data requires substantial modelling efforts of which
deriving the chemical gas composition is one of the key issues, which
needs to be fast and sufficiently accurate over a wide range of
temperatures and pressures. The predictions of the molecular
composition of the observed gases depends critically on the
availability and precision of thermo-chemical data.

In the first approximation, the gas is assumed to be in Local
Thermodynamical Equilibrium (LTE), so the gas composition can be
calculated from the thermodynamical principle of minimisation of the
system Gibbs free energy \citep[e.g.][]{White1958,Eriksson1971}.
Thermo-chemical equilibrium is a part of the set of LTE assumptions
used in particular to compute the abundances of molecules. Pioneering
work was done here for the atmospheres of cool stars where molecules
are the main opacity carriers \citep[e.g.][]{Tsuji1965, Auer1968,
  Gustafsson1971}. To date, applications of thermo-chemical
equilibrium are widespread in astronomy, for example in gas giant
planet atmospheres, brown dwarfs and low-mass dwarf stars
\citep{Allard1995, Tsuji1996, Lodders2002, Visscher2010, Marley2015}.

Deviations from thermo-chemical equilibrium are generally expected to
occur in diluted gases, in particular due to UV-fields, X-rays or
cosmic rays as, for example, in protoplanetary discs
\citep[e.g.][]{Semenov2011,Woitke2016}, in the upper layers of
planetary atmospheres \citep[e.g.][]{Moses2014,Rimmer2016}, or in case
of fast hydrodynamical processes where the dynamical timescale becomes
smaller than the chemical timescale, for example in interstellar
shocks \citep[e.g.][]{Hollenbach1989}.

\citet{Griffith1999} and \citet{Cooper2006} argue for strong chemical
dis-equilibrium effects and higher CO abundances in BD and exoplanet
atmospheres.  Concerning the atmospheres of cool giant planets,
\citet{Moses2014} have shown that thermo-chemical equilibrium only
prevails in the deepest layers ($p$\,$>$\,a~few bar), whereas
transport-induced quenching and photochemistry are likely to be
important in the upper layers. However, recent kinetic chemical models
\citep{Visscher2006, Visscher2010, Zahnle2009, Line2010, Moses2011,
  Kopparapu2012, Venot2012} have shown that in hot exoplanet
atmospheres ($T\!\ga\!1200\,$K), the chemical timescales are in fact
quite short, and hence thermo-chemical equilibrium prevails. For the
hottest planets, \citet{Line2013} and \citet{Oreshenko2017} have shown
that at a pressure of 100\,mbar CH$_4$, CO, H$_2$O and H$_2$ should be
close to chemical equilibrium even when fast atmospheric mixing is
assumed.  \citet{Line2013} have pointed out that spectral observations
probe layers in exoplanet atmospheres where quenching by atmospheric
mixing can be relevant, but not photochemistry.

At low temperatures $\rm(\la 2000\,K)$, the formation of solids and
liquids becomes an essential part of the problem of finding the
equilibrium composition. The condensates selectively consume most of
certain elements (e.g.\ Al, Si, Mg, Fe), whereas others remain in the
gas phase down to substantially lower temperatures (e.g.\ S,
Cl). The abundant elements H, He, C, N, O remain mostly
gaseous before even they, eventually, will condense e.g.\ in form of
water, ammonia and methane ices like on Titan. The remaining gas phase
element abundances can differ by orders of magnitude from the initial,
undepleted (e.g. solar) abundances before condensation
\citep[e.g.][]{Woitke2004,Juncher2017}, which is essential to
understand the spectroscopic appearance of cool and cold objects like
AGB stars and their winds \citep{Gail1984, Sedlmayr1995, Helling1996,
  Woitke2001, Jeong2003}, or the formation of clouds in brown dwarfs
and planetary atmospheres \citep{Lunine1986, Tsuji1996, Marley1999,
  Allard2001, Burrows2002, Tsuji2005, Marley2002, Helling2008}.


Since dust grains can easily de-couple from the gas phase (for
example rain-out in planetary atmospheres, gravitational settling in
protoplanetary discs, acceleration by radiation pressure in AGB star
winds), the condensed elements may be carried away, gather
in other places (like oceans) or re-evaporate in other places
(e.g.\ comet tails). These effects are very likely to cause most 
peculiar local element abundances, with large effects on gas 
phase abundances and spectroscopic appearance. Such effects
are rarely studied in detail \citep[see, however,][]{Booth2017}. 
Instead, researchers tend to rely on known sets
of element abundances, in particular the solar element abundances
\citep[e.g.][]{Asplund2009} and those derived from Earth rock
analysis.

Deviations of element abundances from these known sets have mostly
been treated by reducing all metal abundances by the same factor,
i.e.\ by changing the metallicity [M/H], \citep[e.g.][]{Helling2009,
  Witte2009, Crossfield2013, Hu2014, Morley2017} or by changing the
carbon-to-oxygen ratio \citep[e.g.][]{Madhusudhan2011,
  Madhusudhan2012, Helling2017, Eck2017}. \citet{Gaidos2000} was one
of the earliest works suggesting the importance of C/O in an exoplanet
context.  \citet{Molliere2015} have studied the impact of C/O on the
$(p,T)$-structure and emergent spectra of irradiated planets.
Planetary atmospheres with very different element abundances have been
studied by \citep{Mbarek2016, Mahapatra2017}, considering an
atmosphere made of evaporated rock.


In order to derive the composition of a gas in thermo-chemical
equilibrium, different approaches are used by different groups working
on stellar, brown dwarf and planetary atmospheres. Three major schools
are: (i) use of law of mass action and molecular equilibrium constants
based on Gibbs free energy data
\citep{Tsuji1973,Stock2008,Bilger2013}, (ii) use of law of mass action
and molecular equilibrium constants based on partition functions
\citep{Allard1995, Gustafsson2008, Barklem2016, Eck2017}, and (iii)
minimisation of the system Gibbs free energy \citep{Lodders2003,
  Ackerman2001, Miller2009, Mbarek2016, Blecic2016}.
\citet{Helling2008} provide a summary of approaches used in brown
dwarf and planetary atmospheres.

Approaches (i) and (ii) usually lead to a system of $N$ algebraic
equations for $N$ unknowns, which can be solved by any root-finding
algorithms, for example by the Newton-Raphson method. In contrast, the
Gibbs free energy minimisation technique (iii) requires special
numerical minimisation algorithms \citep{Spang1962}, for example the
Dantzig-simplex method, the Hessian-conjugate gradient method, or the
Lagrangian steepest-descent method \citep{Blecic2016}. The Gibbs free
energy minimisation approach can easily be generalised to include
equilibrium condensation, which is the reason why this approach
  was introduced originally. However, finding the root of a coupled
non-linear equation system $\vec{F}(\vec{x})=0$ can be done
numerically in much more efficient ways than finding the minimum of
$F=\sum F_i^2(\vec{x})\to$\,min, because the vector $\vec{F}$ has much
more information than the scalar $F$. Therefore, Gibbs free energy
minimisation codes tend to be considerably slower.

Beyond thermo-chemical equilibrium models for gas phase and
condensates, hybrid methods have been developed where the 
  condensation is treated time-dependently, but the concentrations of
gaseous molecules are calculated in thermo-chemical equilibrium
\citep[e.g.][]{Gail1984, Sedlmayr1995, Woitke2004, Woitke2006,
  Hofner2007, Hofner2016, Helling2013}, following the idea that dust
formation is by far the slowest process, causing the first deviations
from LTE. 

The probably most common approach used in the retrieval community,
however, is not to solve any equations at all, but to treat the
molecular concentrations as free parameters which are then adjusted to
match observations (see e.g.\ Table 1 in \citealt{Helling2008}). This
is partly done for reasons of feasibility, as the retrieval technique
requires to run tens of thousands of models, but also to allow the
method to retrieve non-equilibrium effects whereas enforcing
equilibrium would rather be considered as a
limitation. \citet{Benneke2015}, \citet{Line2016} and
\citet{Oreshenko2017} have shown that it is feasible to include
equilibrium chemistry in exoplanet retrieval methods.

This paper summarises in Sect.~\ref{s:cpe} the theoretical background
of our approach to compute the abundances of molecules and condensates
in thermo-chemical equilibrium.  Section~\ref{s:tcd} gives an overview
of the thermo-chemical data available for molecules and condensates.
We assess the level of (dis-)agreement between various data sources.
Section~\ref{s:ggchem} introduces our thermo-chemical equilibrium
code.  Section~\ref{s:results} present our results concerning the
spectroscopically most active molecules, a benchmark test, and the
condensation sequence of elements. We demonstrate that equilibrium
condensation leads to a significant increase of the C/O ratio in the
gas phase where phyllosilicates play a particular
role. Section~\ref{s:tung} re-addresses the question of the first
condensate in space, which is of particular interest for dust
formation in AGB star winds. Section~\ref{s:sc} contains our summary
and conclusions.



\section{Chemical and phase equilibrium}
\label{s:cpe}

\subsection{Molecular equilibrium}
\label{sec:eqgas}
Let us consider a molecule $\mol$ made of three elements A,B,C, where 
$a,b,c$ are the stoichiometric factors. Guldberg's law of mass action
\citep[e.g.][]{Berline1969} is given by
\begin{equation}
 \frac{p_\mol}{\pst} = 
     \left(\frac{p_{\rm A}}{\pst}\right)^a
     \left(\frac{p_{\rm B}}{\pst}\right)^b
     \left(\frac{p_{\rm C}}{\pst}\right)^c
     \exp\left(-\frac{\dG}{RT}\right)  \ ,
\end{equation}
where the $p_i = n_i kT$ are the partial pressures [dyn/cm$^2$], $n_i$
the particle densities [cm$^{-3}$] and $\pst$ is a standard pressure.
$\dG$ is the Gibbs free energy of formation [J/mol] of the molecule 
at standard pressure from neutral atoms at the same temperature
\begin{eqnarray}
  \dG &\!\!\!\!=\!\!\!\!& \GG(\mol,T) \nonumber\\
      &\!\!\!\!-\!\!\!\!& a\,\GG({\rm A},T) 
                        - b\,\GG({\rm B},T) 
                        - c\,\GG({\rm C},T) \ .
\end{eqnarray}
The equilibrium constants $k_p$ are introduced as
\begin{equation}
  p_\mol = k_p(\mol,T)\;p_{\rm A}^a\,p_{\rm B}^b\,p_{\rm C}^c \ ,
  \label{eq:pmol}
\end{equation} 
where
\begin{equation}
  k_p(\mol,T) = \big(\pst\big)^{1-a-b-c}
  \exp\left(-\frac{\dG}{RT}\right) \ .
  \label{eq:kp}
\end{equation} 
As we are using cgs-units in this paper, the $k_p$ have units $\rm
(dyn/cm^2)^{1-a-b-c}$. $T$ is the gas temperature [K], $R$ the
ideal gas constant [J/mol/K] and $k$ the Boltzmann constant [erg/K].

To determine the chemical composition of the gas, we
solve the element and charge conservation equations as
\begin{equation}
  \ek\,\nH = \sum_i s_{i,k}\,n_i  \ ,
  \label{eq:conserve}
\end{equation}
where $\ek$ are the element abundances normalised to
hydrogen ($\epsilon_{\rm H}\!=\!1$), $\nH$ is the total hydrogen nuclei
density $\nH=\sum_i s_{i,{\rm H}}\,n_i$. $n_i$ denote all gas particle
densities including free electrons, neutral and charged atoms, and
neutral and charged molecules. $s_{i,k}$ is the stoichiometric factor
of element $k$ in gas particle $i$. We assume charge neutrality by 
the inclusion of the charge as an additional element 'el' with zero abundance 
($\epsilon_{\rm el}\!=\!0$) where $s_{i,{\rm el}}\!=\!0$ for neutrals, 
$s_{i,{\rm el}}\!=\!+1$ for ions and $s_{i,{\rm el}}\!=\!-1$ for cations.
The free electron has $s_{{\rm el},{\rm el}}\!=\!-1$.
The gas density $\rho$\,[g/cm$^3$] is given by
\begin{equation}
  \rho = \sum_i m_i\,n_i = \nH \sum_k m_k\,\ek \ ,
  \label{eq:dens}
\end{equation}
where the second part of Eq.\,(\ref{eq:dens}) follows from
Eq.\,(\ref{eq:conserve}), and the total gas pressure is given by
\begin{equation}
  p = \sum_i n_i\,kT = n(\,\rho,\!T)\,kT = \frac{\rho\,kT}{\mu(\,\rho,\!T)}
  \ .
  \label{eq:ptot}
\end{equation}
$m_i$ are the gas particle masses and $m_k$ are the
masses of the elements. The total particle
density $n\!=\!\sum_i n_i$ and the mean molecular
weight $\mu=\sum_i m_i\,n_i / n$ are results of the
computations and hence depend on density and temperature.
The total hydrogen nuclei particle density $\nH$ is always
proportional to the mass density $\rho$, whereas $n$ is not,
causing the gas to deviate from an ideal gas.

After elimination of all molecular particle densities from
Eq.\,(\ref{eq:conserve}) by using the $k_p(T)$-data
(see Eq.\,\ref{eq:pmol}), Eq.\,(\ref{eq:conserve}) becomes a system of
non-linear algebraic equations with $K$ unknowns, namely the atomic
partial pressures and the electron partial pressure, for example
$\{p_k\,|\,k\!=\!{\rm H},...\,,\!{\rm W},{\rm el}\}$ where H is
hydrogen and W is tungsten. All results depend solely on $\nH$, $T$ and
$\ek$.

If a solution is requested for a given pressure and temperature, an
iteration is performed where the mean molecular weight $\mu$ is
initially guessed, Eqs.\,(\ref{eq:ptot}) and (\ref{eq:dens}) are used
to compute $\nH$ from $p$ and $T$, the equilibrium chemistry is solved
with $\nH$ and $T$, and finally $\mu$ is re-calculated. We find this
iteration to converge after 1-5 calls of the equilibrium chemistry.

For high temperatures, $T\!\ga\!1000$\,K, the solution is easy to find
numerically as all particle concentrations stay within $10^{\,-\,{\rm
    a\ few\ ten}}$. However, for $T\!\to\!500$\,K, the results become
increasingly more extreme, for example the electron concentration
approaches $10^{-30}$, and for $T\!\to\!100$\,K the problem turns into
a numerical nightmare with some $k_p\!>\!10^{\,+\,\rm several\ 1000}$ and
some atom and electron concentrations
$<\!10^{\,-\,\rm{several\ hundred}}$. We overcome these problems by
switching to quadruple precision arithmetics at low temperatures and
by applying an iterative procedure according to the hierarchical order
of elements to provide extremely good initial guesses of the unknowns
for the final Newton-Raphson iteration. We explain the details of the
numerical approach of the new {\sc GGchem}-code in App.~\ref{AppC}.

\subsection{Condensed phases}
\label{sec:eqdust}

The supersaturation ratio of a condensate $j$ is given by 
\begin{equation}
  \Sj = \frac{\pj}{\pvap_j(T)} \ ,
  \label{eq:super1}
\end{equation}
where $\pvap_j(T)$ is the vapour pressure
\begin{equation}
  \pvap_j(T) = \pst \exp\left(\frac
      {\GG(j[{\rm cond}],T)-\GG(j,T)}{RT}\right) \ .
  \label{eq:pvap}
\end{equation}
In phase equilibrium we have
\begin{equation}
 \Sj \ \left\{\begin{array}{ll}
      < 1 & \mbox{condensate is unstable and not present,}\\
      = 1 & \mbox{condensate is stable and present,}\\
                               \end{array}\right.
  \label{eq:phaseEQ}
\end{equation}
whereas $\Sj\!>\!1$ (supersaturation) is not possible in phase
equilibrium.  Equation~(\ref{eq:phaseEQ}) states that the partial
pressure of any molecule is limited in phase equilibrium, if a
corresponding condensed phase exists. Once $\pj$ exceeds that
threshold (its vapour pressure), that molecule condenses\footnote{The
  actual process of condensation is not considered in the frame of
  chemical and phase equilibrium.} and leaves the gas phase, which
reduces a few gas phase element abundances $\ek$.  This reduction
causes not only $\pj$ to drop, but also affects on all other partial
pressures in the gas (usually reduces them). This process continues
until phase equilibrium is established, see appendix~B in
\citep{Helling2008}.

\begin{figure*}
\vspace*{-2mm}
\begin{tabular}{cccc}
\hspace*{-4mm}
\includegraphics[width=55mm,height=56mm,trim=10 0 0 0, clip]{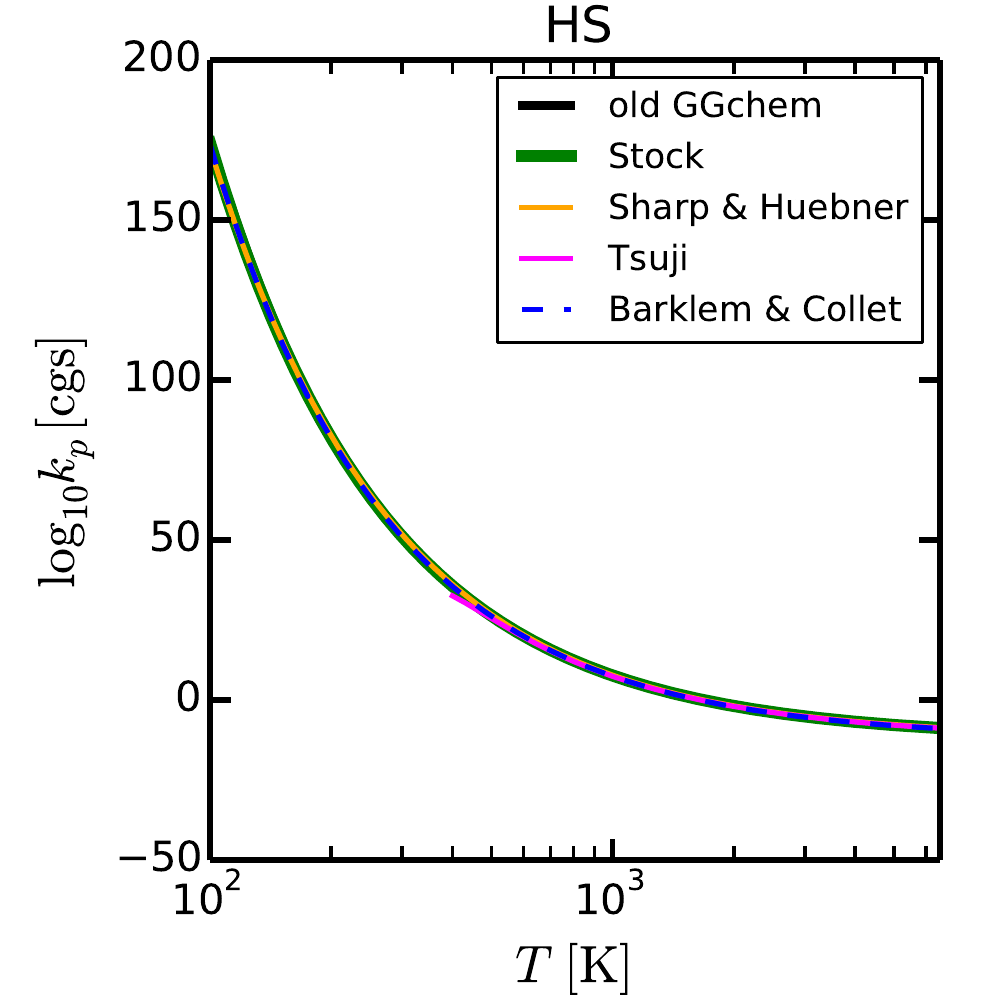} &
\hspace*{-7.5mm}
\includegraphics[width=41mm,height=44mm,trim=10 -14 2 0, clip]{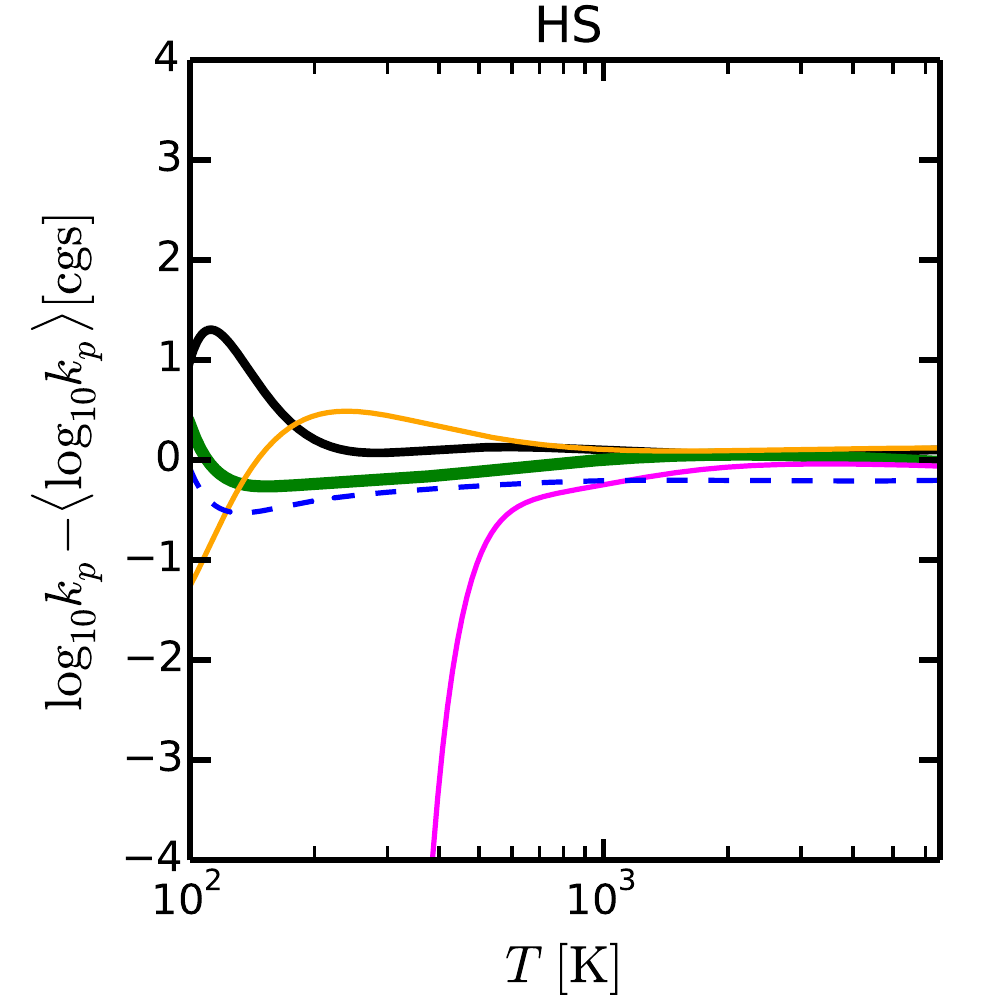} &
\hspace*{-6mm}
\includegraphics[width=55mm,height=56mm,trim=5 0 2 0, clip]{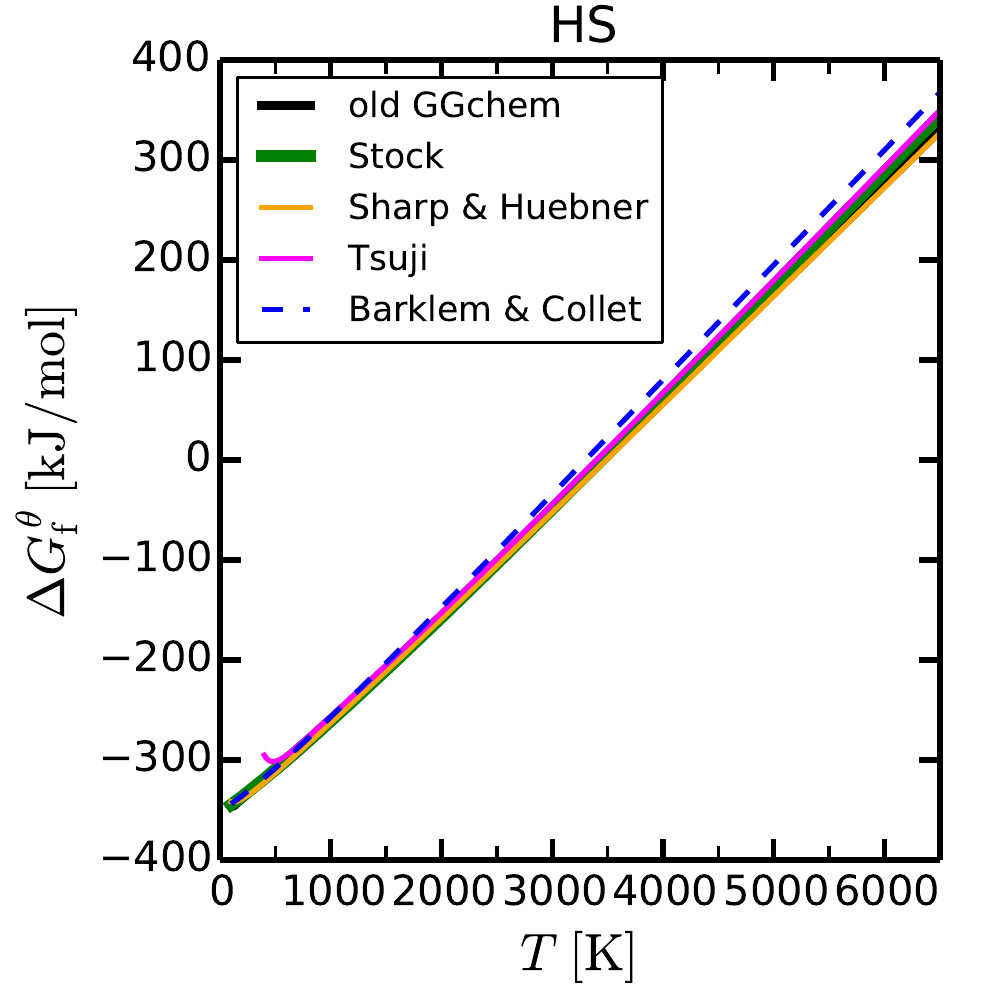} &
\hspace*{-7mm}
\includegraphics[width=41mm,height=44mm,trim=3 -14 0 0, clip]{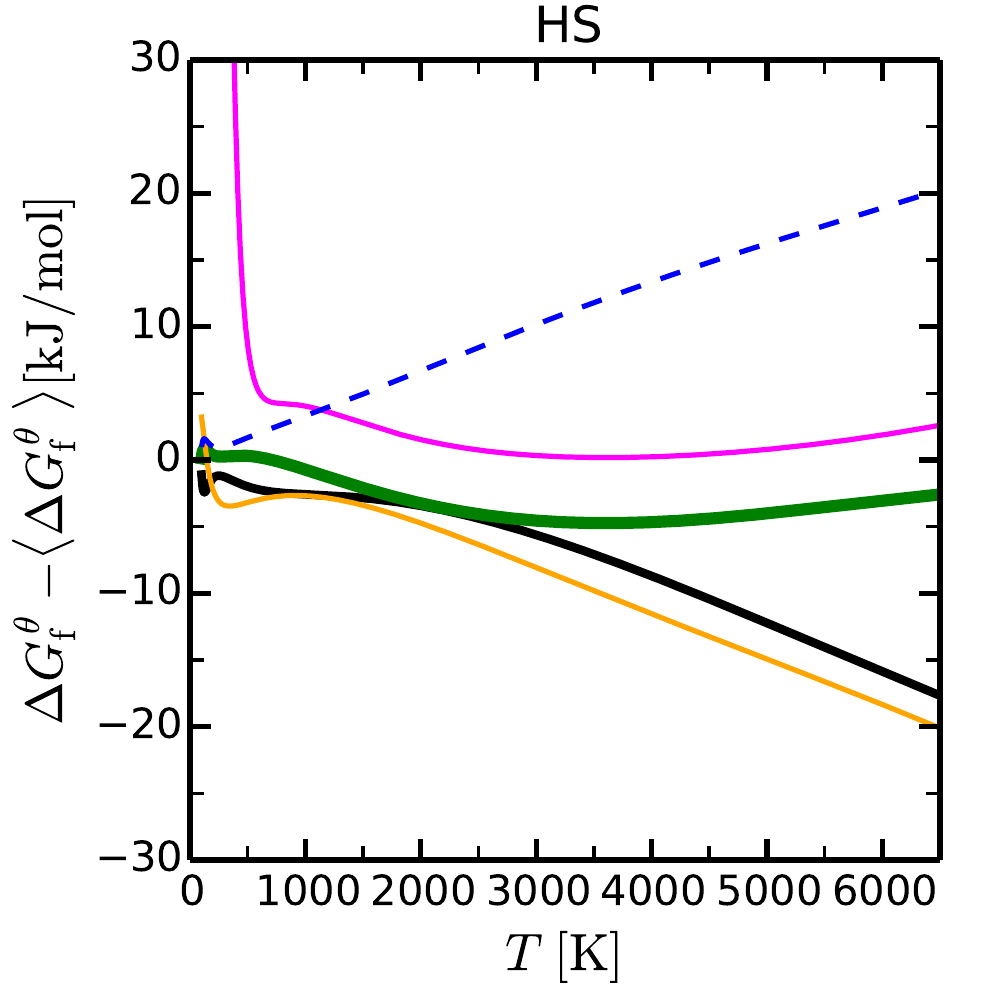} \\
\hspace*{-4mm}
\includegraphics[width=55mm,height=56mm,trim=10 0 0 0, clip]{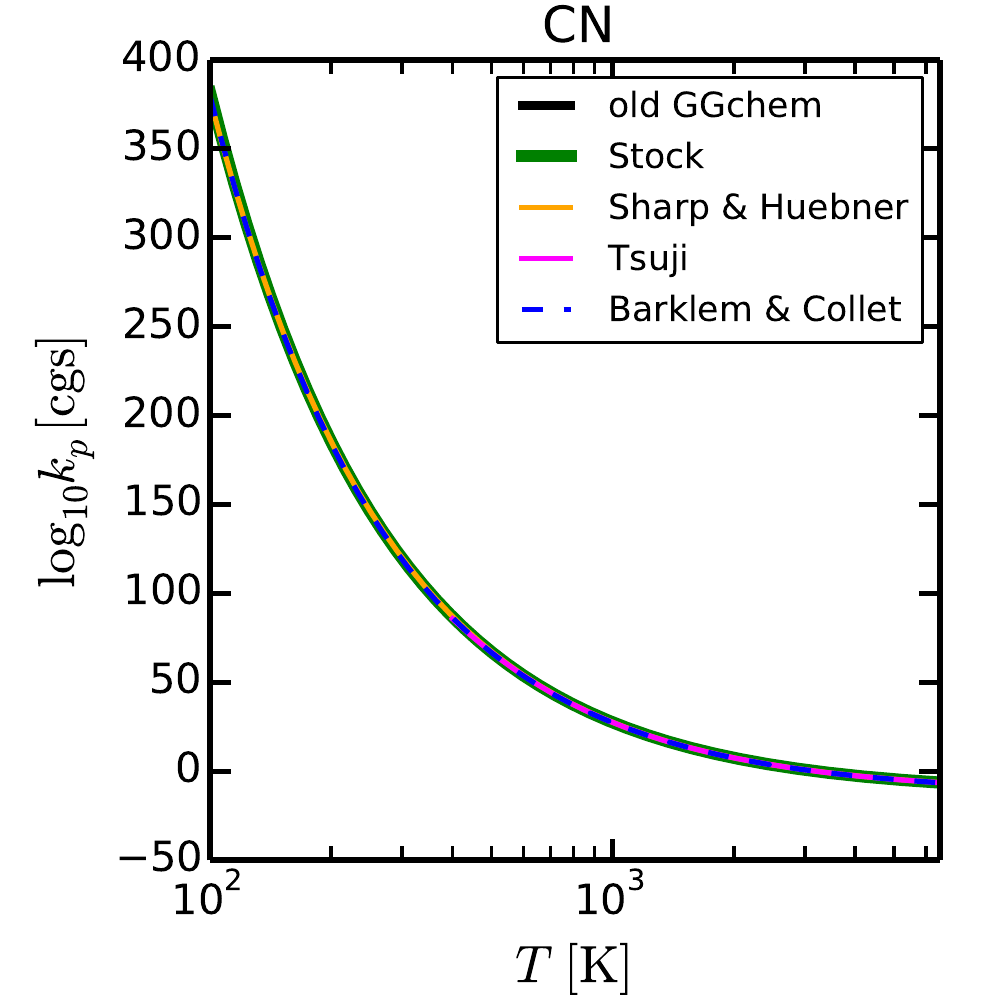} &
\hspace*{-7.5mm}
\includegraphics[width=41mm,height=44mm,trim=10 -14 2 0, clip]{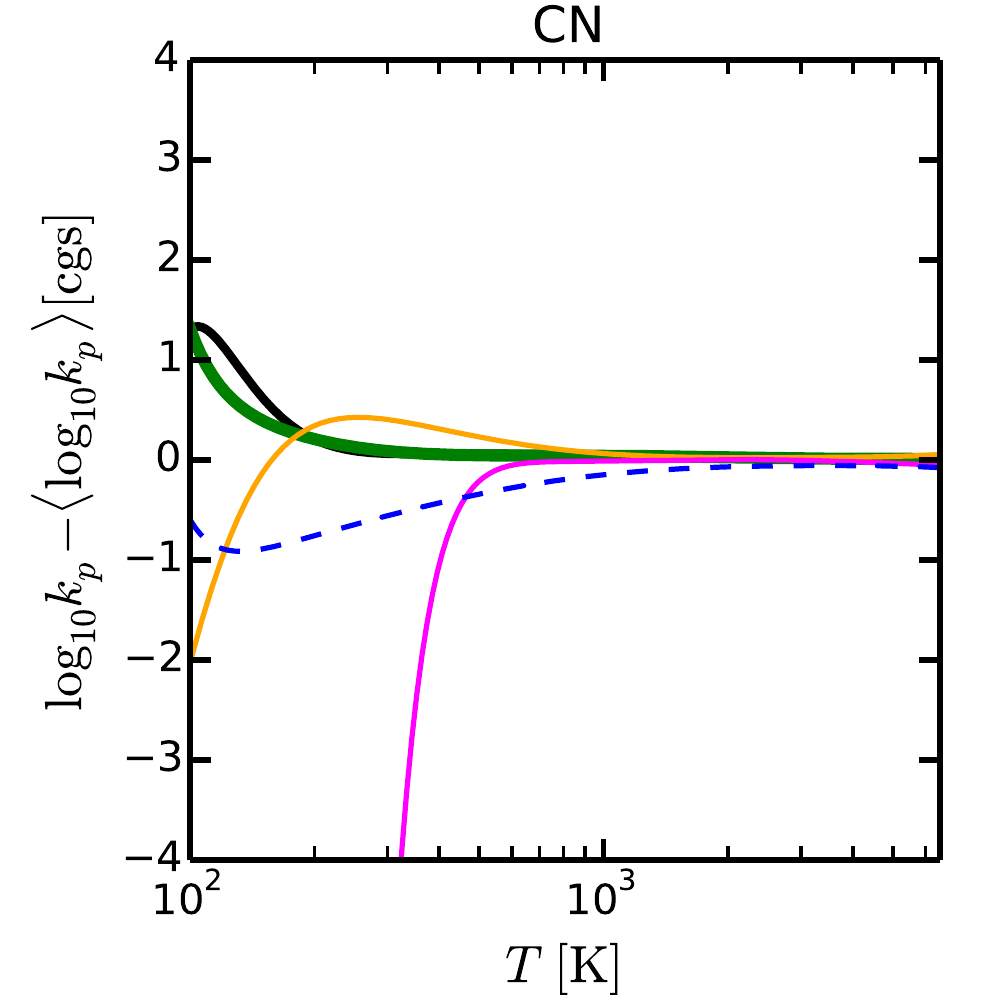} &
\hspace*{-6mm}
\includegraphics[width=55mm,height=56mm,trim=5 0 2 0, clip]{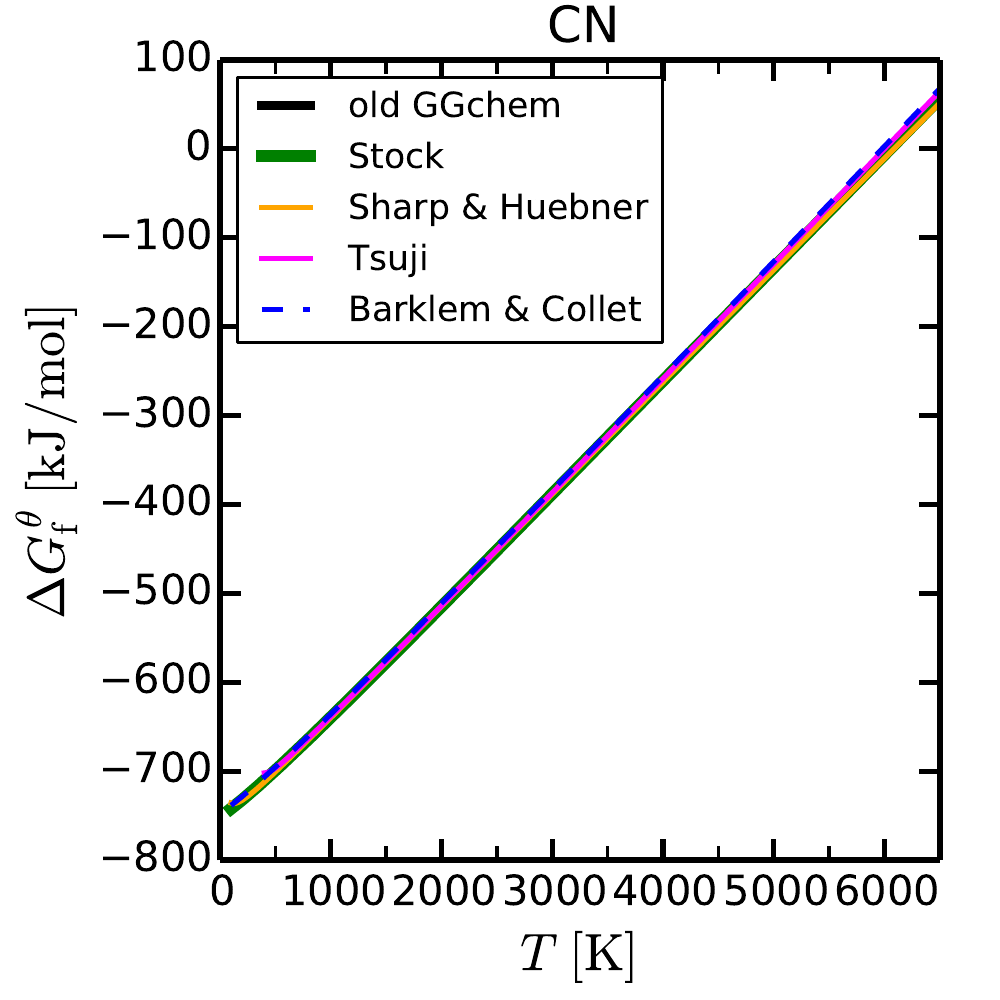} &
\hspace*{-7mm}
\includegraphics[width=41mm,height=44mm,trim=3 -14 0 0, clip]{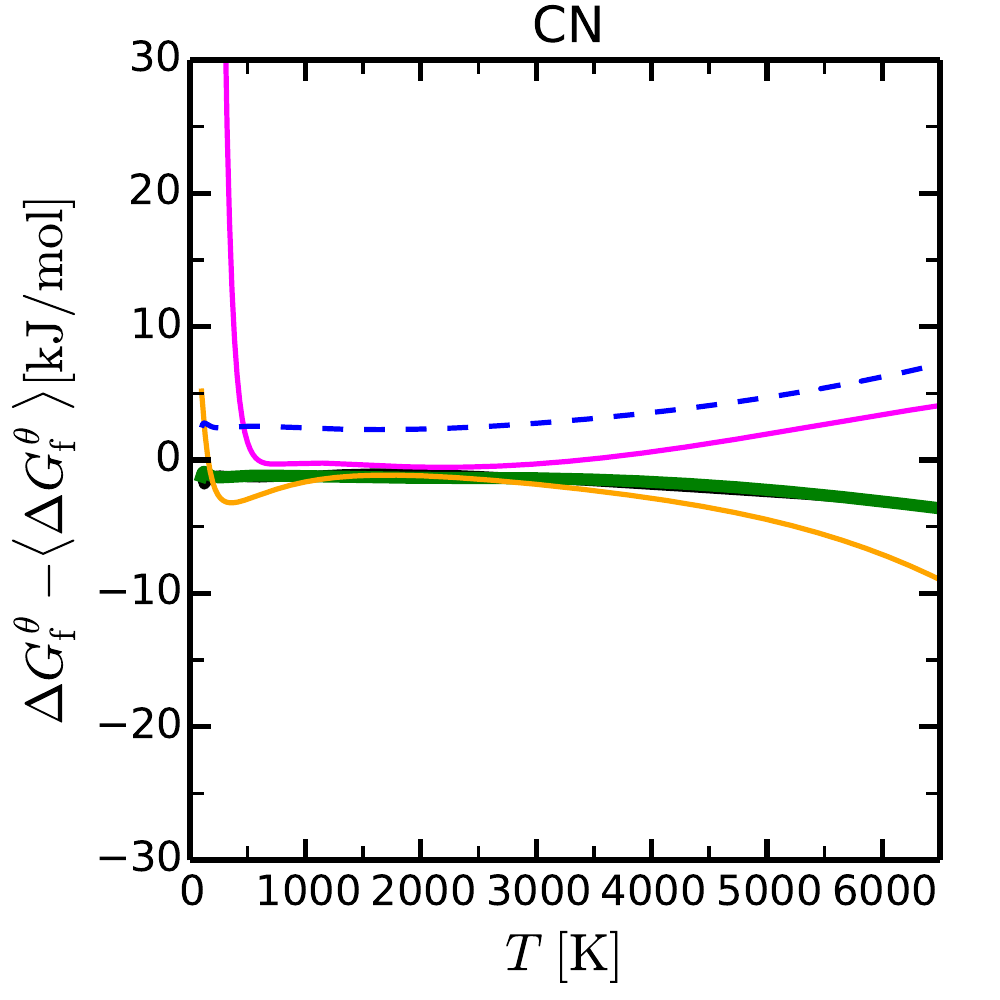} 
\\[-2mm]
\end{tabular}
\caption{Comparison of $k_p(T)$ data for molecules HS (mercapto) and
  CN (cyanogen) used by several groups.  The second column of smaller
  plots shows the deviations of $\log_{\rm 10}k_p$ from the mean
  values (computed without the Tsuji data). On the right side, the
  data has been converted to $\dG$\,[kJ/mol] (see Eq.\,\ref{eq:conv}),
  before computing the deviations from the mean values. The large
  deviations at small temperatures for the \citep{Tsuji1973} and
  \citep{Sharp1990} data are due to extrapolation errors beyond the
  valid fit-range.  HS is classified as belonging to the group {\sl
    ``data disagrees''}, and CN to {\sl ``data disagrees at low
    temperatures''} in our comparison catalogue \citep{Worters2017}
  since even without the Tsuji data, there are relatively large
  differences to the new \citep{Barklem2016} data.}
\label{fig:kps}
\vspace*{-2mm}
\end{figure*}

For condensates which have no corresponding molecule in
the gas phase (which is true for most minerals),
Eq.\,(\ref{eq:super1}) is not applicable. However, we can consider the
fictive nominal molecule of that condensate, for example $j\!=\!\mol$
with unknown $\GG(\mol,T)$, and apply Eqs.\,(\ref{eq:super1}) and
(\ref{eq:pvap}). Using also Eqs.\,(\ref{eq:pmol}) and (\ref{eq:kp}), 
we find the generalised supersaturation ratio to be
\begin{equation}
  S_\mol = \left(\frac{p_{\rm A}}{\pst}\right)^a
          \left(\frac{p_{\rm B}}{\pst}\right)^b
          \left(\frac{p_{\rm C}}{\pst}\right)^c
      \exp\left(-\frac{\dG}{RT}\right)
\label{eq:genS}
\end{equation}
where
\begin{eqnarray}
  \dG &\!\!\!\!=\!\!\!\!& \GG(\mol[{\rm cond}],T) \nonumber\\
      &\!\!\!\!-\!\!\!\!& a\,\GG({\rm A},T) - b\,\GG({\rm B},T) - c\,\GG({\rm C},T)
  \label{eq:dGsolid}
\end{eqnarray}
is the Gibbs free energy of formation of the condensed phase $\mol$ at
standard pressure from atoms in the gas phase at temperature $T$. The
unknown Gibbs free energy $\GG(\mol,T)$ of the fictive molecule
cancels out.

The addition of Eq.~(\ref{eq:phaseEQ}) to chemical equilibrium
models\linebreak ($\to$ {\sl equilibrium condensation} or {\sl phase
  equilibrium} models) leads to a considerable complication of the
mathematical structure of the problem to solve. The case
differentiation in Eq.~(\ref{eq:phaseEQ}) means that we do not know
a-priori which condensates to consider as only those which finally
result to be present contribute to the number of equations to be
solved (in form of $\Sj\!=\!1$), whereas the supersaturation ratios of
all other condensates do not matter as long as they stay
$<1$. Clearly, at very high temperatures, where no condensates
  are stable at all, the problem falls back to the pure gas phase
equilibrium discussed in the previous section.

Our numerical method in {\sc GGchem} is based on a Newton-Raphson
iteration with nested calls of the equilibrium chemistry with reduced
gas phase element abundances $\ek$ due to condensation, see
App.~\ref{AppD}. This reduction becomes extremely large at low
temperatures -- up to hundreds of orders of magnitudes.  The
additional unknowns to be solved for are the reduced gas phase element
abundances as affected by the condensation of the selected solid and
liquid phases. This selection may change during the iteration. Albeit
being somewhat sophisticated, we find that this method is fast
and very accurate down to 100\,K when using quadruple precision
arithmetics.

\section{Thermo-chemical data}
\label{s:tcd}

\subsection{Molecular equilibrium constants}
\label{sec:kp}

The $k_p$-data of the various ions and molecules must usually be
fitted as function of temperature in some way, before they can be used
in models. An alternative approach was recently proposed by
\citet{Blecic2016}, who use all available NIST-JANAF $\dG$ data-points
directly by means of internal spline-fits.  Various fit-functions have
been proposed as summarised below. From Eq.\,(\ref{eq:kp}) we have the
general relation between $\dG$ and $k_p$ as
\begin{equation}
  \ln k_p = (1-n)\ln\pst - \frac{\dG}{RT}  \ .
  \label{eq:conv}
\end{equation}
where $n=a+b+c$ is the sum of stoichiometric coefficients in the
molecule, for example $n\!=\!3$ for H$_2^-$ and $n\!=\!1$ for H$_2^+$.

\citet{Tsuji1973} has used the following fit-function
\begin{equation}
  \log_{10} k_p^{\rm Tsu}(T) 
  = -a_0 - a_1\theta - a_2\theta^2 - a_3\theta^3 - a_4\theta^4
\end{equation}
for 335 molecules, where $\theta=5040/T$ and $a_i$ are the fit
coefficients. All $k_p$ formulae listed here are valid in cgs units.

\citet{Gail1986} have used
\begin{equation}
  \ln k_p^{\rm Gail}(T) 
  = a_0 + a_1\theta + a_2\theta^2 + a_3\theta^3 + a_4\theta^4
\end{equation}
which is the functional form that was used in the old {\sc GGchem}
code until recently
\citep[e.g.][]{Helling2006,Bilger2013,Helling2017}. The old data
collection had 205 molecules.\linebreak  
\indent\citet{Sharp1990} have used
\begin{equation}
  \ln k_p^{\rm S\&H} = (1-n)\ln\pst
     -\frac{a_0/T + a_1 + a_2\,T + a_3\,T^2 + a_4\,T^3}
           {R_{\rm cal}T} \ ,
\end{equation}
fitting directly the $\dG$\,[cal/mol] of 184 molecules. Here, the
standard pressure is $\pst\!=\!1\,$atm and $R_{\rm
  cal}\!=\!1.987\,$cal/mol/K.  

A new approach was presented by \citet{Stock2008}
\begin{equation}
  \ln k_p^{\rm St} = (1-n)\ln\pst
      + \left(\frac{a_0}{T} + a_1\,\ln T + a_2 + a_3\,T +
      a_4\,T^2\right)
  \label{eq:Stock}
\end{equation}
who fitted the dimensionless quantity $-\dG/(RT)$ for 924 molecules,
to be used with $\pst\!=\!1\,$bar. This functional form has a very
smooth and stable behaviour towards low temperatures. It also gives
automatically more weight to the data points at low temperatures, so
we have adopted this functional form and most of their data in the new
{\sc GGchem} code.

\citet{Barklem2016} have recently published partition functions for
291 diatomic, and charged diatomic molecules, from which $k_p$ is
derived, but without providing a fit formula. Note that their data is
in SI units, and the definition of their $k_p$ is different from ours,
in particular for the charged molecules. We have done all necessary
conversions (see App.~\ref{AppA}) and fitted the Barklem \& Collet
data with a Stock-function.

Figure~\ref{fig:kps} compares the fitted data for two example
molecules found in all 5 datasets. The deviations are as large as
10\,kJ/mol for HS and about 5\,kJ/mol for CN. At $T\!=\!300\,$K, an
uncertainty of $\pm$\,6\,kJ/mol (0.06\,eV) translates into an
uncertainty in $k_p$ of about one order of magnitude, same for
$\pm$\,60\,kJ/mol (0.6\,eV) at $T\!=\!3000\,$K.  The complete
comparison catalogue \citep{Worters2017} includes 2782 individual
datasets for 1155 molecules, where similar plots as shown in
Figure~\ref{fig:kps} can be found for all molecules for which we found
at least two different data sources.  We observe astonishingly large
deviations in some cases. The reasons for these deviations are not
entirely clear, but some factors could be
\begin{itemize}
  \item different primary data sources,
  \item extrapolation errors,
  \item the ``art of fitting'', and human errors.
\end{itemize} 
\citet{Gail1986}, \citet{Sharp1990} and \citet{Stock2008} have all
based their fits on the JANAF database, but using different
editions. Gail \& Sedlmayr have used the 2$^{\rm ed}$ edition
\citep{Stull1971}, Sharp \& Huebner have used the 3$^{\rm rd}$ edition
\citep{Chase1982,Chase1986}, and Stock has used the 4$^{\rm th}$
edition \citep{Chase1998}. All three works include some Tsuji-data for
molecules missing in JANAF. \citet{Stock2008} has refitted the
Tsuji-data for $T\!>\!1000\,$K with his fit-function that is more
reliable concerning extrapolation towards lower temperatures. Despite
using more or less the same original data sources, some molecules show
considerable deviations even at medium temperatures
\citep[see][]{Worters2017}. We conclude that errors of about half an
order of magnitude for $k_p$ can easily occur, especially at low
temperatures, depending on which fit-formula is chosen and how the fit
was obtained in detail.  \citet{Tsuji1973} and \citet{Barklem2016}
have used other, independent data sources.  Some obvious differences
occur in particular when different molecular dissociation energies are
assumed.  We also find a few outliers, where one data source is off by
many orders of magnitude with respect to all others. Although we
cannot exclude human errors on our side, our comparisons have been
made in a highly automated way, so these outliers are quite puzzling.

\citet{Tsuji1973} and \citet{Sharp1990} recommend to use their fits
only for $T\!\ga\!1000\,$K, where the Sharp\,\&\,Huebner fit-function
behaves somewhat more robustly in extrapolations towards low
temperatures. This is because $\dG(T)$ is an approximately linear
function of $T$ as shown in Fig.~\ref{fig:kps}. The fit term
$\dG\!\sim\!1/T$ causes the extrapolation problems at low $T$ which is
avoided in the \citep{Stock2008}-fits, which are therefore much more
reliable for extrapolations towards low $T$.  The old {\sc GGchem}
fits \citep{Gail1986} should not be applied for $T\!\la\!500\,$K, and
behave in a similar way as the Tsuji fits, namely with strong
extrapolation errors toward low temperatures, because of a
$\dG\!\sim\!1/T^3$ term.  

Appendix~\ref{AppA} highlights a few more details of
this comparison and Table~\ref{tab:kp} lists our selected set of
$k_p(T)$ functions for 568 molecules composed of 24 elements, which
can be safely applied to temperatures $100-6000$\,K. We could use more
elements and molecules, but the data listed in Table~\ref{tab:kp} have
been carefully checked and benchmarked against the TEA code
\citep{Blecic2016}, see Sect.~\ref{sec:bench}.

\subsection{Condensed phase thermo-chemical data}
\label{sec:dGcond}

\begin{figure}
\vspace*{1mm}\hspace*{-3mm}
\includegraphics[width=91mm,trim=15 280 95 5, clip]{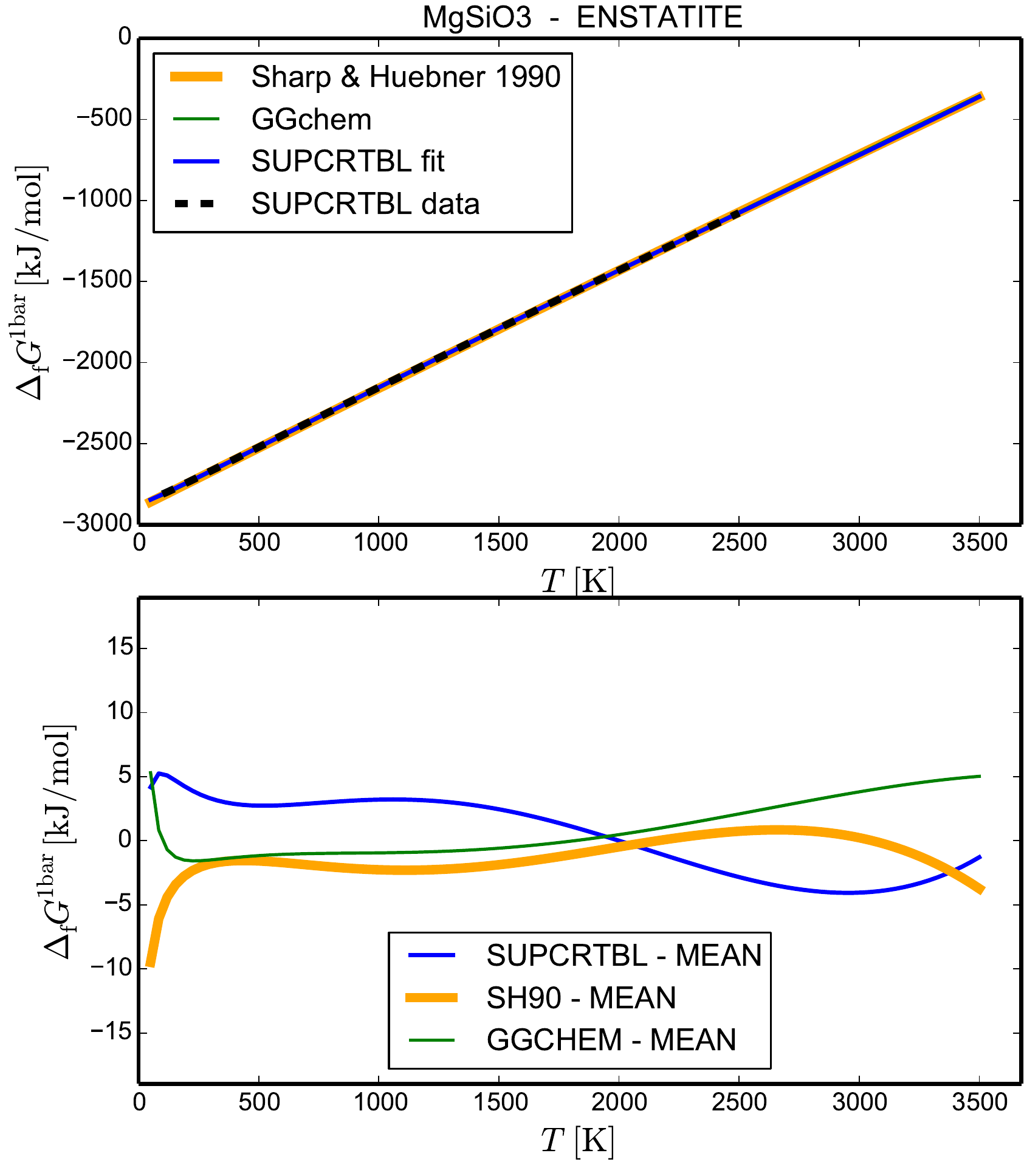}\\[-6mm] 
\caption{Comparison of condensed phase Gibbs free energy data from
  \citep[][orange]{Sharp1990}, fits to the NIST-JANAF database done by
  the authors of this paper for {\sc GGchem} (green), and fits to the
  SUPCRTBL data \citep[][blue]{Zimmer2016}. The SUPCRTBL data points have
  been generated for temperatures 100\,K to 2500\,K according to the
  equations and parameters given in Zimmer et al.\ (black dashed
  curve). The lower plot shows the deviations [kJ/mol] with respect to
  the mean of all three data sources.}
\label{fig:dGfits}
\vspace*{-1mm}
\end{figure}

We have used two main sources for thermo-chemical data of condensed
phases in this paper, the NIST-JANAF database
\citep{Chase1982,Chase1986}\footnote{\url{http://kinetics.nist.gov/janaf}},
and the geophysical SUPCRTBL database \citep{Zimmer2016,Johnson1992}.
When extracting Gibbs free energy data $\dG(T)$ and/or vapour pressure
data $\pvap(T)$ from these sources, it is important to understand the
reference states of the elements with respect to which the data is
presented, see more details in App.~\ref{AppB}. NIST-JANAF uses
temperature-dependent reference states.  When subtracting the
$\dGNIST$ data, the reference state Gibbs free energies cancel out, so
we can use
\begin{eqnarray}
  &&\dG = \dGNIST(\mol[{\rm cond}],T) \\
  && - a\,\dGNIST({\rm A},T) 
     - b\,\dGNIST({\rm B},T) 
     - c\,\dGNIST({\rm C},T) \nonumber\ ,
\end{eqnarray}
or, respectively,
\begin{eqnarray}
  \pvap_\mol(T) = \pst \exp\bigg(\frac{1}{RT}\Big[
     && \hspace*{-6mm} \dGNIST(\mol[{\rm cond}],T) \\[-2mm]
     && \hspace*{-8mm} -\dGNIST(\mol,T)\Big]\bigg)\nonumber
\end{eqnarray}
with $\pst\!=\!1\,$bar. The $\dGNIST(T)$ data is available in the
$7^{\rm th}$ column of the NIST-JANAF files at temperature points
mostly separated by 100\,K. The Gibbs free energy data can be directly
compared to \citet{Sharp1990}, and the vapour pressure data can be
compared to e.g.\ \citet{Yaws1999}, \citet{Weast1971} and
\citet{Ackerman2001}.

The geophysical SUPCRTBL database provides condensed phase Gibbs free
energy data $\dGSU(T,p)$ via formulas with tabulated coefficients
which partly have a direct physical meaning. In this database, the
Gibbs free energy of formation is defined in a different way, namely
with respect to the standard states of the elements.  These standard
states are the most stable forms of the pure elements at standard
pressure $\pst\!=\!1\,$bar and room temperature
$\Tref\!=\!298.15$\,K. These are atomic gases for the noble gases He,
Ne, Ar, ...; diatomic gases $\rm\frac{1}{2}H_2$, $\rm\frac{1}{2}N_2$,
$\rm\frac{1}{2}O_2$, $\rm\frac{1}{2}F_2$, $\rm\frac{1}{2}Cl_2$ for H,
N, O, F, Cl; liquids for Br and Hg; and crystalline solids for all
other elements. In order to use these data for our purpose, we need to
subtract the Gibbs free energy of formation of the free atoms at
temperature $T$ from the standard states at reference temperature
$\Tref$. The respective thermo-chemical data of the atoms are not
available in SUPCRTBL, however we can use the NIST-JANAF database to
provide this link, see further details in App.~\ref{AppB}.  To fit
the resulting $\dG(T)$ functions, we have generated 100 $\dG$ data
points at log-equidistant temperature points between 100\,K and
2500\,K.

While fitting the data points from either the NIST-JANAF or the
SUPCRTBL database by smooth functions, we have tried various options
and carefully selected those fits which perform best concerning
precision and possible interpolation and extrapolation issues, see
App.~\ref{AppB}. If the corresponding molecule of a condensate is
known to exist as a free molecule in the gas, it is generally
advantageous to use $\pvap(T)$ fits, because the values to be fitted
only correspond to a few 100\,kJ/mol. However, if no such molecule
exists (which is true for most minerals) we need to fit the $\dG$
directly, with values of several 1000\,kJ/mol, where the relative
precision of the fits becomes crucial.  A particular challenge is
to properly fit pairs of solid and liquid phase data, which are quite
similar to each other, yet their intersection point should match the
melting point, see Fig.~\ref{fig:pvapfit}.  Table~\ref{tab:SOLID}
lists our fit formula and fit coefficients for 97 condensates, mostly
fitted by us to the NIST-JANAF database, with a few additions from
other sources. The fits done by ourselves generally have precisions
$<\!1\,$kJ/mol, much better in most cases.  The melting points are
matched to within $<20\,$K (or $<\!1\%$), much better in most cases,
see \ref{tab:liquid}.  Table~\ref{tab:SUdata} lists the $\dG$ data for
160 condensates extracted from the SUPCRTBL database.

Figure~\ref{fig:dGfits} shows a comparison between $\dG$-fits for
MgSiO$_3$ {\sl(enstatite)} obtained from three sources. The reader is
welcome to study an auxiliary document \citep{Woitke2017} which
contains additional figures like Fig.~\ref{fig:dGfits} for all 121
condensates extracted from the SUPCRTBL database. The different data
generally agree well with each other, within about 5\,kJ/mol, although
there are a few exceptions, remarkably CaAl$_2$Si$_2$O$_8$
{\sl(anorthite)} which seems off by about 100\,kJ/mol in
\citep{Sharp1990}.  The Sharp \& Huebner fits can safely be
extrapolated to about 500\,K, but may have extrapolation artefacts
below.  All new fits to the NIST-JANAF data can be used from about
3500\,K down to 100\,K. This does require some extrapolations for a
small number of condensates where no such NIST-JANAF data exist, but
we have carefully checked that our fits continue smoothly and at least
do not produce any artefacts.  The behaviour of the various fit
functions is similar to the molecular $k_p$-fits as described in
Sect.~\ref{sec:kp}, the Stock-fit function is most robust towards low
temperatures, but slightly less accurate at high temperatures.

\section{The GGchem code}
\label{s:ggchem}

{\sc GGchem} is an abbreviation for the German word {\sl
  ``Gleich-Gewichts-Chemie''} which means equilibrium chemistry. The
code has been originally developed by H.-P.~Gail in the 1970s
\citep[e.g.][]{Gail1986}. Further works on the code have been done by
C.~Dominik, Ch.~Helling and P.~Woitke in Berlin until 2005. For this
paper, we have completely re-written the code and given it a modern
FORTRAN-90 code architecture. We have updated the thermo-chemical
input data (Sect.~\ref{sec:kp}), and have added a quadruple precision
version and a new pre-iteration scheme (App.~\ref{AppC}) to
improve code stability and range of applicability down to 100\,K. We
have added the equilibrium condensation part (App.~\ref{AppD}),
and have added and revised the condensed phase thermo-chemical data
(Sect.~\ref{sec:dGcond}).  The code is publicly available including
all thermo-chemical data\footnote{\url{https://github.com/pw31/GGchem}}.

\begin{figure*}
\vspace*{-3mm}
\hspace*{1mm}
\resizebox{185mm}{!}{
\begin{tabular}{ccc}
\hspace*{-6mm}
\includegraphics[height=48mm,trim=10 64 0 10, clip]{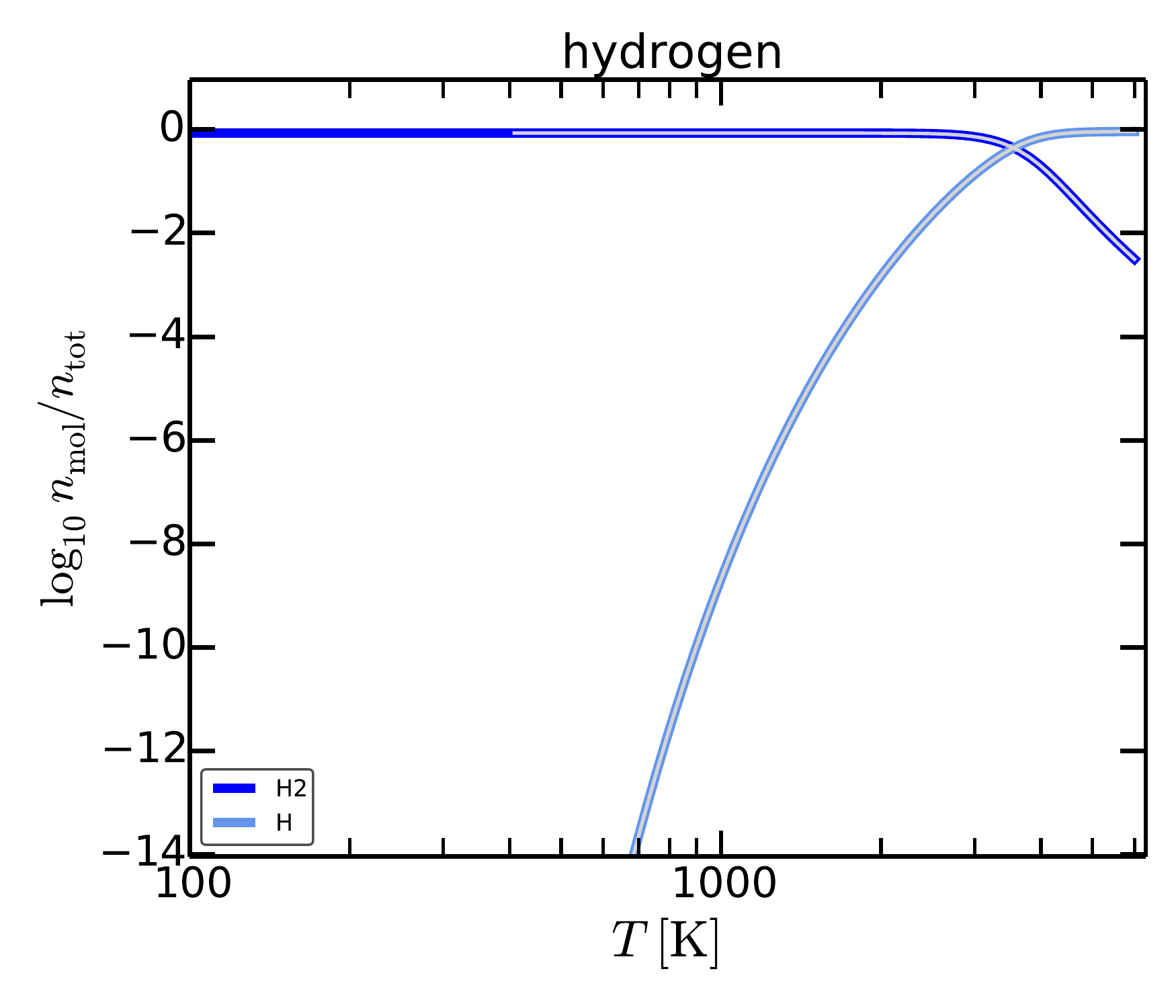} &
\hspace*{-6mm}
\includegraphics[height=48mm,trim=43 64 0 10, clip]{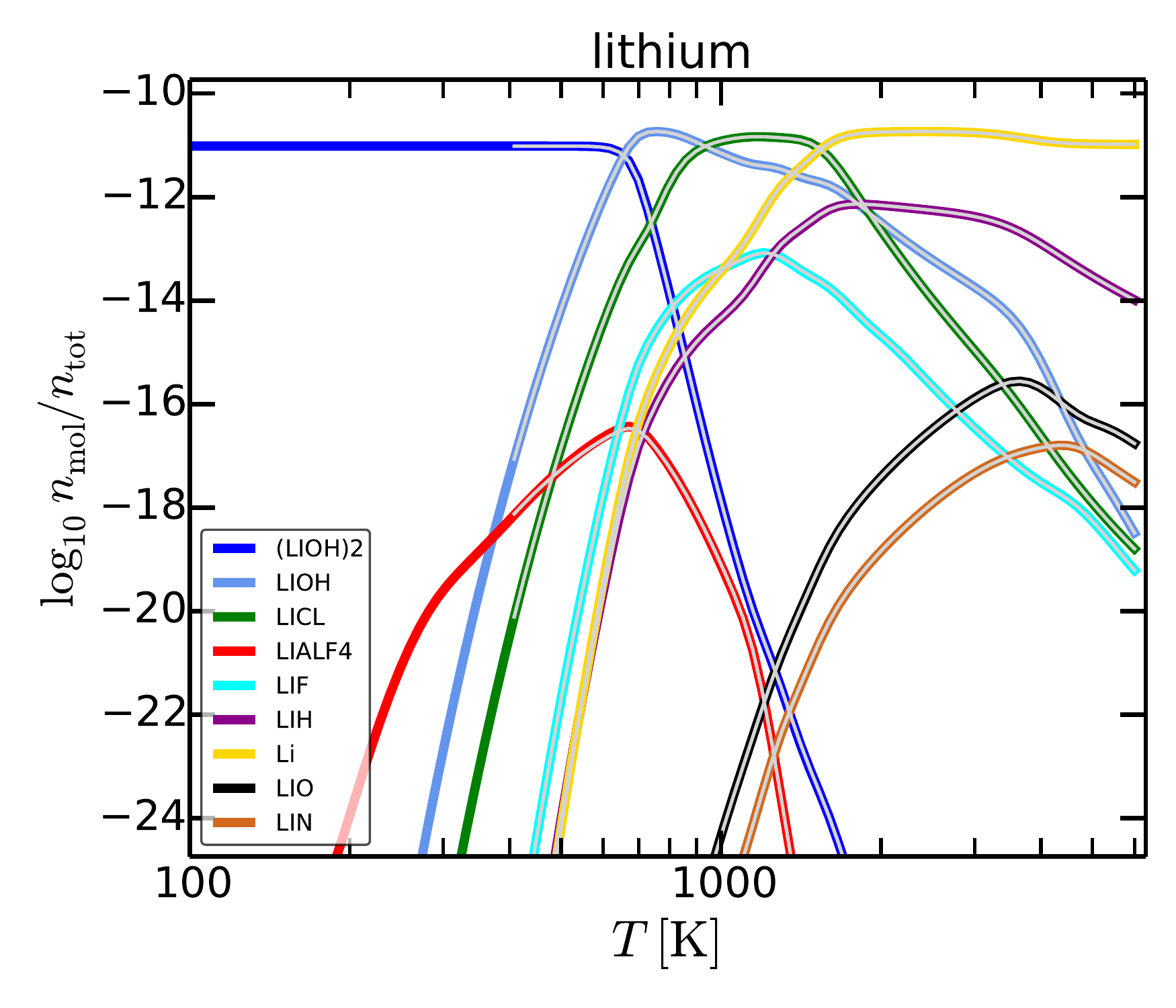}  &
\hspace*{-6mm}
\includegraphics[height=48mm,trim=43 64 0 10, clip]{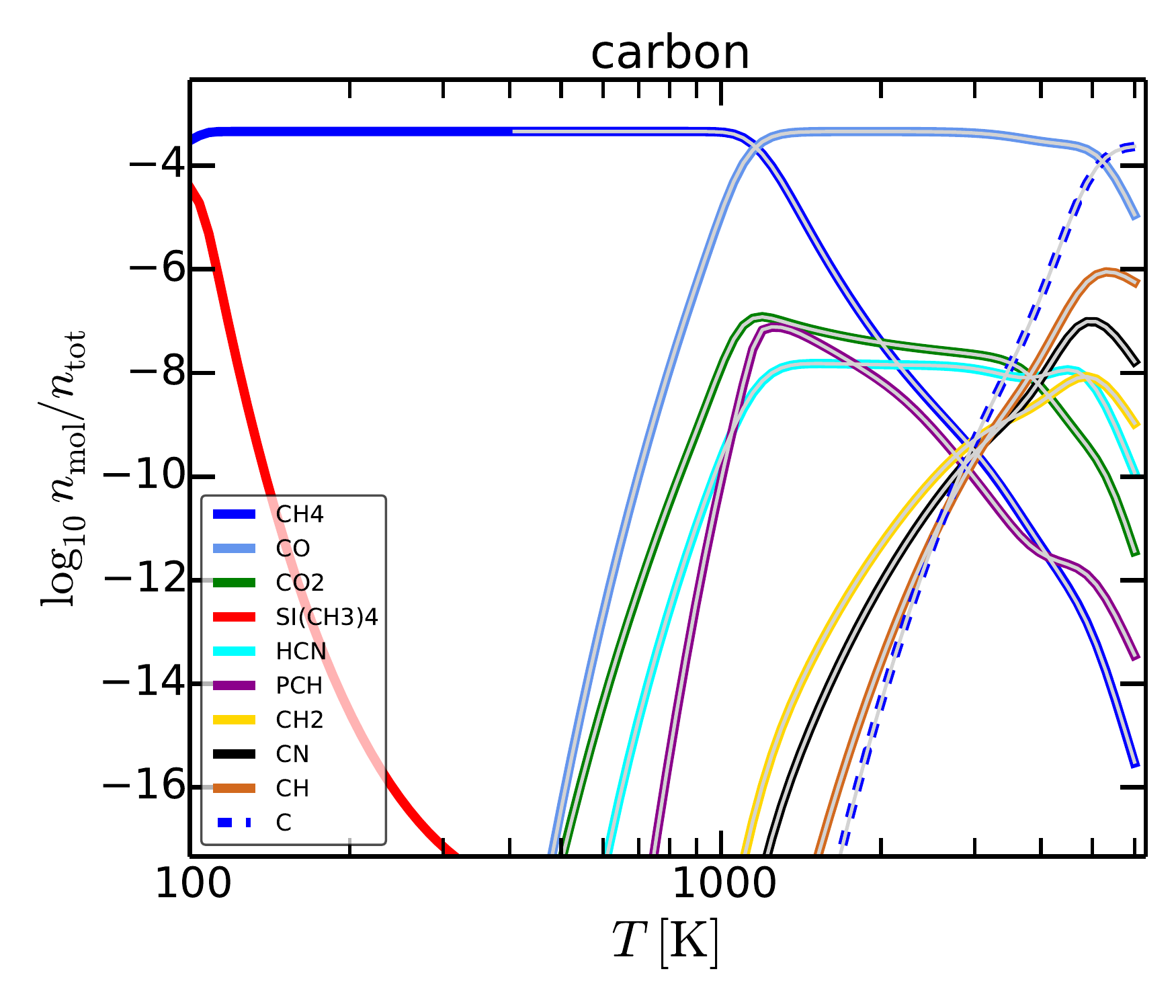} 
\\[-1mm]
\hspace*{-6mm}
\includegraphics[height=48mm,trim=10 64 0 10, clip]{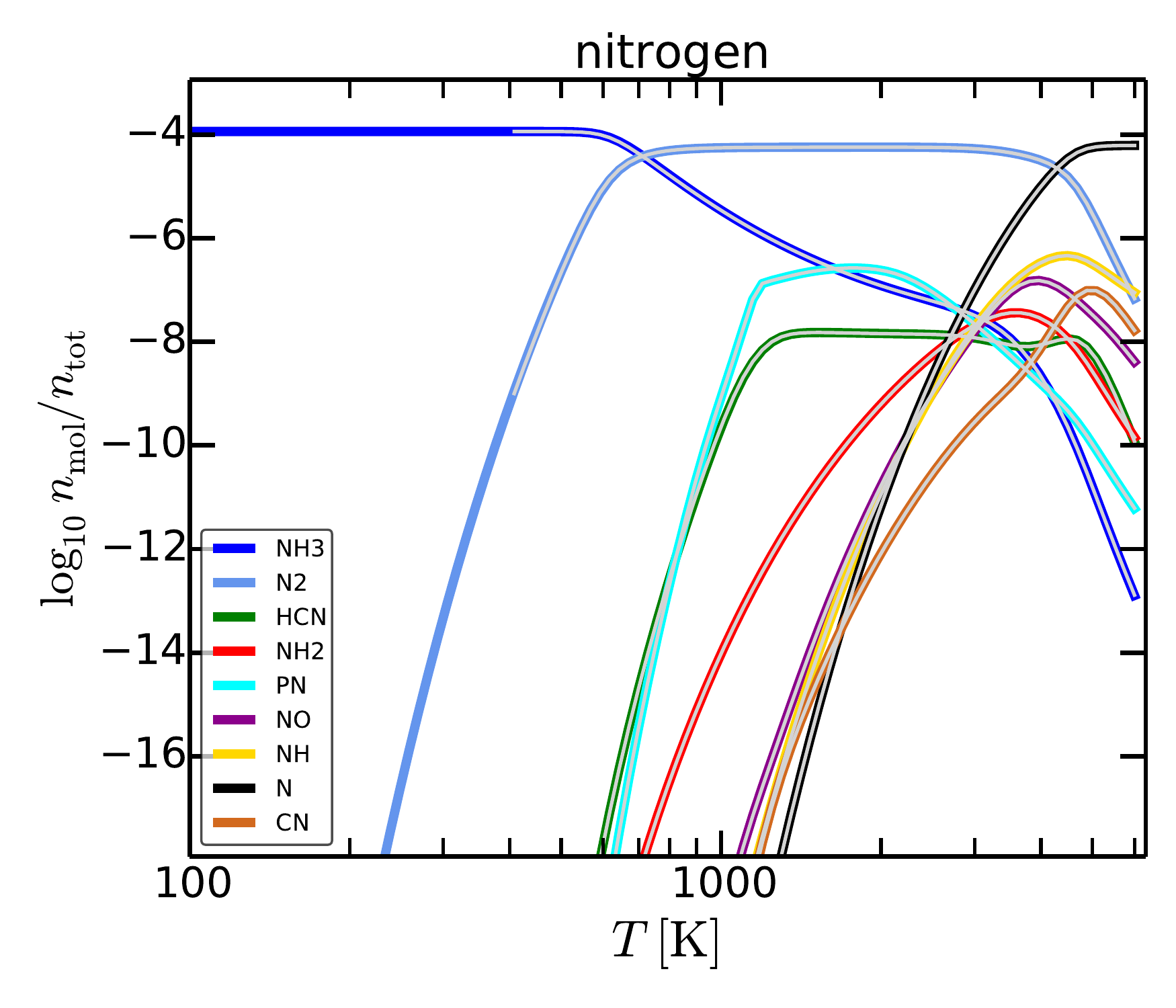} &
\hspace*{-6mm}
\includegraphics[height=48mm,trim=43 64 0 10, clip]{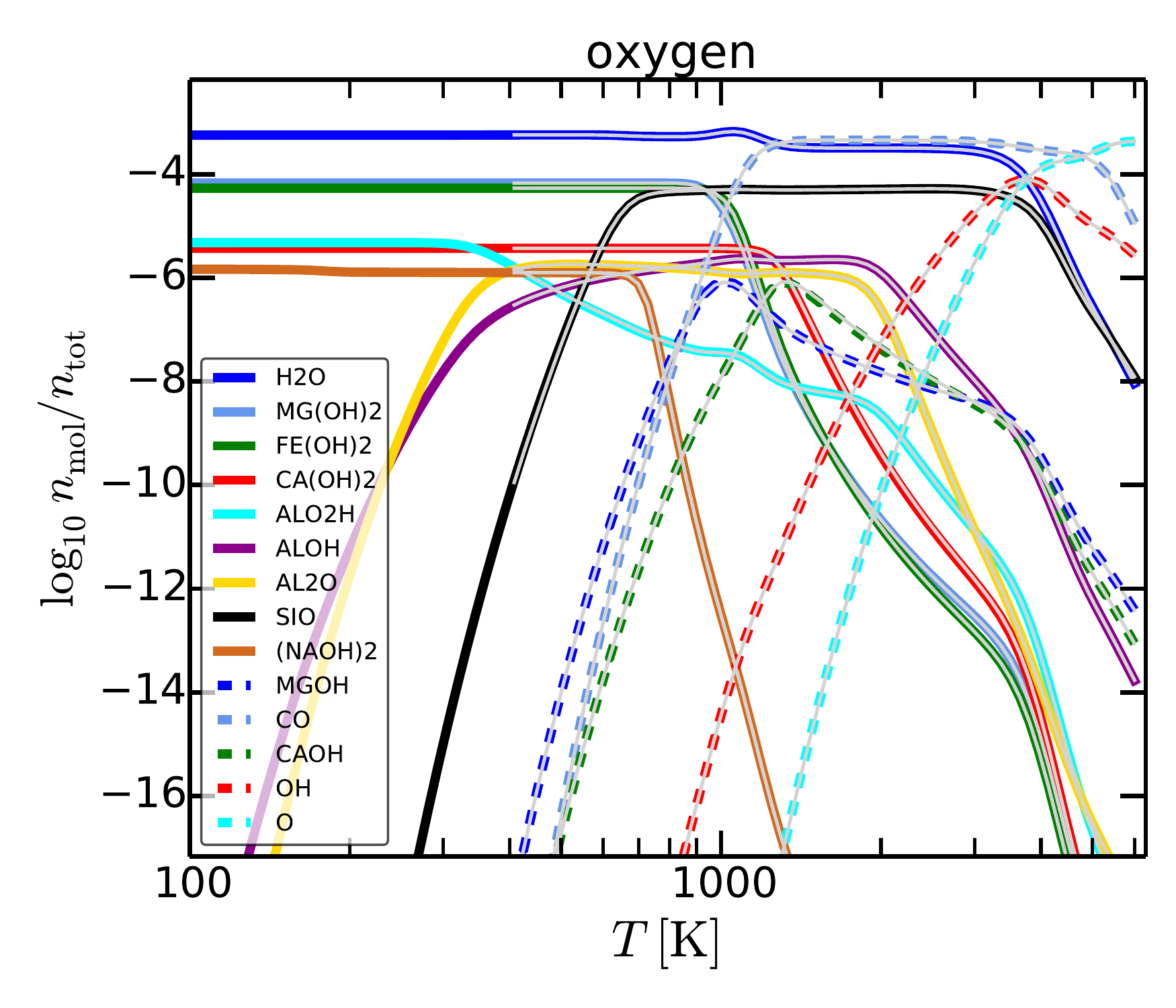}   &
\hspace*{-6mm}
\includegraphics[height=48mm,trim=43 64 0 10, clip]{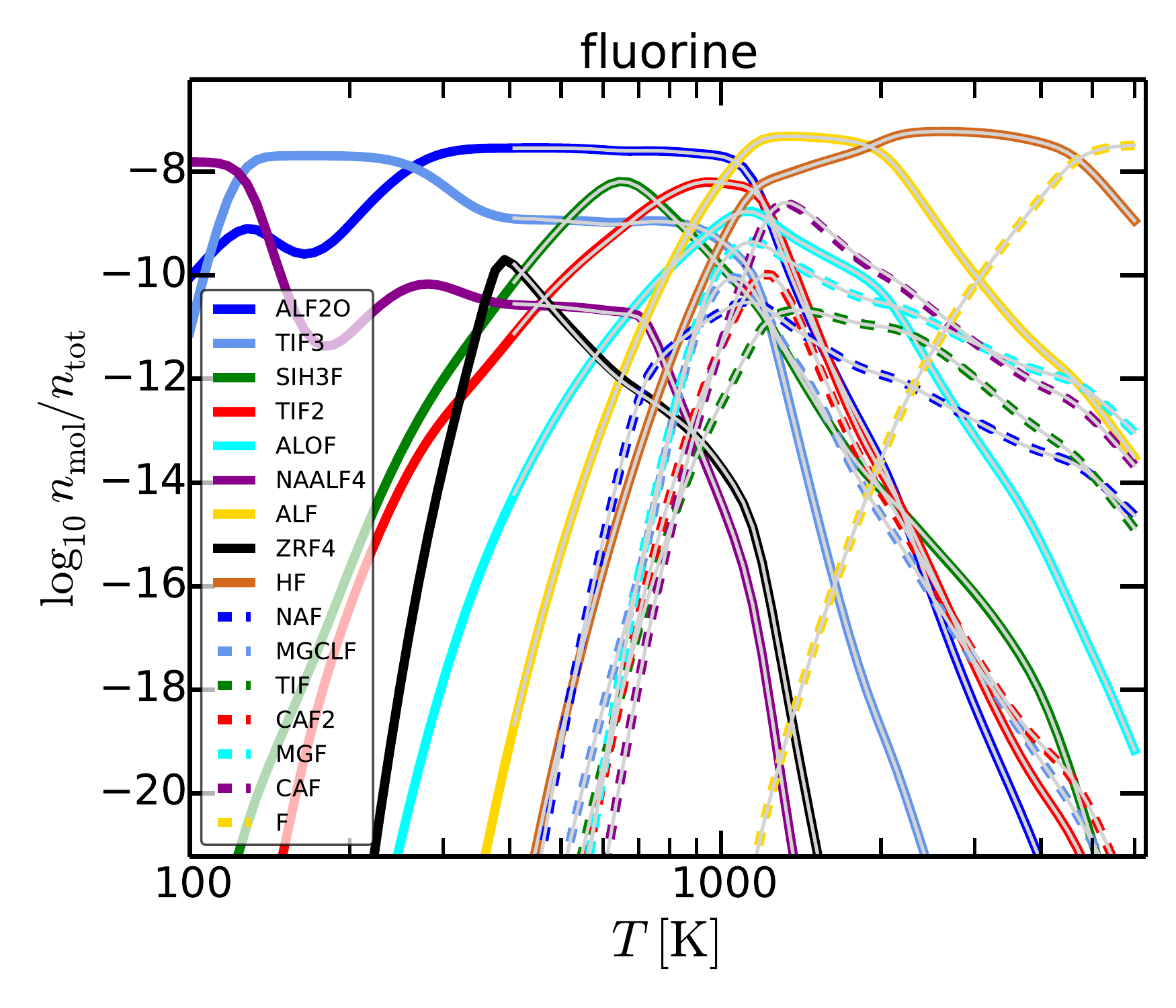} 
\\[-1mm]
\hspace*{-6mm}
\includegraphics[height=54mm,trim=10 17 0 10, clip]{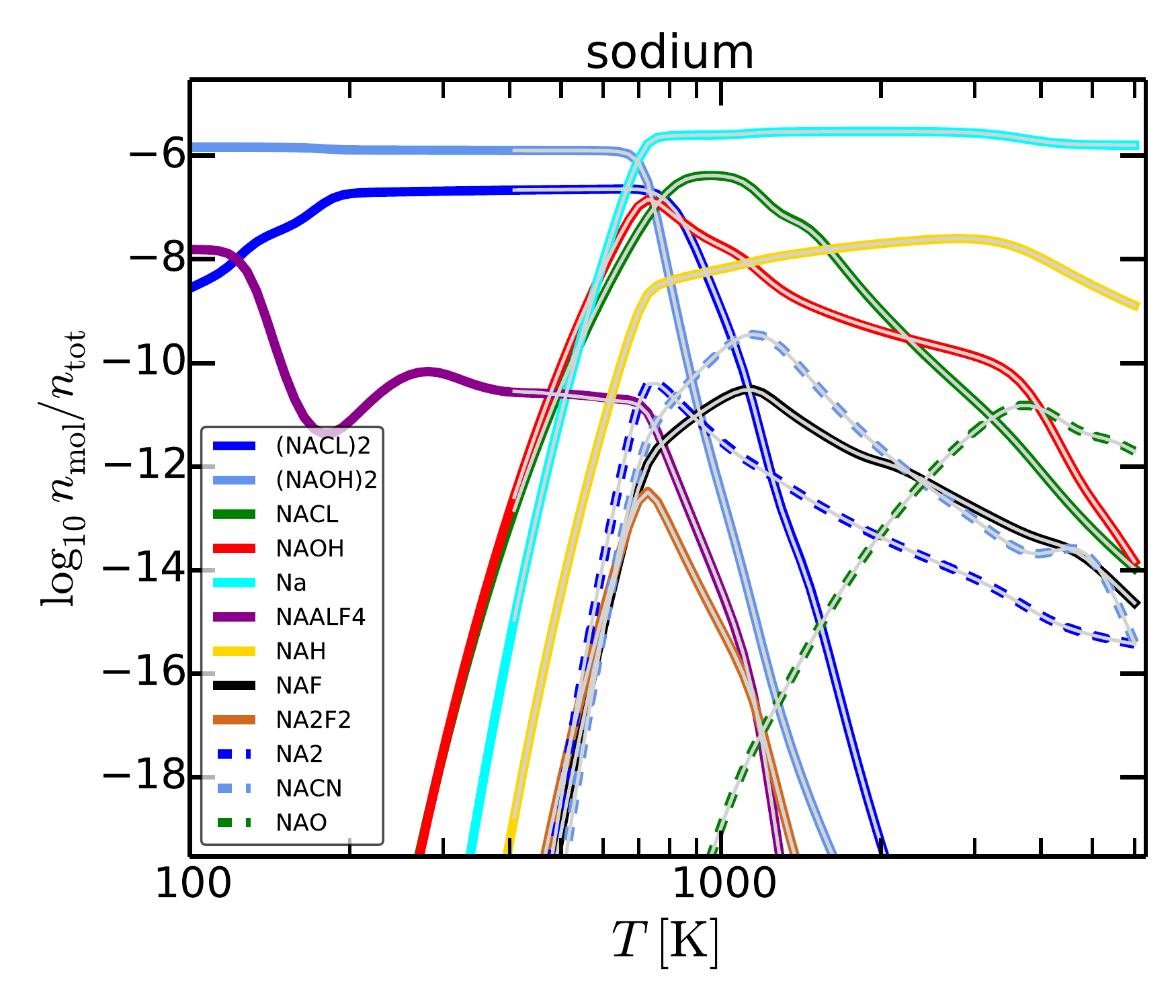} &
\hspace*{-6mm}
\includegraphics[height=54mm,trim=43 17 0 10, clip]{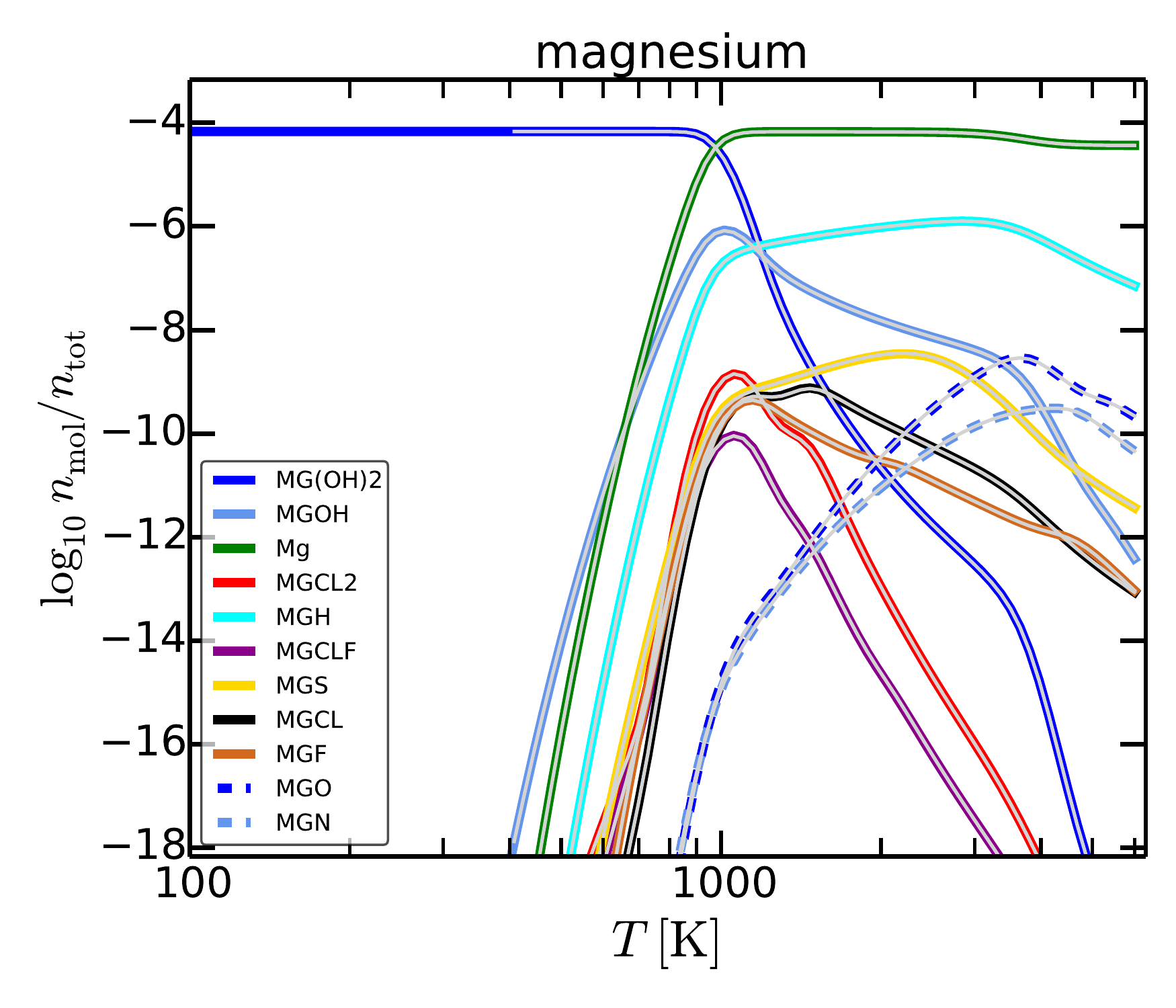} &
\hspace*{-6mm}
\includegraphics[height=54mm,trim=43 17 0 10, clip]{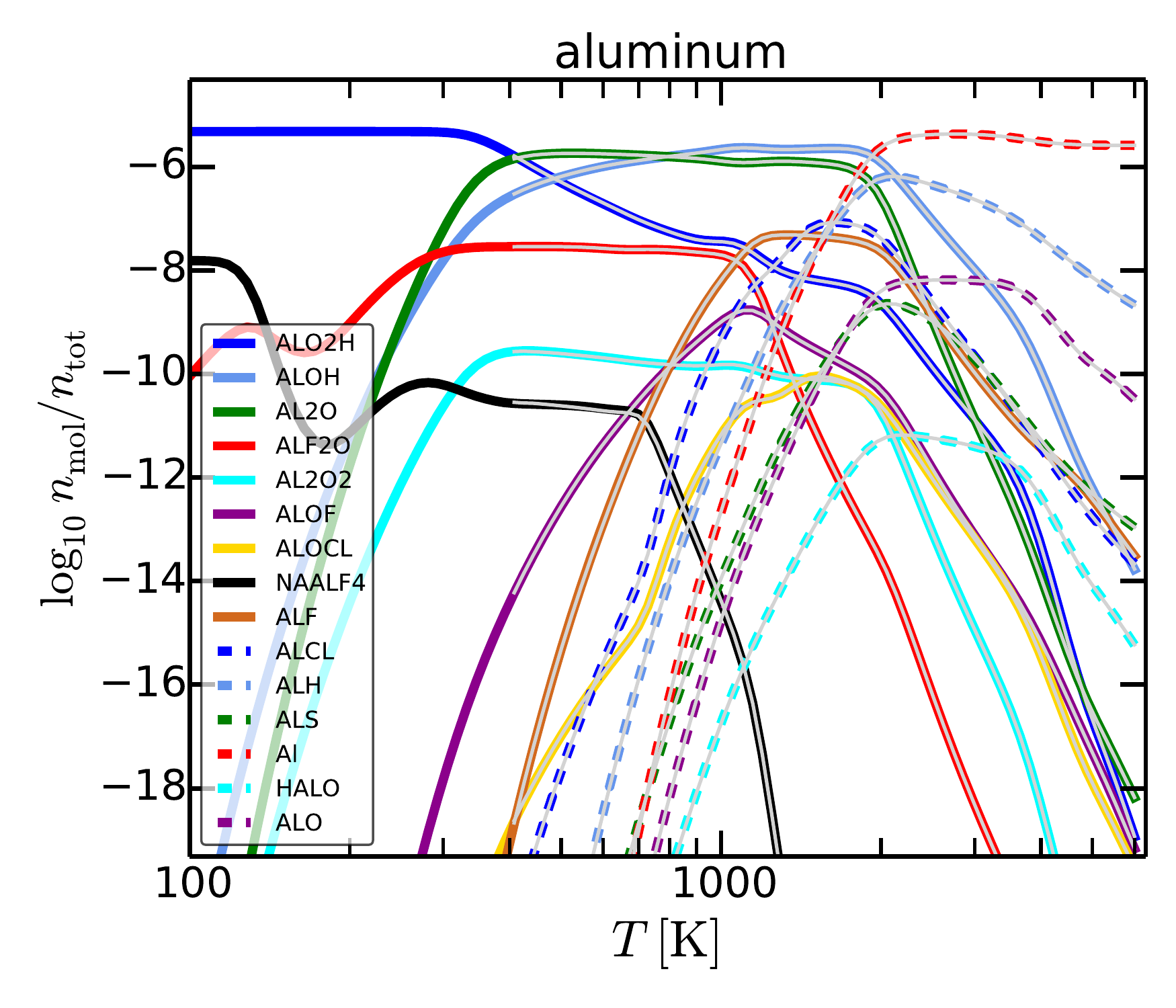} 
\\[-2mm]
\end{tabular}}
\caption{Benchmark test against TEA \citep{Blecic2016} at
  constant $p\!=\!1$\,bar and $T\!=\!100\,...\,6000$\,K, showing
  mixing ratios $n_i/n$. The coloured full and dashed curves are the
  {\sc GGchem} results. The TEA results are overplotted with narrow
  grey lines. Only the most important molecules are shown, i.e.\ those
  which reach a certain threshold concentration, depending on
  element. The TEA results stop at 400\,K as TEA has
  difficulties converging at lower temperatures, even using 2000
  iterations. The following molecules with notable concentrations
  are missing in TEA: SiH$_2$, SiH$_3$; CaH; TiS, TiH, TiN, TiC; CrH,
  CrS; MnH, MnS, MnCl, MnF, MnO; FeH; NiH, NiF,
  NiO, and the following molecules have been omitted by us in the TEA
  model to improve its convergence: Si(CH$_3$)$_4$, SiCH$_3$Cl$_3$ and
  Ni(CO)$_4$. See continuation on next page.}
\label{fig:TEA}
\end{figure*}
\addtocounter{figure}{-1}
\begin{figure*}
\vspace*{-3mm}
\begin{tabular}{ccc}
\hspace*{-6mm}
\includegraphics[height=48mm,trim=10 64 0 10, clip]{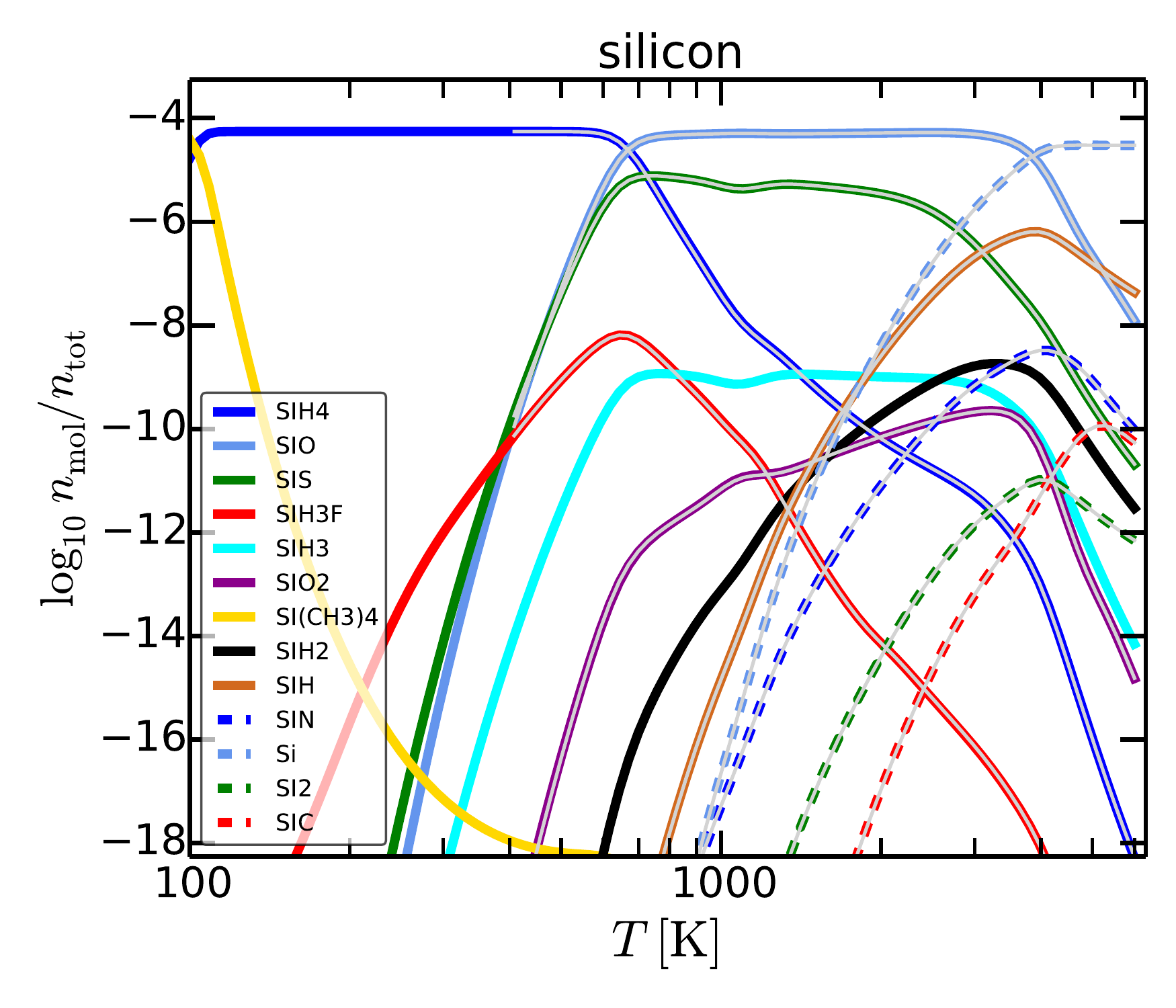} &
\hspace*{-6mm}
\includegraphics[height=48mm,trim=43 64 0 10, clip]{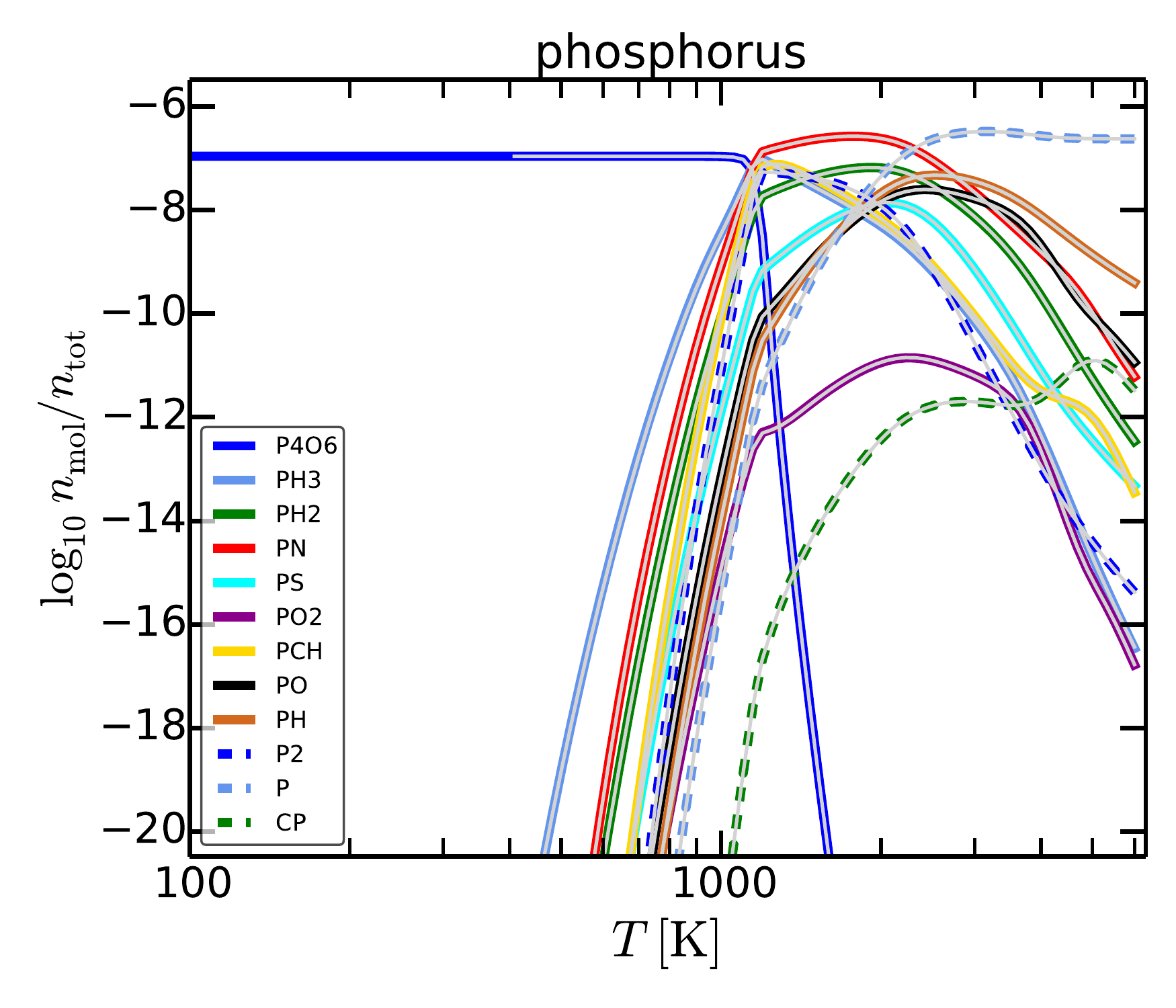} &
\hspace*{-6mm}
\includegraphics[height=48mm,trim=43 64 0 10, clip]{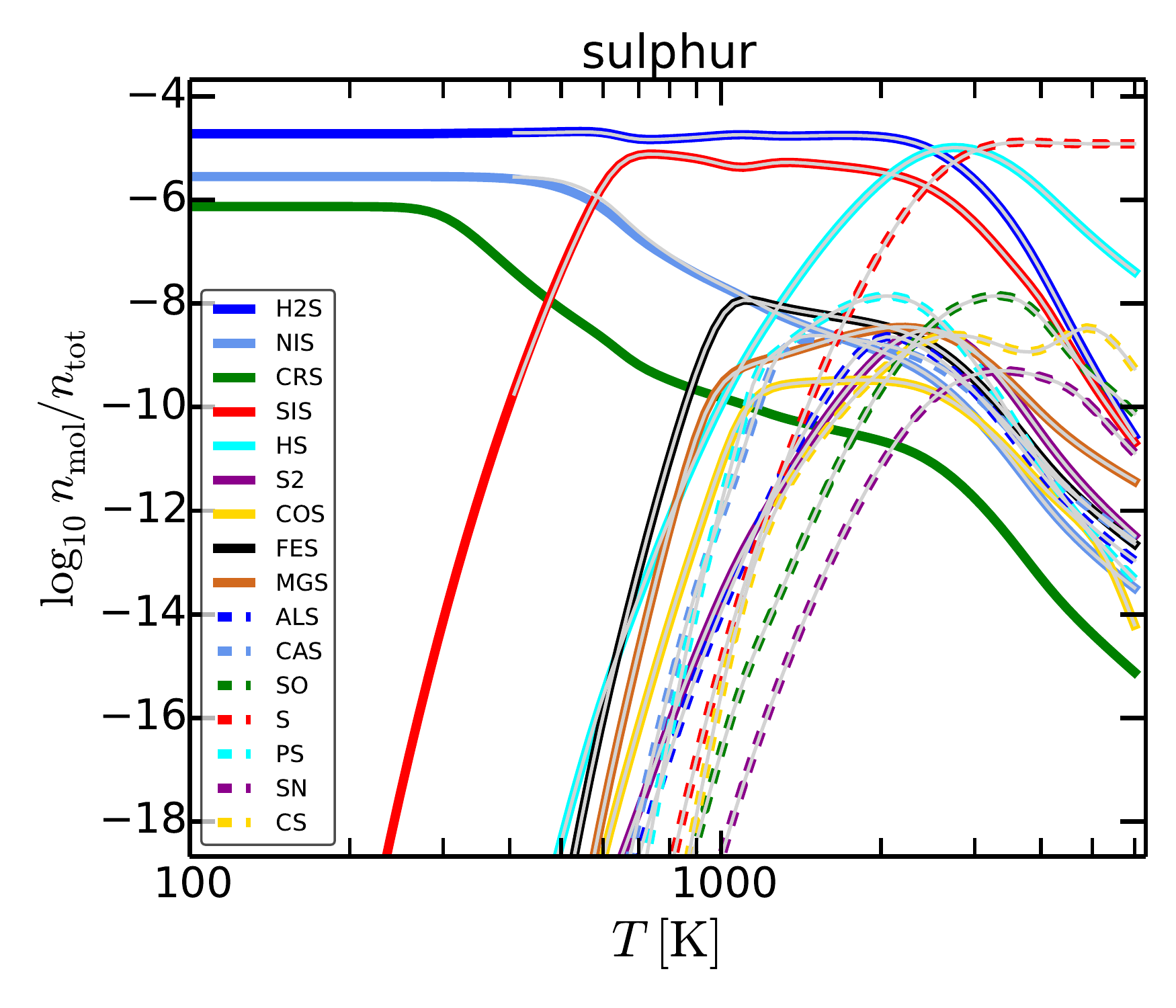} 
\\[-1mm]
\hspace*{-6mm}
\includegraphics[height=48mm,trim=10 64 0 10, clip]{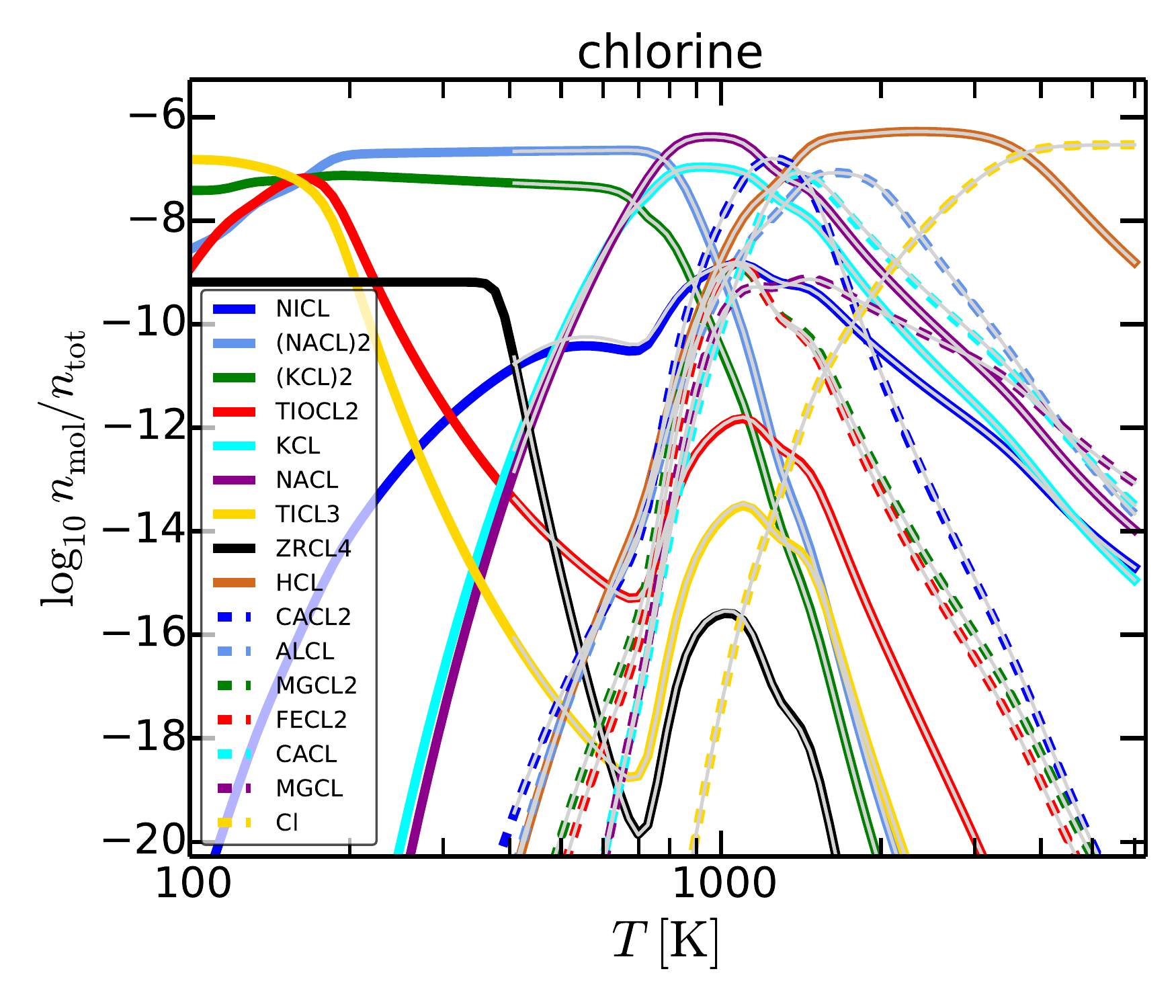} &
\hspace*{-6mm}
\includegraphics[height=48mm,trim=43 64 0 10, clip]{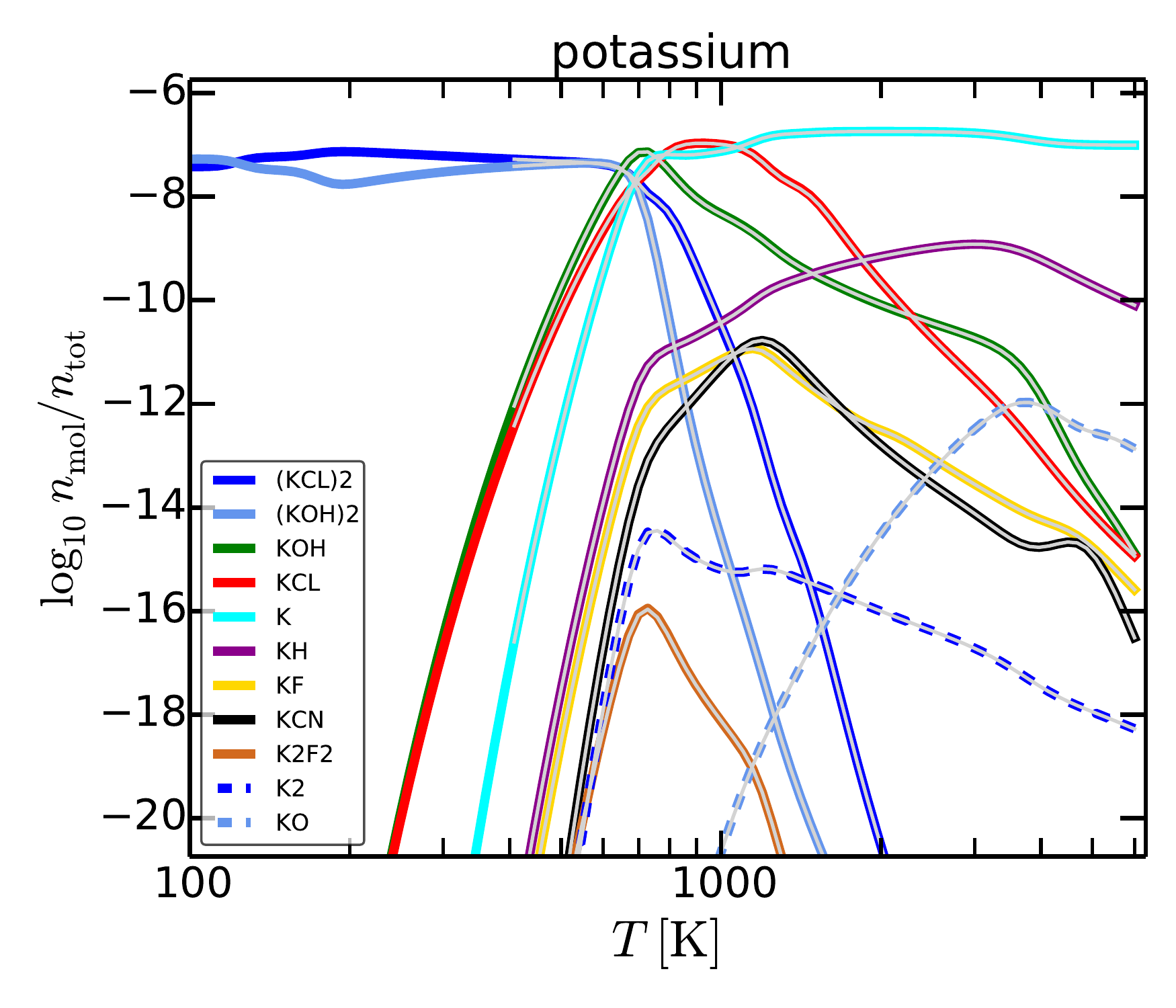} &
\hspace*{-6mm}
\includegraphics[height=48mm,trim=43 64 0 10, clip]{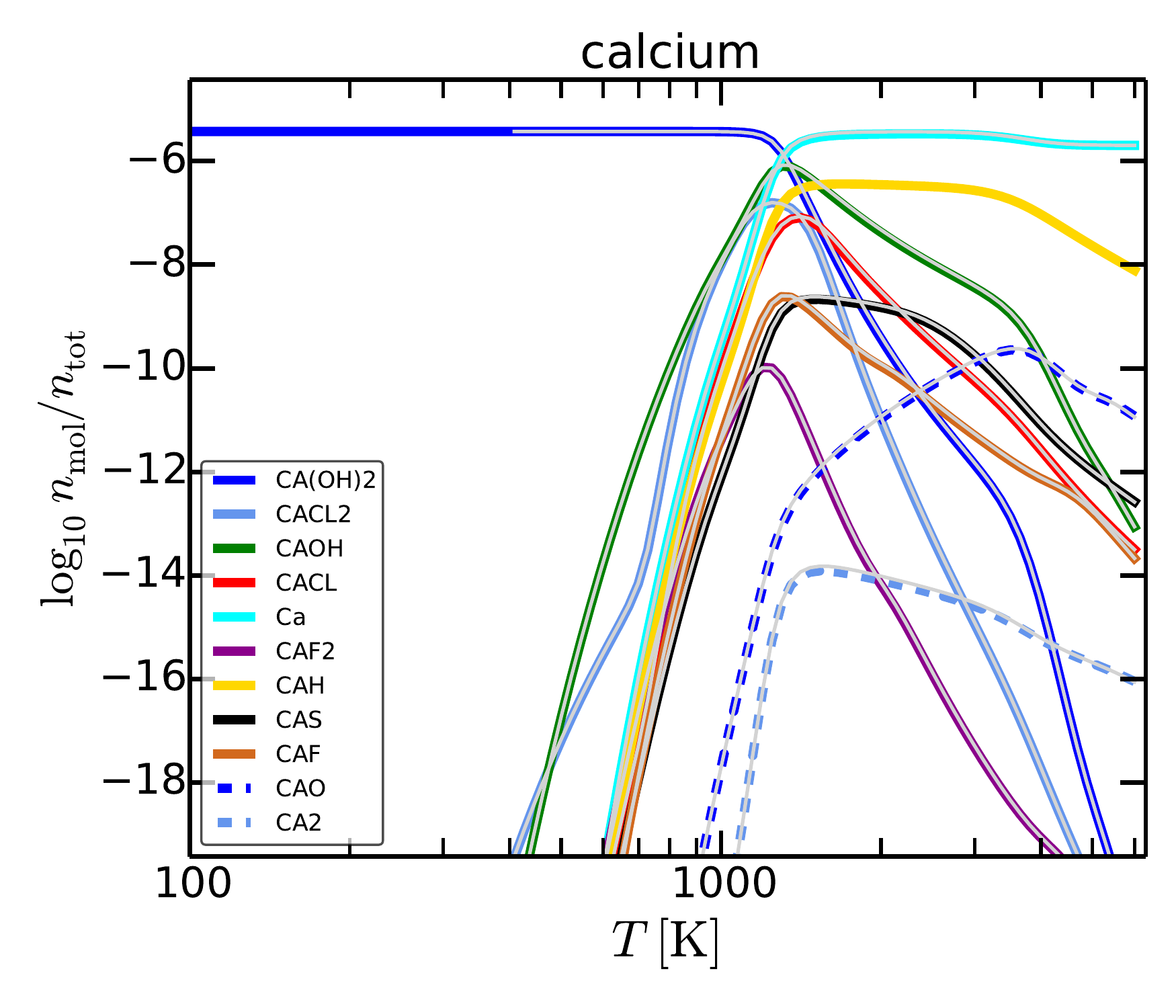} 
\\[-1mm]
\hspace*{-6mm}
\includegraphics[height=48mm,trim=10 64 0 10, clip]{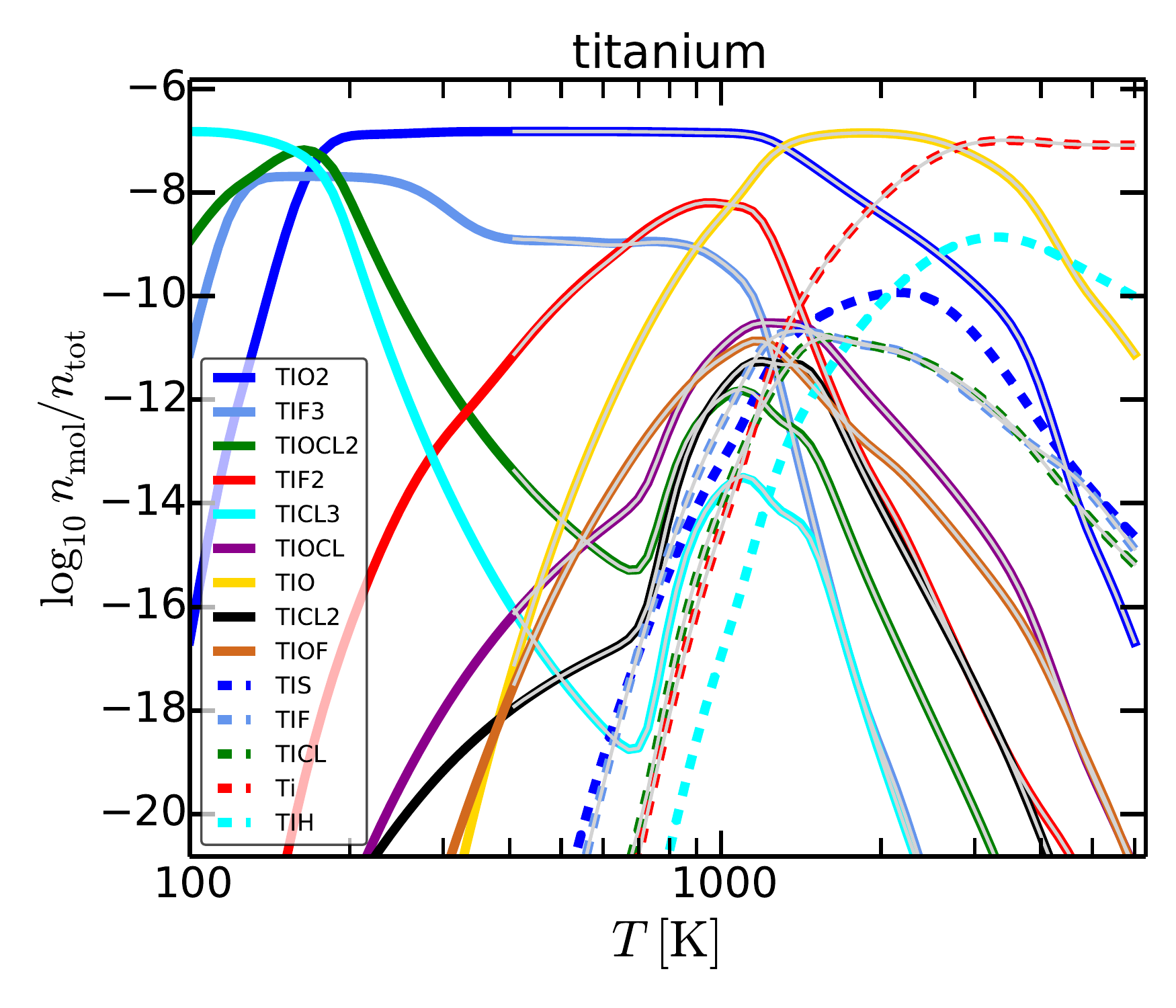} &
\hspace*{-6mm}
\includegraphics[height=48mm,trim=43 64 0 10, clip]{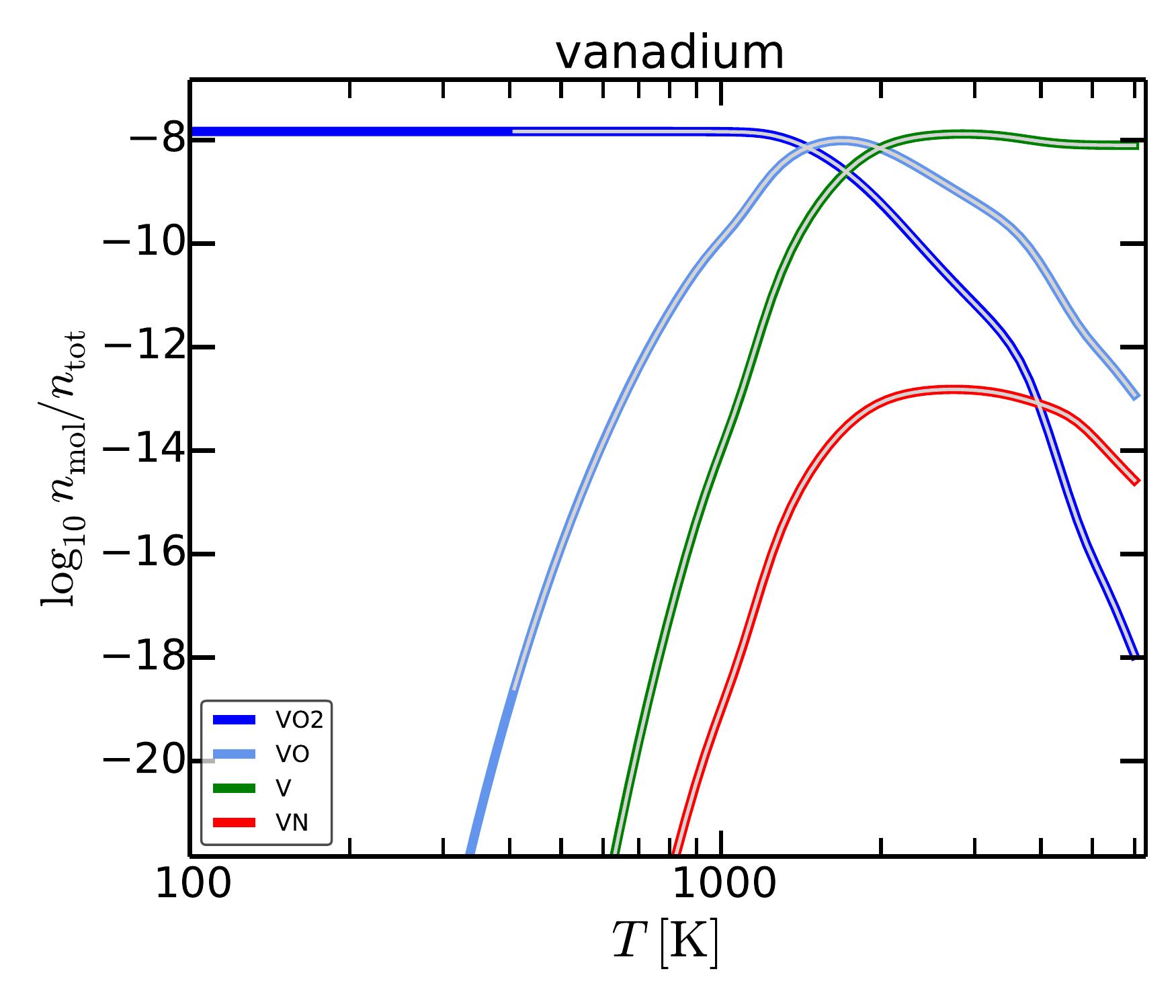} &
\hspace*{-6mm}
\includegraphics[height=48mm,trim=43 64 0 10, clip]{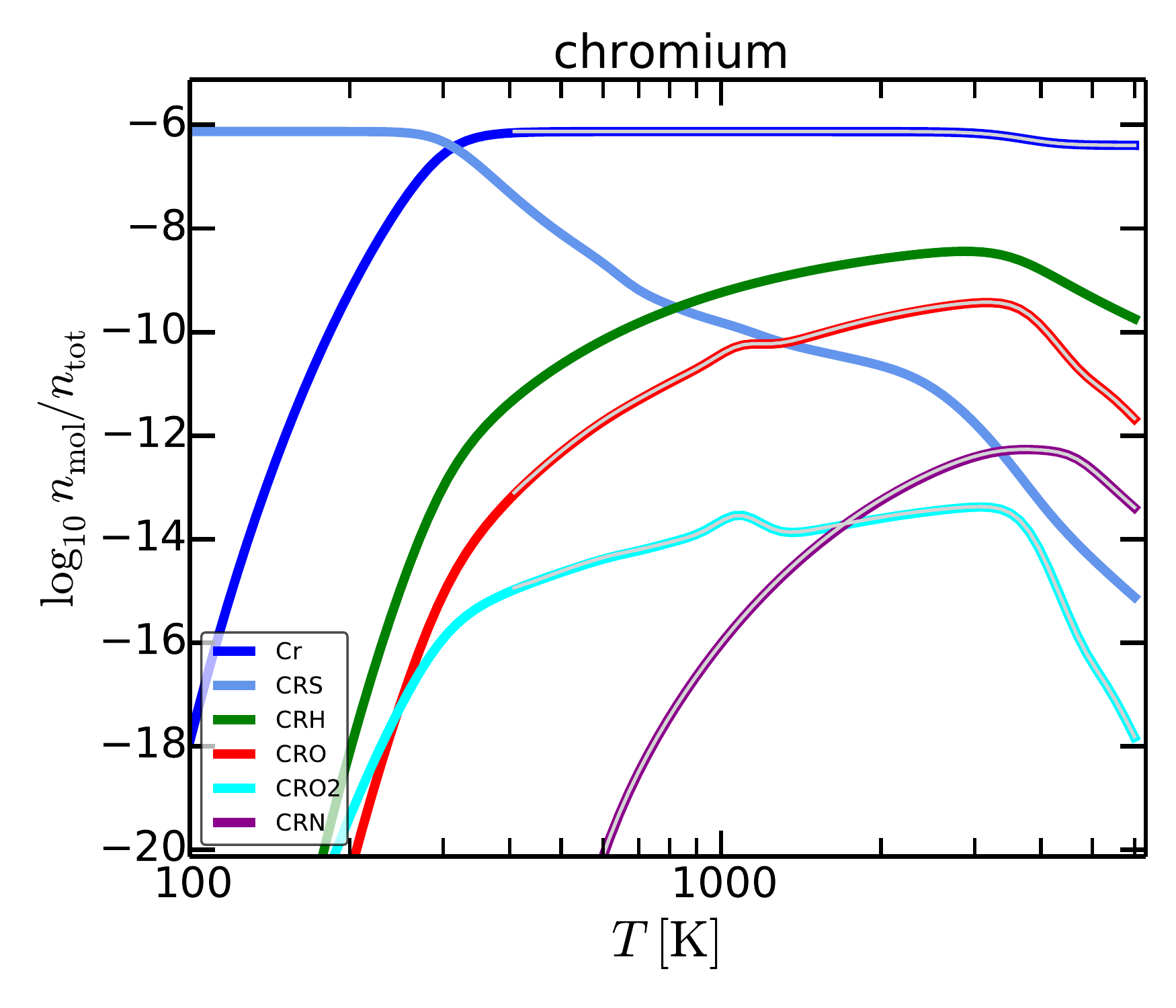} 
\\[-1mm]
\hspace*{-6mm}
\includegraphics[height=48mm,trim=10 64 0 10, clip]{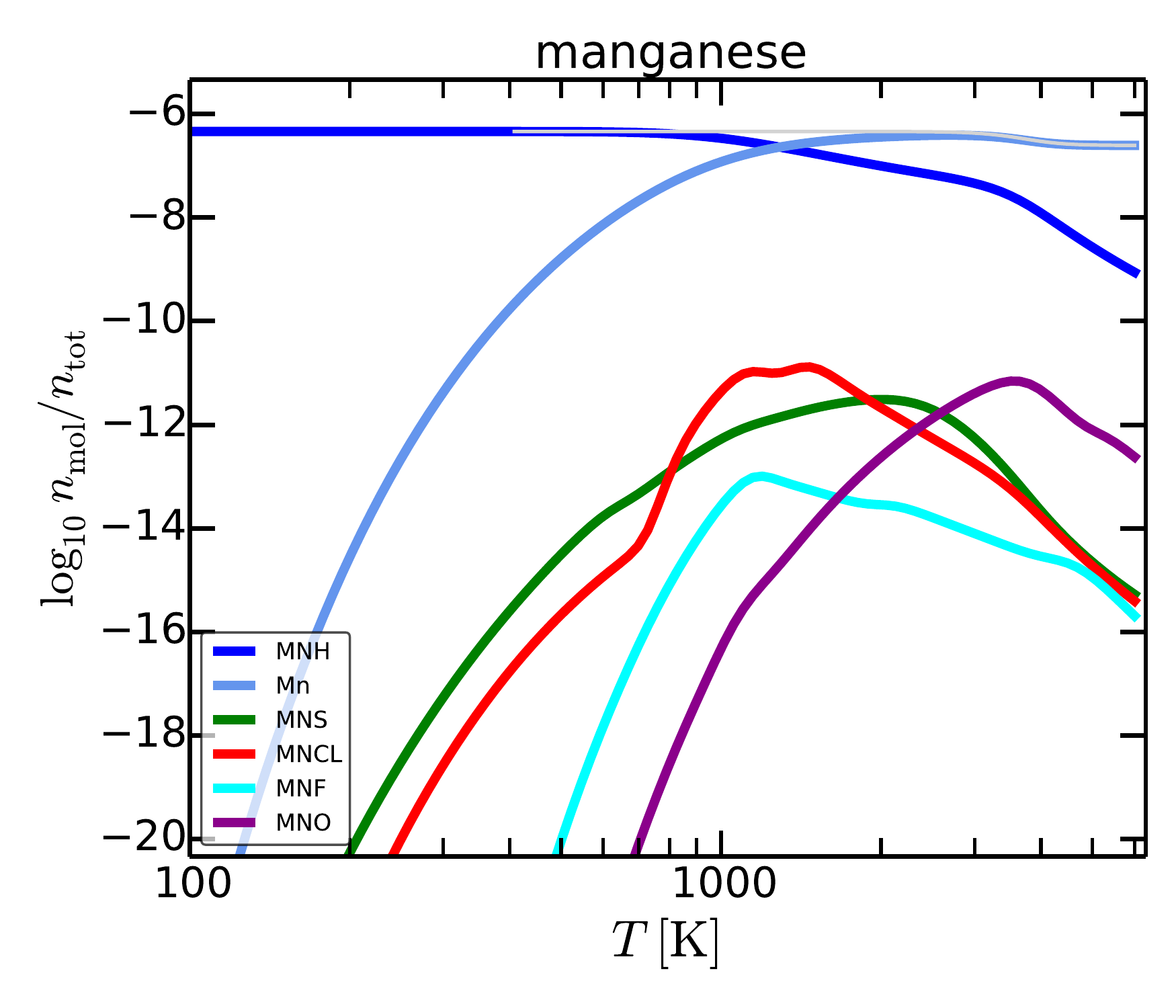} &
\hspace*{-6mm}
\includegraphics[height=48mm,trim=43 64 0 10, clip]{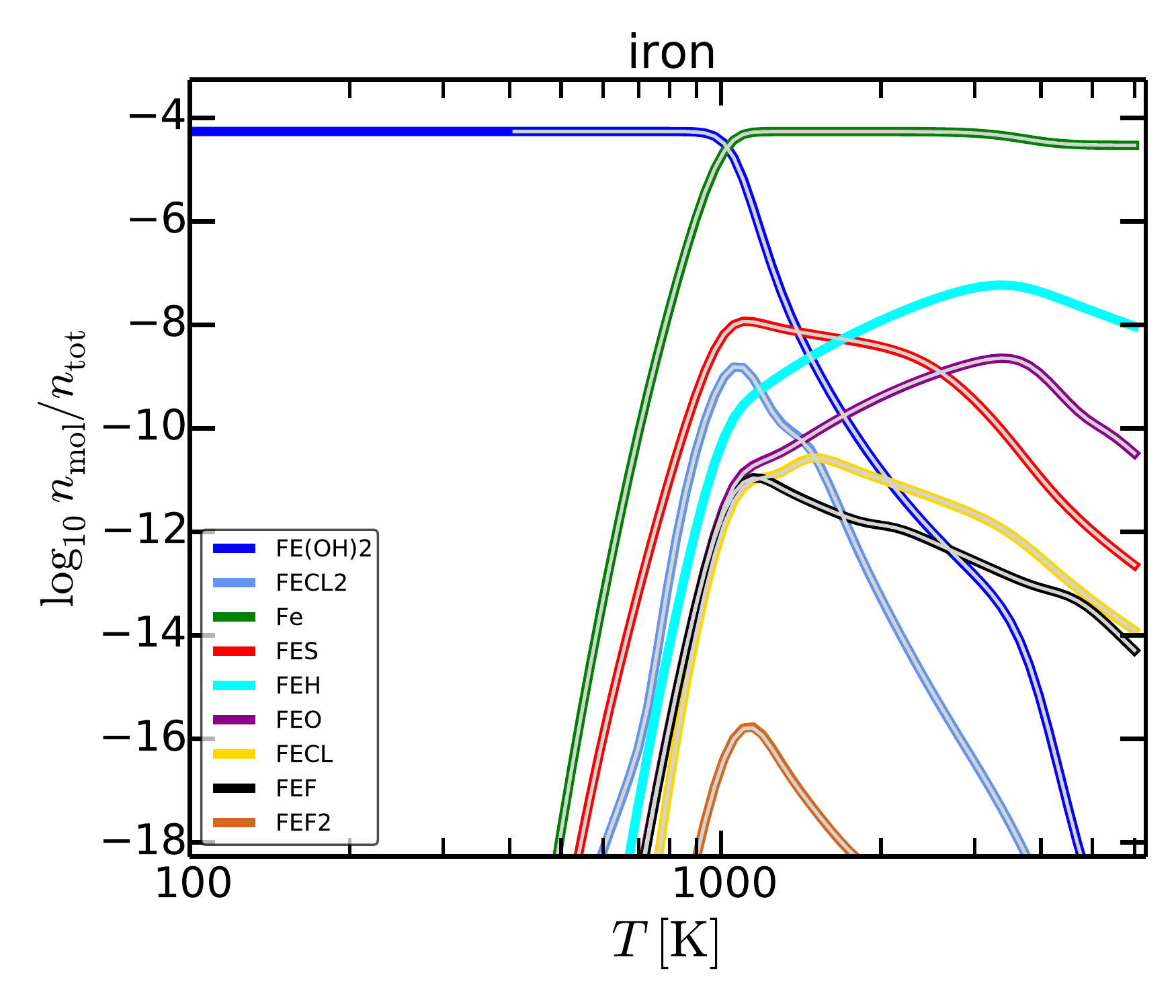} &
\hspace*{-6mm}
\begin{minipage}{60mm}
{\ }\\[-40mm]\hspace*{-1.5mm}
\includegraphics[height=54.3mm,trim=43 17 0 10, clip]{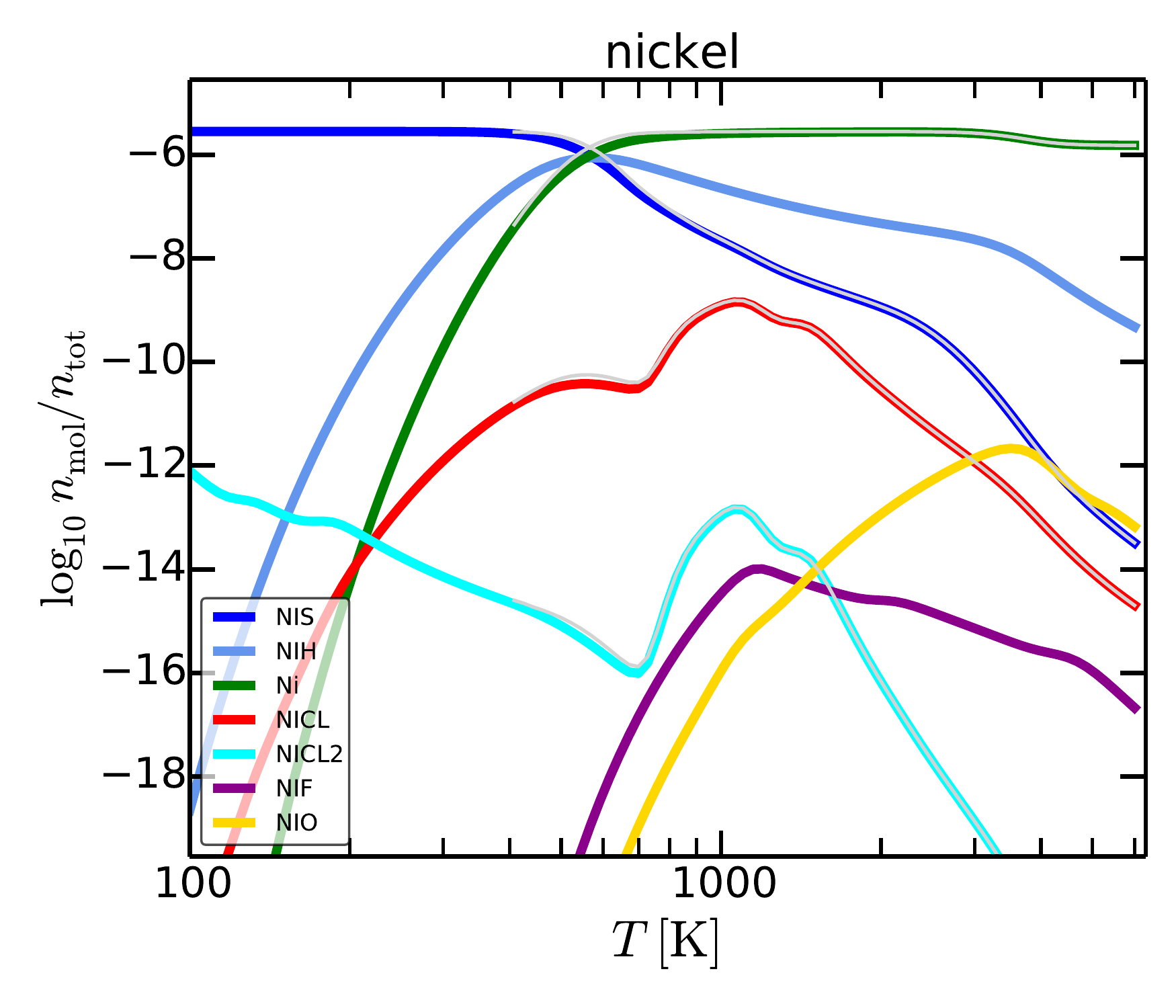}
\end{minipage}
\\[-6mm]
\hspace*{-6mm}
\includegraphics[height=55mm,trim=10 17 0 10, clip]{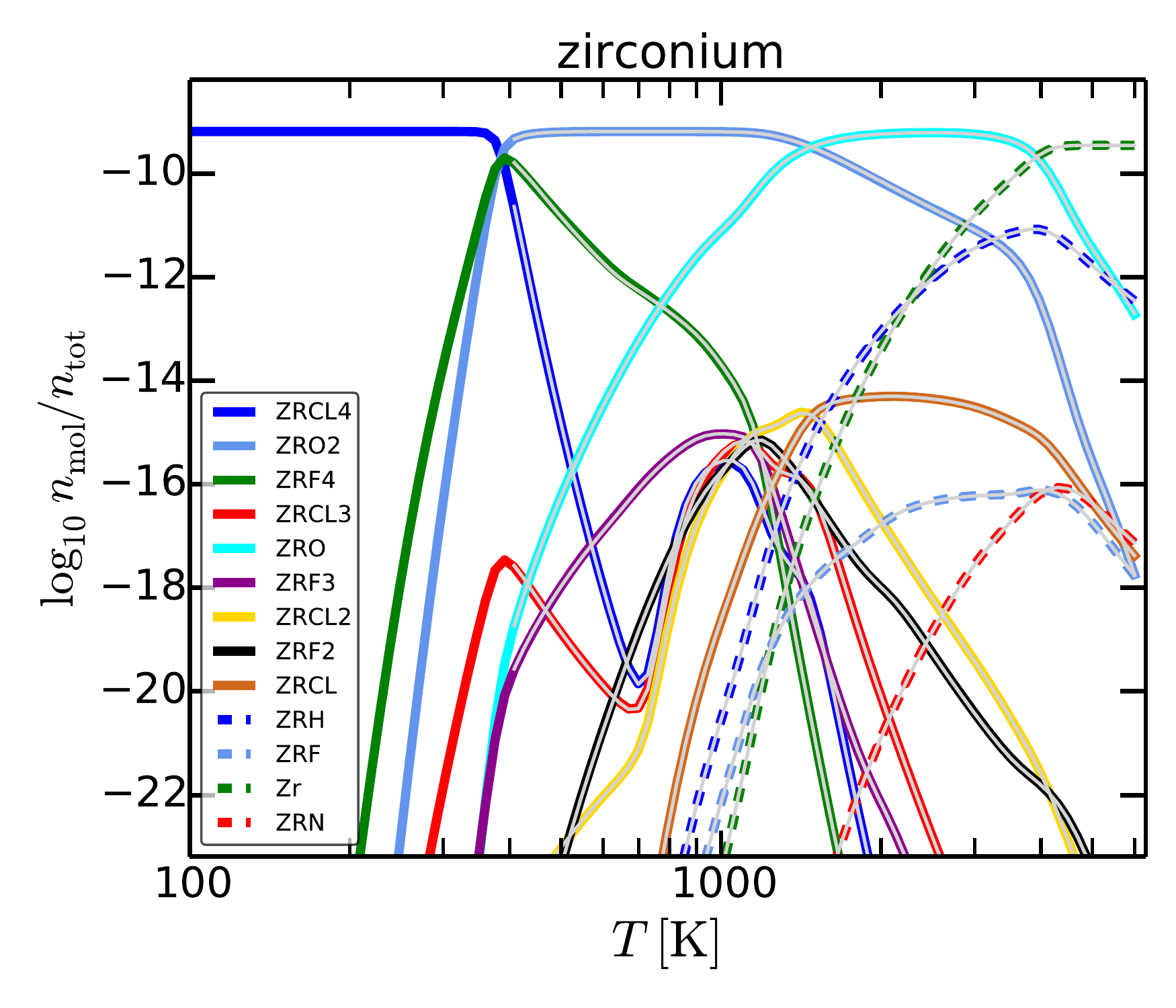} &
\hspace*{-6mm}
\includegraphics[height=55mm,trim=43 17 0 10, clip]{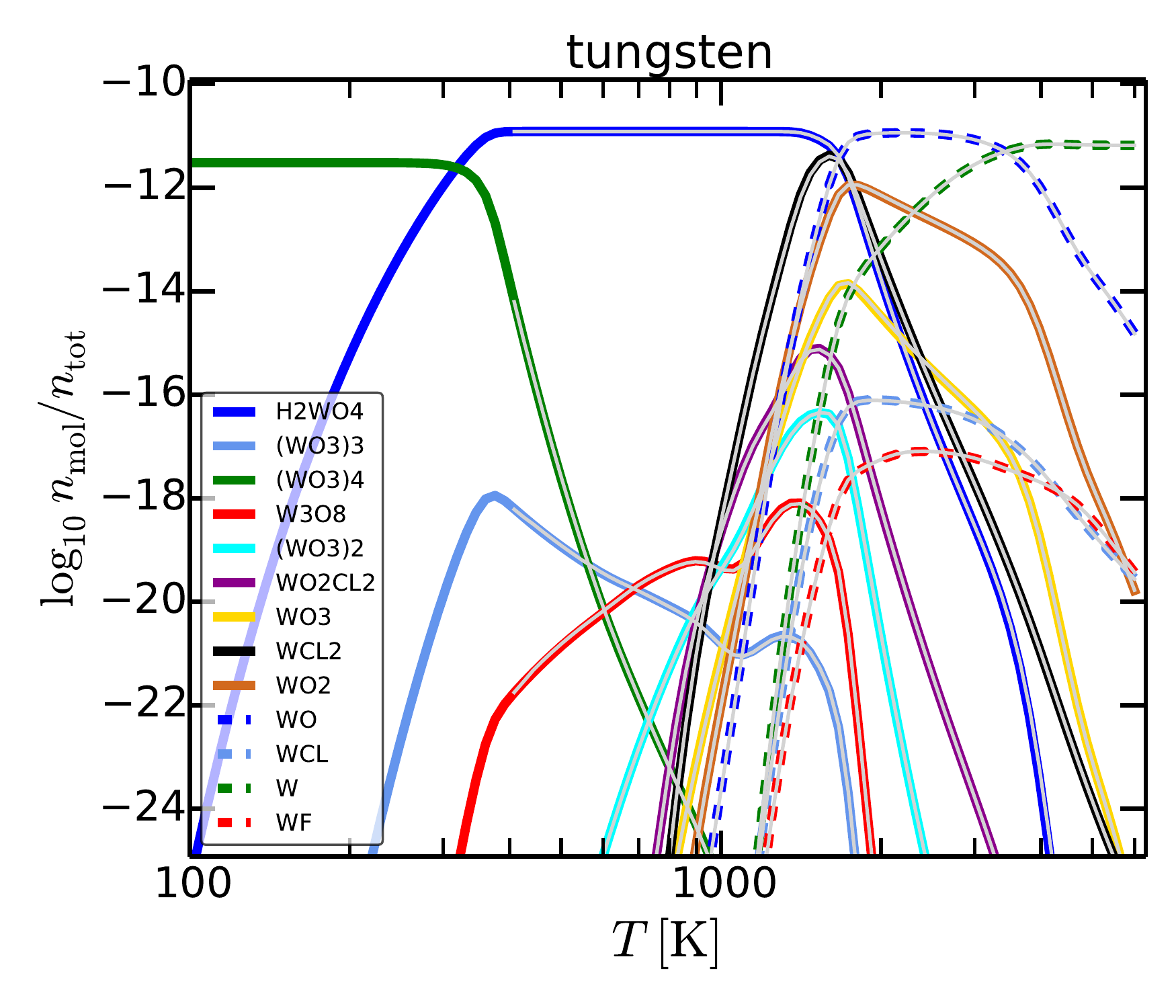} 
\\[-2mm]
\end{tabular}
\caption{{\bf (continued):}\ \ The differences in the selection/availability of
  molecular species explain all visible deviations between {\sc GGchem} and TEA.}
\end{figure*}

{\sc GGchem} allows the user to select elements, molecules and
condensates via versatile input files. All elements from hydrogen to
zirconium (atomic number 40) are supported, with additional options
to include tungsten (atomic number 74) and charges. The element
abundances are currently pre-compiled with options for solar
abundances \citep{Asplund2009}, meteorite abundances or Earth crust
abundances.  The default choice is to select all molecules and
condensates which are composed of the selected elements from provided
data-files, but the user is welcome to develop their own input files,
e.g.\ a customised small model optimised for speed,
with just a few elements and hand-selected species. There are options
to just solve the pure gas phase chemical equilibrium or the combined
problem with equilibrium condensation. The gas mass density $\rho$ or
the total gas pressure $p$ can be chosen as input parameter, besides
the temperature $T$. 

\section{Results}
\label{s:results}

\subsection{TEA benchmark test}
\label{sec:bench}

To benchmark our results, we use the public Thermo-chemical
Equilibrium Abundances (TEA) code
\citep{Blecic2016}\footnote{\url{https://github.com/dzesmin/TEA}}.
TEA follows the Gibbs free energy minimisation method as described by
\citet{White1958} and \citet{Eriksson1971}, using a Lagrangian
optimisation scheme with Lambda correction algorithm. The Gibbs free
energies of the molecules are computed from the thermo-chemical data
in the NIST-JANAF tables \citep{Chase1998}, for most molecules given
between 100\,K and 6000\,K, separated between 100\,K, provided by
Dr.~Thomas C.~Allison in October 2012 through private communication.
The programs uses internal spline interpolations to compute the Gibbs
free energy between these temperature points. The NIST-JANAF data is
available for 600 gaseous molecular species (84 elements).

TEA has been developed in particular to determine the abundances of the major,
most-abundant and spectroscopically most active gaseous species expected to
be present in hot, giant exoplanetary atmospheres, as these species
have a dominant influence on the planetary spectra in the optical and
infrared \citep[e.g.][]{Seager2010, Burrows2014}. TEA has been benchmarked
against CEA (Chemical Equilibrium with Applications), see
\citep{Gordon1994,McBride1996}, as well as against
several analytical codes \citep{Burrows1999, Heng2016, Heng2016b}.
These tests typically involve a few elements with a few dozens 
of molecules between 500\,K and 4000\,K, where TEA
is found to converge robustly within reasonable times 
(2-3 CPU-sec per $(p,T)$-point). 

For our benchmark test between TEA and {\sc GGchem} we have selected
the following 24 elements of astrophysical interest: H, He, Li, C, N,
O, F, Na, Mg, Al, Si, P, S, Cl, K, Ca, Ti, V, Cr, Mn, Fe, Ni, Zr and W
with solar abundances \citep{Asplund2009}. The task was to compute the
chemical composition of the gas at $p\!=\!1\,$bar from $T\!=\!6000\,$K
down to 100\,K (or as low as possible), with all available molecules.

Since the TEA-code can presently only handle neutral gas species, we
have disregarded all charged species and switched off equilibrium
condensation in {\sc GGchem} for this test. The results are shown in
Fig.~\ref{fig:TEA}. For practically all depicted molecules, the {\sc
  GGchem} results (coloured lines) agree very well with the TEA
results, to a precision better than the linewidth of about
$\sim\!0.05$\,dex.  All visible deviations can be explained by some
molecules not available in NIST-JANAF, and are hence missing in TEA,
or some very large molecules that we have de-selected on purpose in order to
improve the convergence of TEA. The missing molecules are listed in
the caption of Fig~\ref{fig:TEA}. {\sc GGchem} uses 421
molecular species plus 24 neutral atoms for this test, whereas TEA
uses altogether 400 species including the 24 neutral atoms. The agreement
between TEA and {\sc GGchem} is very reassuring, however, TEA failed
to converge within 2000 iterations for temperatures below 400\,K.

{\sc GGchem} needs about 0.004 CPU-sec per point on a 2.8\,GHz Linux
Laptop. The time consumption is mainly caused by the internal matrix
operations which scale as $K^3$ where $K$ is the number of
elements. For the lower temperatures in this benchmark test,
{\sc GGchem} uses quadruple precision arithmetics. For temperatures
$>\!1000$\,K, the code switches to double precision which takes
only 0.0013 CPU-sec per point for 24 elements.  To compute the
100 temperature points requested for this benchmark test, the total
computational time consumption of {\sc GGchem} is 0.71\,CPU-sec, where
about half of it is initialisation.  This is in sharp contrast to TEA
which needs about 45\,CPU-min per point with 2000 iterations,
resulting in altogether 74 CPU-hours to complete this benchmark test.
Clearly, we have pushed TEA to its limits in this benchmark test,
TEA should normally be run with about 100 iterations where it performs
as stated above. However, using TEA with 2000 iterations in this
benchmark test to obtain results for as low as possible temperatures, 
we conclude that {\sc GGchem} is more than $10^{\,5}$ times faster
than TEA.

\subsection{Abundance of major molecules}

\begin{figure}
\vspace*{0mm}
\hspace*{-2.5mm}
\includegraphics[width=94mm,trim=20 20 60 10, clip]{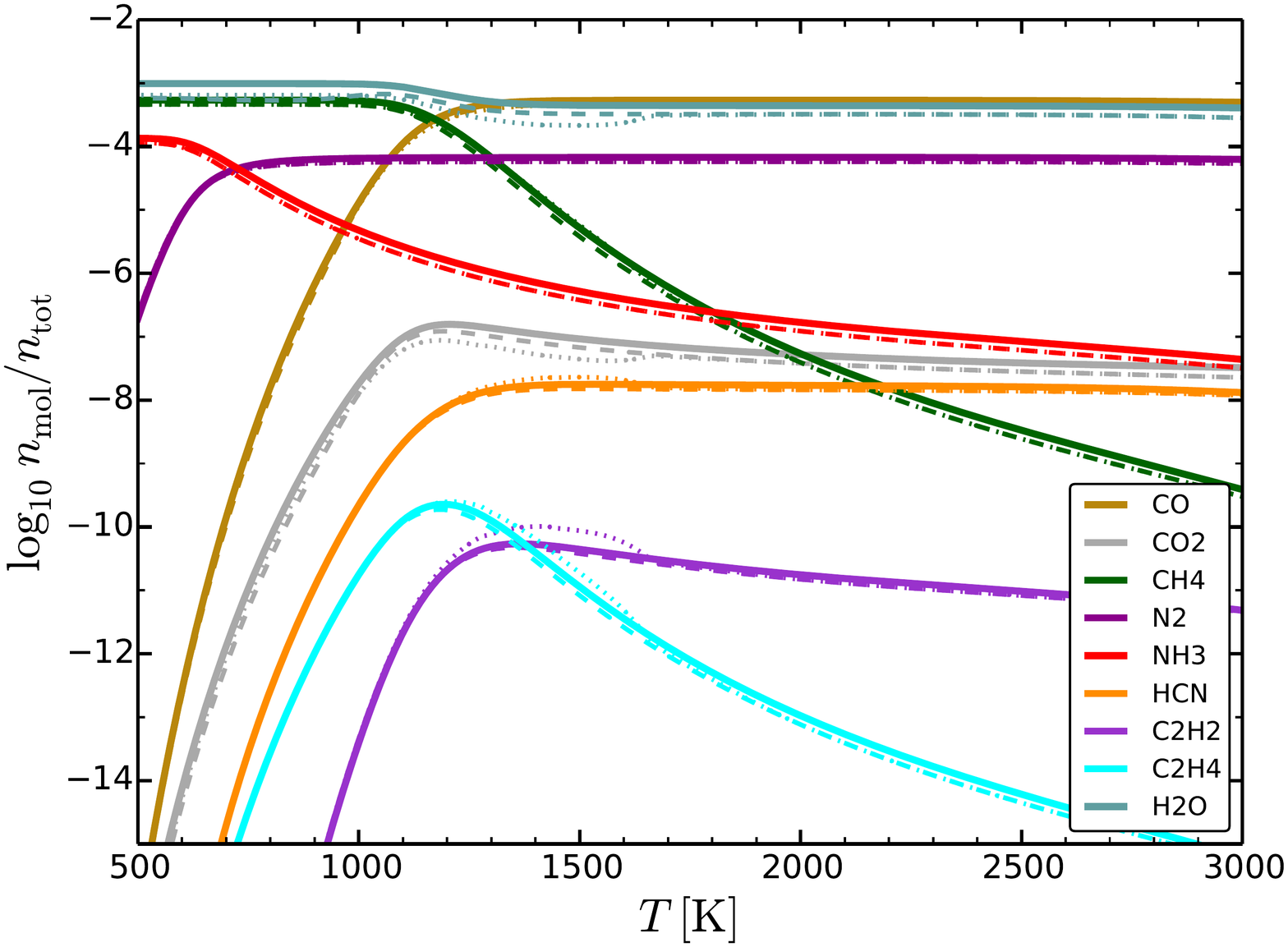}\\[-6mm] 
\caption{Concentrations of major molecules at 1\,bar according to a
  simplified 9-molecule model in chemical equilibrium (full lines), in
  comparison to a model with 24 elements and 445 molecules (dashed)
  and a full model with equilibrium condensation (dotted).}
\label{fig:mainmol}
\vspace*{-1mm}
\end{figure}

In a first application, we have used the simplified setup of
\citet{Heng2016} with 4 elements (H, C, N, O) and 9 molecules (H$_2$,
H$_2$O, CO, CO$_2$, N$_2$, NH$_3$, HCN, C$2$H$_2$, C$_2$H$_4$) with
solar abundances \citep{Asplund2009}.  The results shown in
Fig.~\ref{fig:mainmol} can be directly compared to the upper part of
figure~1 in \citep{Heng2016}. {\sc GGchem} needs about 0.12\,milli
CPU-sec per point in this setup.

Figure~\ref{fig:mainmol} compares these results to two other {\sc
  GGchem} models. The dashed lines show the results from the full
model with 24 elements and 445 molecules as specified in
Sect.~\ref{sec:bench}, and the dotted lines show the results from a
full model with equilibrium condensation as specified in
Sect.~\ref{sec:eqcond}. The comparison shows that, although the
simplified 9-molecule model results in very reasonable approximations
for the concentrations of the major molecules between 500\,K and
3000\,K, there are some notable deviations around 1000-2000\,K which
are likely to be relevant for the retrieval of atmospheric properties
from spectroscopic observations.  These deviations are due to the
consumption of oxygen by (i) the formation of SiO, Mg(OH)$_2$ and
Fe(OH)$_2$ molecules, and (ii) silicate and phyllosilicate
condensation.  Both effects reduce the concentrations of H$_2$O
  and CO$_2$, and increase C$_2$H$_2$ and C$_2$H$_4$ concentrations,
  by up to 0.5\,dex around 1500\,K. This effect is discussed further
in Sect.~\ref{sec:C/O}.

\begin{figure}
\vspace*{-0.5mm}
\hspace*{-3mm}
\includegraphics[width=92.5mm,trim=20 15 47 13, clip]{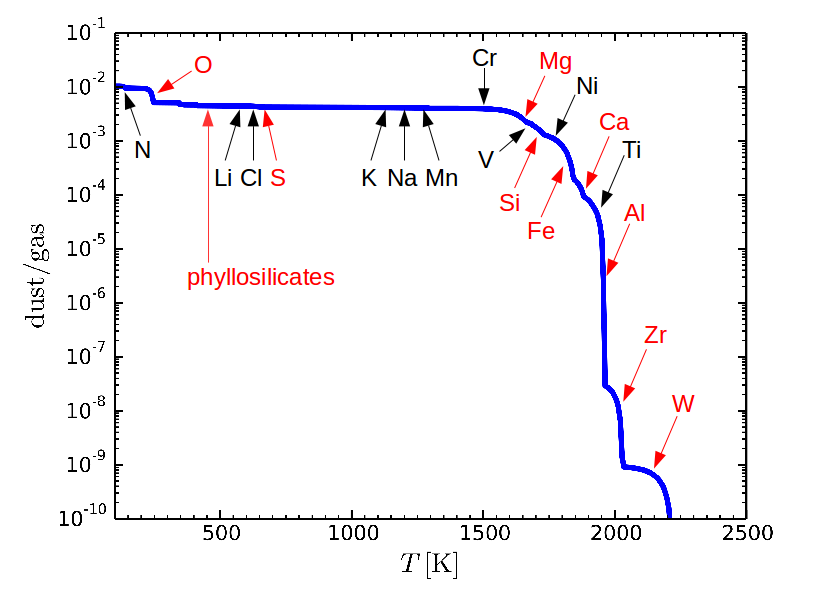}\\[1mm] 
\hspace*{-3mm}
\includegraphics[width=93mm,height=66mm,trim=12 15  7 10, clip]{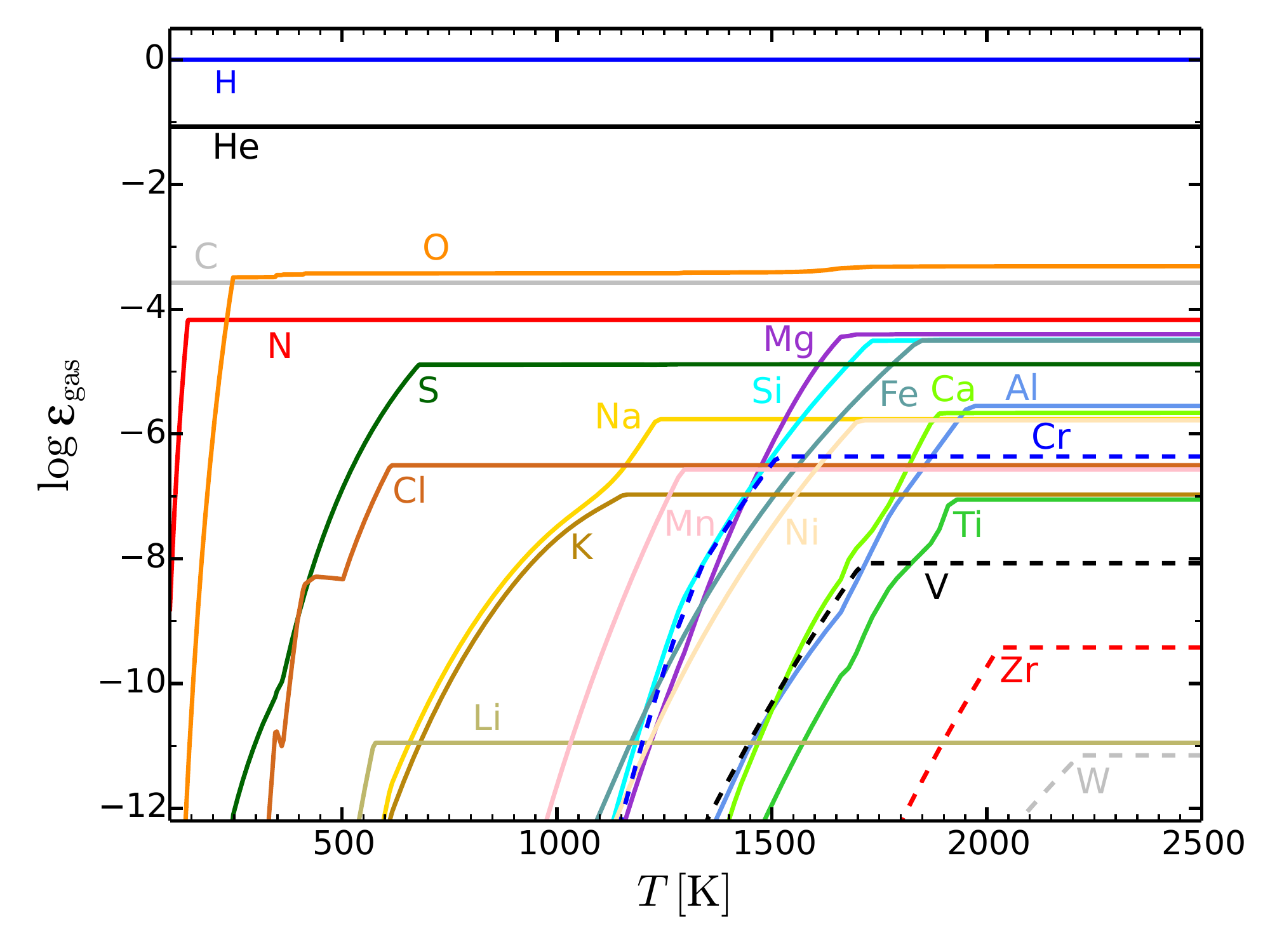}\\[-5mm] 
\caption{{\bf Upper part:} The dust/gas mass ratio in phase
  equilibrium as function of temperature at constant $p\!=\!1$\,bar
  for solar element abundances. The arrows indicate where the elements
  start forming condensed phases in significant amounts.  The elements
  marked in red have a significant influence on the shape of the
  dust/gas$(T)$-curve. The {\bf lower part} shows the remaining
  element abundances in the gas phase.}
\label{fig:dustgas}
\vspace*{-2mm}
\end{figure}

\subsection{The condensation of the elements}
\label{sec:eqcond}

In the following, we study the results of a model with equilibrium
condensation switched on, at constant pressure $p\!=\!1\,$bar, for 22
elements (H, He, Li, C, N, O, Na, Mg, Al, Si, S, Cl, K, Ca, Ti, V, Cr,
Mn, Fe, Ni, Zr, W) and charged species, with solar abundances. We have
omitted fluorine and phosphorus in this model because of our possibly
incomplete data for fluorine and phosphorus condensates. Altogether 388
gas phase and 190 condensed species are taken into account in this
model. Gas phase species include 365 molecules, 22 free atoms and the
free electron. Condensed species include 38 liquids
(Table~\ref{tab:liquid}). The thermo-chemical data of the condensed
species are mainly taken from the SUPCRTBL database
(Table~\ref{tab:SUdata}), completed/replaced by our own fits to
NIST-JANAF data of 92 condensed species (Table~\ref{tab:SOLID}), plus
five further condensed species (CaTiO$_3$[s], MnS[s], NiS[s],
NiS$_2$[s], Ni$_3$S$_2$[s]) taken from Sharp \& Huebner, which are
neither available in NIST-JANAF nor in SUPCRTBL.

Figure~\ref{fig:dustgas} shows the dust/gas mass ratio as function of
temperature in phase equilibrium, calculated as 
\begin{equation}
  \mbox{dust/gas} \;=\; \frac{\rho_{\rm cond}}{\rho}
    \;=\; \frac{1}{\rho}\sum_{j\rm[cond]}\,n_{j\rm[cond]} m_{j\rm[cond]}  \ ,
\end{equation}
where $j\rm[cond]$ is the index of a condensed species,
$n_{j\rm[cond]}$ its particle density (i.e.\ number of condensed units
per volume) and $m_{j\rm[cond]}$ its mass.  Tungsten establishes a
dust/gas ratio of $\sim\!\!10^{-9}$ at $T\!\la\!2250\,$K, zirconium
dioxide brings it to $\sim\!\!10^{-7.5}$ at 2000\,K, and aluminium
increases it to $\sim\!\!10^{-4}$ at 1900\,K. Subsequently, calcium,
iron, silicon and magnesium compounds build up a dust/gas ratio of
$\sim\!\!10^{-2.38}\!\approx\!0.004$ at 1500\,K, which then remains
about constant toward much lower temperatures. The condensation of
titanium, nickel, vanadium, chromium, manganese, sodium and potassium
compounds make only minor contributions to dust/gas$(T)$. The value of
$\rm dust/gas\!\approx\!0.004$ is significantly lower than the
standard value of $\rm dust/gas\!\approx\!1/100$
\citep[e.g.][]{Beckwith1990,Chiang1997,Dalessio1998,Dullemond2002}. At
around 650\,K, sulphur starts to condense which increases dust/gas to
about 0.0045 with further minor contributions by chlorine and
lithium. Just below 500\,K, several minerals start to become hydrated
by the incorporation of either water or hydroxyl groups, forming
so-called phyllosilicates. This additional intake of oxygen atoms
increases the dust/gas ratio to about 0.0052. Around 240\,K and
140\,K, water ice and ammonia condense, respectively, which finally
brings the dust/gas ratio to a value of almost exactly 1/100.

The lower part of Fig.~\ref{fig:dustgas} shows the depletion of the
elements from the gas phase due to condensation. At 1000\,K, for
example, the gas consists mainly of S and Cl besides H,
He, C, N and O, whereas all other elements, which are usually more
abundant like Fe, Si and Mg, are already depleted by more than 8
orders of magnitude. The intake of oxygen in silicates and
phyllosilicates leads to a substantial increase of the C/O ratio in
the gas phase, from
$\rm C/O\!=\!0.55$ at $T\!=\!2500$\,K to $\rm C/O\!=\!0.71$
at $T\!=\!1000$\,K (silicates) and $\rm C/O\!=\!0.83$ at
$T\!=\!300$\,K (phyllosilicates). This increase of C/O may well have
observational consequences in brown dwarf and giant gas planet
spectra. Once water condenses, the C/O ratio becomes very large, for
example $\rm C/O\!>\!10^{\,6}$ at 150\,K, and the gas consists mainly
of CH$_4$ and NH$_3$ (beside H$_2$ and noble gases), similar to the
atmosphere of Titan \citep{Fulchignoni2005}\footnote{Titan's
  atmosphere has little hydrogen, hence it consists mainly of N$_2$
  with just some CH$_4$, whereas oxygen is absent due to water
  condensation.}. Remarkably, at $p\!=\!1\,$bar, not a single carbon
atom has been incorporated into any condensed state in phase equilibrium 
down to $100\,$K for solar abundances.

\begin{figure}
\hspace*{-3mm}
\includegraphics[width=92mm,height=69mm,trim=12 15 10 10, clip]{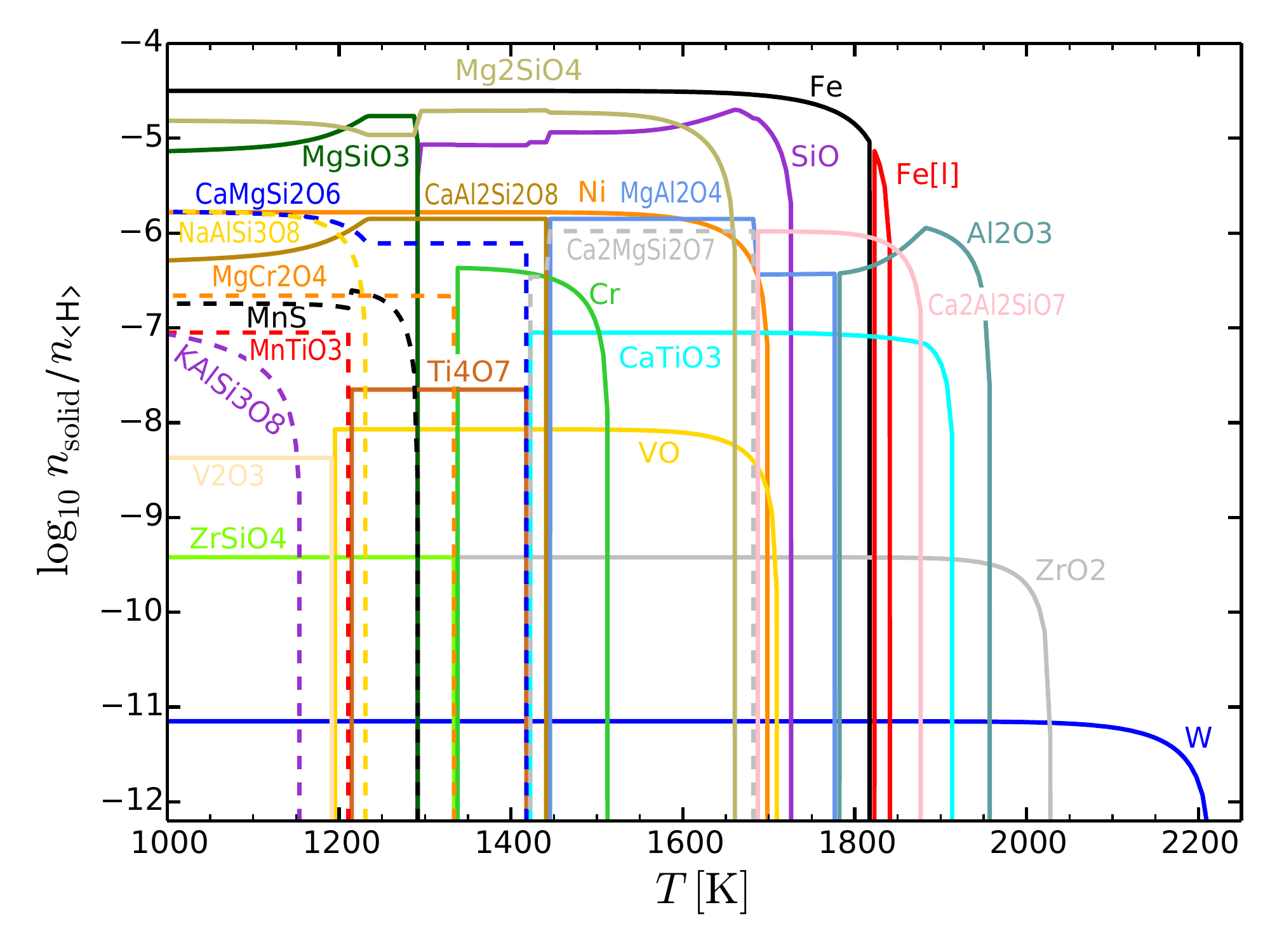}\\[-6mm]
\caption{The onset of condensation at $p\!=\!1$\,bar for solar
  abundances in phase equilibrium. The graph show the concentration of
  the various condensed species with respect to hydrogen nuclei.}
\label{fig:cond1}
\vspace*{-2mm}
\end{figure}

\smallskip\noindent {\bf The onset of condensation:\ \ }
Figure~\ref{fig:cond1} visualises the first stages of condensation in
phase equilibrium at $p\!=\!1\,$bar. W[s] {\sl(crystalline tungsten)} is
found to be the first stable condensate at 2216\,K, followed by
ZrO$_2$[s] {\sl(baddeleyite)} at 2027\,K.  Further phase transitions are
as follows: Al-gas $\to$ Al$_2$O$_3$[s] {\sl(corundum)} at 1957\,K,
Ti-gas $\to$ CaTiO$_3$[s] {\sl(perovskite)} at 1913\,K, Ca-gas $\to$
Ca$_2$Al$_2$SiO$_7$[s] {\sl(gehlenite)} at 1880\,K, Fe-gas $\to$ Fe[l]
{\sl(liquid iron)} at 1841\,K.  These are all type-1
(gas\,$\to$\,condensed) phase transitions, featured by a
smooth, rounded-off increase of the concentration of the new
condensate as function of temperature. This gradual increase is
mirrored by a gradual decrease of one element abundance in the gas
phase, an element which was not affected by condensation before, to
keep the partial pressure equal to the vapour pressure
$\pvap(T)$ of the new condensate.

At 1820\,K and 1777\,K, the first type-2 transitions (condensed
$\to$\,condensed) occur, namely Fe[l] $\to$ Fe[s] {\sl(solid iron)} and
Al$_2$O$_3$[s] $\to$ MgAl$_2$O$_4$[s] {\sl(spinel)}, respectively. These
transitions transform one combination of condensates into
another, without substantial changes of any gas phase
abundances. Type-2 transitions occur suddenly, they do not
change the number of elements affected by condensation, hence the
number of independent variables in the equation system to be solved,
and thus the total number of present condensates (see
App.~\ref{AppD}). Type-2 transitions have a box-like appearance in
Fig.~\ref{fig:cond1}.

\smallskip\noindent {\bf The appearance of the major condensates:\ \ }
The condensation of the elements continues in Fig.~\ref{fig:cond1}
with a number of further type-1 transitions: Si-gas $\to$ SiO[s]
{\sl(silicon monoxide)} at 1729\,K at 1\,bar, V-gas $\to$ VO[s] {\sl(vanadium
  monoxide)} at 1710\,K, Ni-gas $\to$ Ni[s] {\sl(crystalline nickel)}
at 1690\,K, Mg-gas $\to$ Mg$_2$SiO$_4$[s] {\sl(fosterite)} at 1661\,K
and Cr-gas $\to$ Cr[s] {\sl(crystalline chromium)} at 1513\,K. At this
stage, the build-up of the dust/gas ratio to a value of about 0.004 is
mostly completed and there are 11 elements which are strongly reduced
by condensation (see lower part of Fig.~\ref{fig:dustgas}), which sets
the number of simultaneously present condensates to 11 (as explained in
App.~\ref{AppD}).

\smallskip\noindent {\bf Increasing complexity 1:\ \ } A number of
further type-2 transitions occur in Fig.~\ref{fig:cond1} towards lower
temperatures, namely Ca$_2$Al$_2$SiO$_7$[s] $\to$
Ca$_2$MgSi$_2$O$_7$[s] {\sl(akermanite)} at 1685\,K, MgAl$_2$O$_4$[s]
$\to$ CaAl$_2$Si$_2$O$_8$[s] {\sl(anorthite)} at 1440\,K, CaTiO$_3$[s]
$\to$ Ti$_4$O$_7$[s] {\sl(titanium oxide)} at 1420\,K,
Ca$_2$MgSi$_2$O$_7$[s] $\to$ CaMgSi$_2$O$_6$[s] {\sl(diopside)} at
1420\,K, Cr[s] $\to$ MgCr$_2$O$_4$[s] {\sl(picrochromite)} at 1336\,K,
ZrO$_2$[s] $\to$ ZrSiO$_4$[s] {\sl(zircon)} at 1334\,K, SiO[s] $\to$
MgSiO$_3$[s] {\sl(enstatite)} at 1292\,K, Ti$_4$O$_7$[s] $\to$
MnTiO$_3$[s] {\sl(pyrophanite)} at 1211\,K, and VO[s] $\to$
V$_2$O$_3$[s] {\sl(karelianite)} at 1194\,K.  These type-2 transitions
can have some side-effects on other more abundant condensates, so
strictly speaking, each type-2 phase transition affects a combination
of solids where one is formed, one is consumed and a number of other,
more abundant condensates adjust their concentrations. This applies in
particular to MgSiO$_3$[s] and Mg$_2$SiO$_4$[s] which are the main reservoir
for condensed silicon and magnesium, which are affected by most type-2
transitions.

On the left side of Fig.~\ref{fig:cond1}, a few more type-1
transitions take place, namely Mn-gas $\to$ MnS[s] {\sl(alabandite})
at $T\!=\!1290$\,K, Na-gas $\to$ NaAlSi$_3$O$_8$[s] {\sl(albite)} at
$T\!=\!1231$\,K and K-gas $\to$ KAlSi$_3$O$_8$[s] {\sl(microcline)} at
$T\!=\!1154$\,K, increasing the number of simultaneously present
condensates to 14.  At 1000\,K, all three major components of {\sl
  feldspar} are present (KAlSi$_3$O$_8$[s] – NaAlSi$_3$O$_8$[s] –
CaAl$_2$Si$_2$O$_8$[s]) which is the most common mineral on Earth
making up about 41\% of its continental crust
\citep{Anderson2010}. {\sl Feldspar} is the major constituent of {\sl
  basalt}, which is a key component of oceanic crust as well as the
principal volcanic rock on Earth.

\begin{figure}
\vspace*{1mm}
\hspace*{-3mm}
\includegraphics[width=95mm,trim=12 15 10 10, clip]{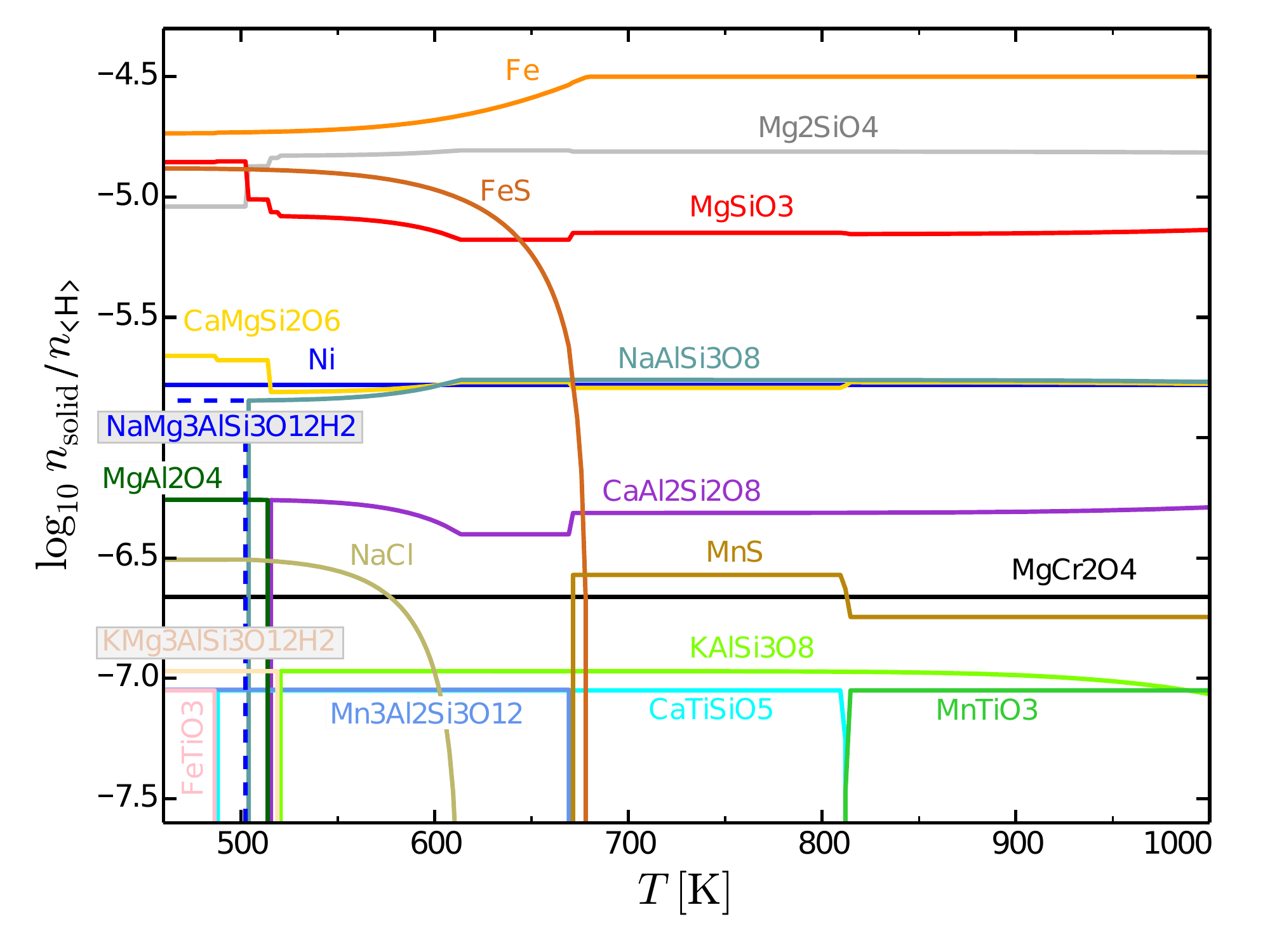}\\[-6mm]
\caption{Concentration of condensed species at medium temperatures
  $\rm 460\,K - 1000\,K$ at $p\!=\!1$\,bar for solar abundances. Note
  the changed scaling of the $y$-axis in comparison to Fig.~\ref{fig:cond1}.}
\label{fig:cond2}
\vspace*{-1mm}
\end{figure}

\smallskip\noindent {\bf Increasing complexity 2:\ \ } 
Figure~\ref{fig:cond2} continues to show the model results down to
460\,K. Most prominently, we see three more type-1 transitions: 
S-gas $\to$ FeS[s] {\sl(troilite)} at 678\,K,
Cl-gas $\to$ NaCl[s] {\sl(halite)} at 613\,K and 
Li-gas $\to$ LiCl[s] {\sl(lithium-cloride)} at 573\,K,
increasing the number of simultaneously present condensates to 17. Further
type-2 transitions occurs as well, in particular MnTiO$_3$[s]
$\to$ CaTiSiO$_5$[s] {\sl(sphene)} at 814\,K, MnS[s] $\to$
Mn$_3$Al$_2$Si$_3$O$_{12}$[s] {\sl(spessartine)} at 670\,K,
KAlSi$_3$O$_8$[s] $\to$ KMg$_3$AlSi$_3$O$_{12}$H$_2$[s] {\sl(phlogopite)} at
520\,K, CaAl$_2$Si$_2$O$_8$[s] $\to$ MgAl$_2$O$_4$[s] {\sl(spinel)} at
514\,K, NaAlSi$_3$O$_8$[s] $\to$ NaMg$_3$AlSi$_3$O$_{12}$H$_2$[s]
{\sl(sodaphlogopite)} at 502\,K, and CaTiSiO$_5$[s] $\to$ FeTiO$_3$[s]
{\sl(ilmenite)} at 486\,K.  KMg$_3$AlSi$_3$O$_{12}$H$_2$ and
NaMg$_3$AlSi$_3$O$_{12}$H$_2$ are the first two phyllosilicates, with
additional intake of hydrogen and oxygen.

\begin{figure}
\hspace*{-3mm}
\includegraphics[width=92mm,trim=12 15 10 10, clip]{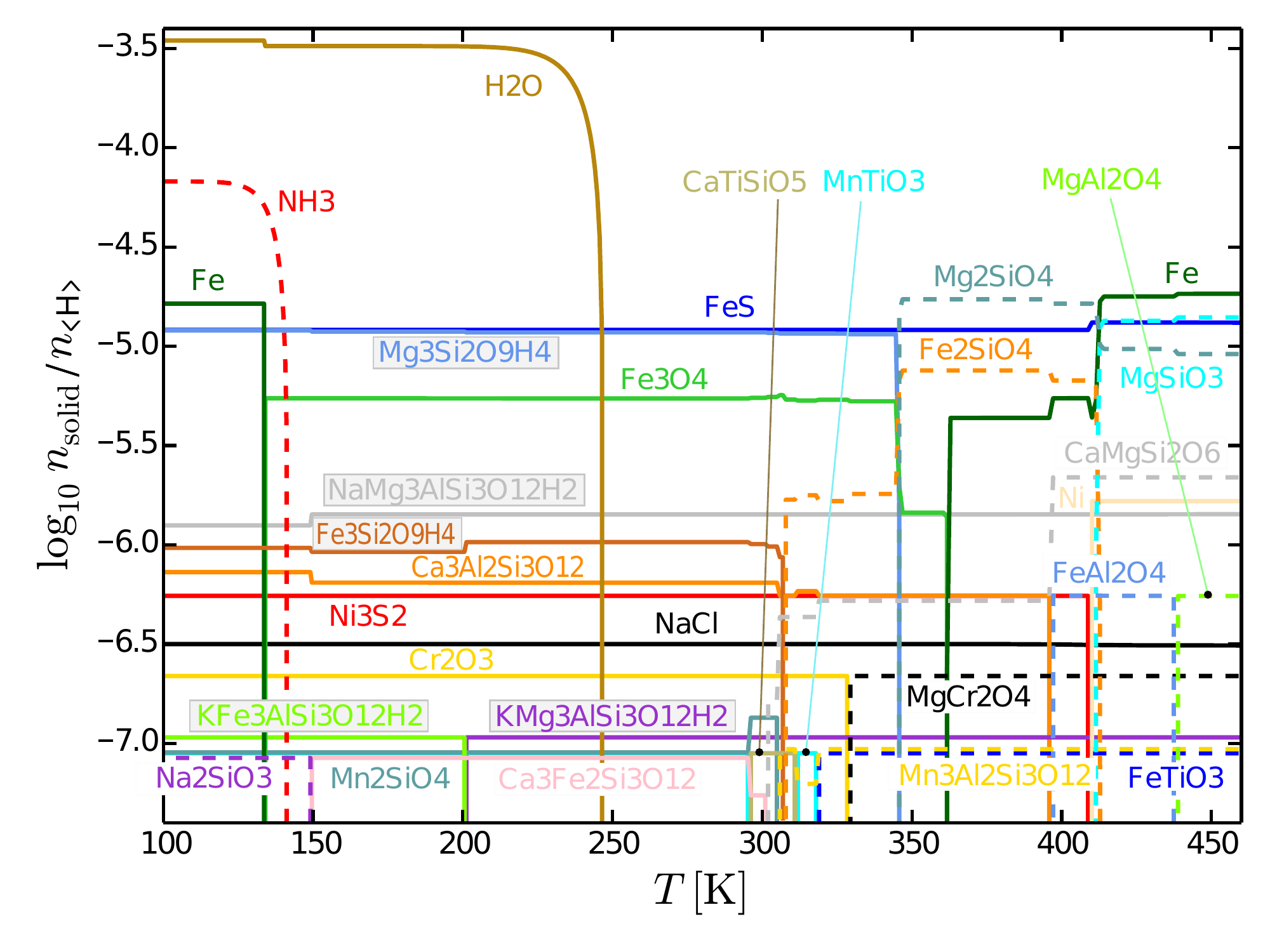}\\[-6mm]
\caption{Concentration of condensed species at temperatures
  $\rm 100\,K - 460\,K$ at $p\!=\!1$\,bar for solar abundances.}
\label{fig:cond3}
\end{figure}

\smallskip\noindent {\bf The formation of phyllosilicates:\ \ } 
Figure~\ref{fig:cond3} shows the results down to
100\,K. Again, a number of complex type-2 transformations occur,
where most of the condensates are replaced by phyllosilicates, step by
step.  The most significant phyllosilicate is Mg$_3$Si$_2$O$_9$H$_4$[s]
{\sl(lizardite)} which replaces Mg$_2$SiO$_4$[s] at 345\,K. At 270\,K,
the main condensates are Mg$_3$Si$_2$O$_9$H$_4$[s], FeS[s], Fe$_3$O$_4$[s]
{\sl(magnetite)}, NaMg$_3$AlSi$_3$O$_{12}$H$_2$[s],
Fe$_3$Al$_2$Si$_3$O$_{12}$[s] {\sl(almandine)}, Ca$_3$Al$_2$Si$_3$O$_{12}$[s]
{\sl(grossular)} and Ni$_3$S$_2$[s] {\sl(heazlewoodite)}, whereas the
more simple condensates commonly known from higher temperatures (fosterite,
enstatite, solid Fe, etc.) are all gone.

\smallskip\noindent {\bf The condensation of water and ammonia:\ \ }
The condensation of the elements is completed by two major type-1
transitions: O-gas $\to$ H$_2$O[s] {\sl(water ice)} at 247\,K and
N-gas $\to$ NH$_3$[s] {\sl(ammonia ice)} at 141\,K. We do not observe
the condensation of CO[s] or CO$_2$[s] {\sl(dry ice)} in our
equilibrium models with solar abundances, although at 1\,bar, CO$_2$
is well-known to deposit to a solid at\linebreak
-79$\rm^{\circ}C\!=\!195\,$K on Earth. This is because the Earth
atmosphere is oxygen-rich and relatively hydrogen-poor. For solar
abundances, all carbon is locked up in gaseous CH$_4$.  At 100\,K, all
elements except hydrogen, carbon and noble gases have condensed into
solid phases, and the composition of the gas phase is extremely simple
-- just H$_2$, He, CH$_4$, and noble gases. The next major phase
transition to occur would be C-gas $\to$ CH$_4$[s] {\sl(methane ice)},
but this transition occurs below 100\,K and our models do not
reach these temperatures.  We note that we are currently lacking
organic liquids/solids in our collection of condensed species, which
might change some of these conclusions.


\begin{figure}
\vspace*{-0.5mm}
\hspace*{-3mm}
\includegraphics[width=94mm,trim=10 531 0 0, clip]{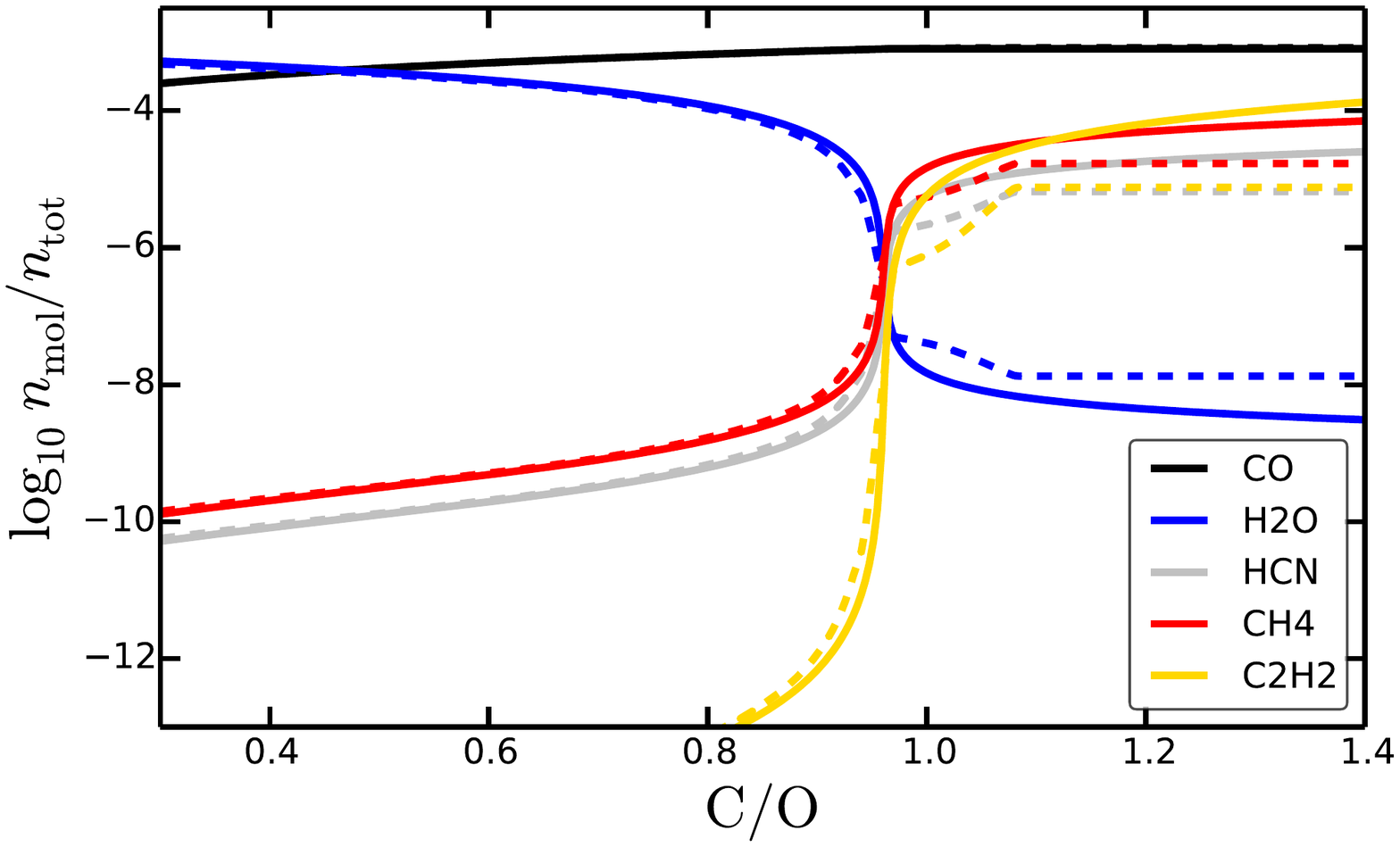}\\[-30mm]
\hspace*{13mm}\begin{minipage}{4cm}
$\small\boxed{p\!=\!10\,{\rm mbar}, \ \ T\!=\!1500\rm\,K}$
\end{minipage}\\[23.5mm]
\hspace*{-3mm}
\includegraphics[width=94mm,trim=0 490 0 12, clip]{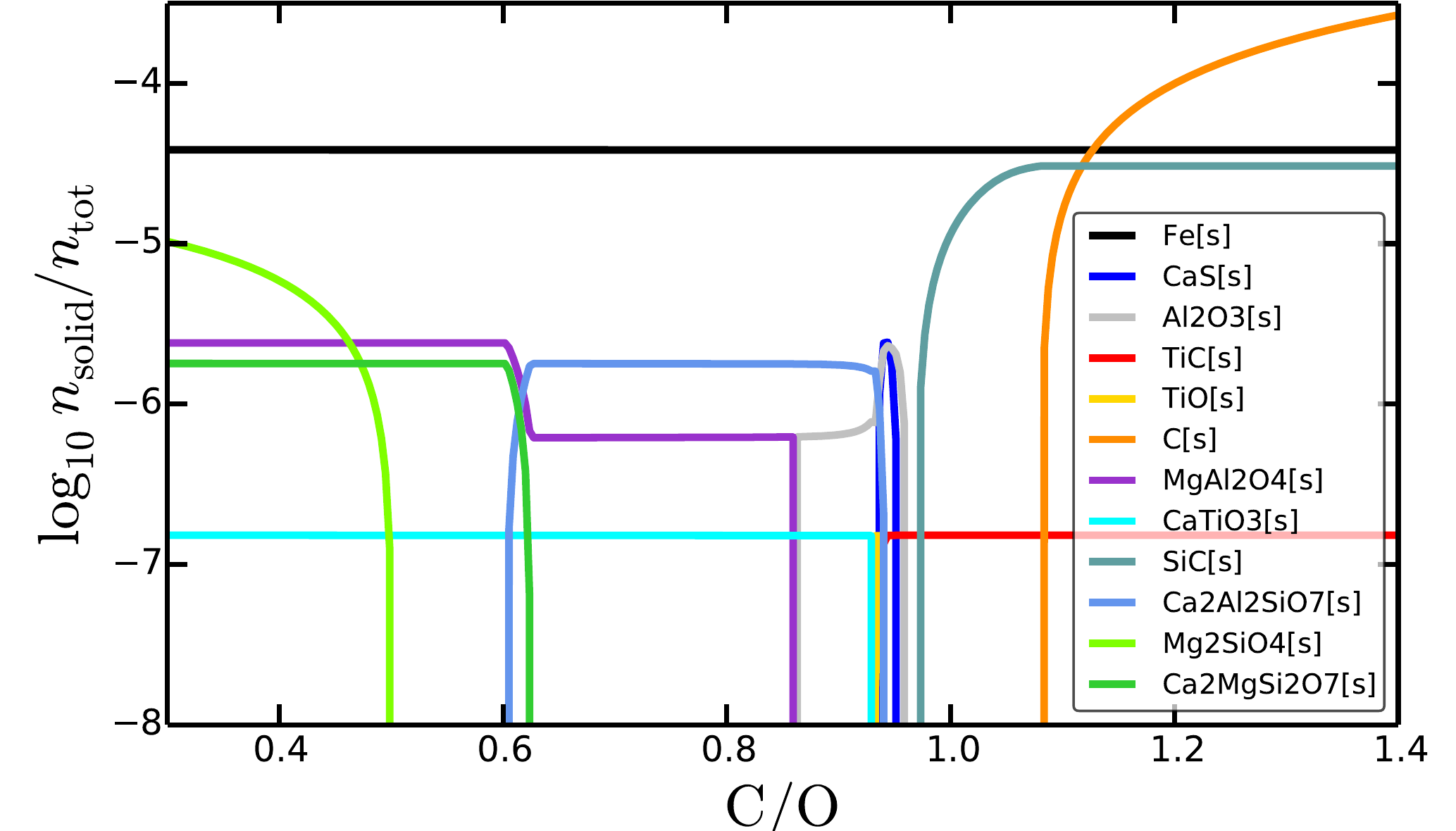}\\[-4mm]
\caption{The impact of CO-blocking on the composition of the gas at
  constant pressure and temperature. In the upper plot, the full lines
  represent the results from a pure gas phase model without
  condensation (24 elements and 445 molecules, see
  Sect.~\ref{sec:bench}), and the dashed lines show the results from
  the model including equilibrium condensation
  (Sect.~\ref{sec:eqcond}). $\epsilon_{\rm O}$ is a fixed value here,
  whereas $\epsilon_{\rm C}$ varies.}
\label{fig:CzuO_dep}
\vspace*{-2mm}
\end{figure}

\subsection{CO-blocking} 
\label{sec:CO-block}

Figure~\ref{fig:CzuO_dep} demonstrates the well-known impact of the
carbon-to-oxygen ratio ${\rm C/O}=\epsilon_{\rm C}/\epsilon_{\rm O}$
on the composition of the gas phase, and on the condensation.  The
figure is designed to match the results of \citet{Molliere2015},
although here (in contrast to Molli{\`e}re et al.) we are plotting
particle concentrations.  At temperatures between about $850\,$K and
$4000\,$K at 10\,mbar, CO is the dominant gas species in the
carbon-oxygen system with a dissociation energy of about
11.1\,eV. Therefore, CO almost completely consumes all carbon when
$\rm C/O\!\la\!1$ or all oxygen when $\rm C/O\!\ga\!1$, an effect
known as {\sl CO-blocking} \citep[see
  e.g.][]{Sedlmayr1995,Sedlmayr1997,Beck1992}.  The effect is
particularly strong on hydro-carbon molecules like C$_2$H$_2$ {\sl
  (acetylene)} which are suppressed by more than 6 orders of magnitude
if $\rm C/O\!<\!1$.

Closer inspection of our results (Fig.~\ref{fig:CzuO_dep}) shows that
the transition actually takes place around $\rm C/O\!\approx\!0.96$,
which agrees well with the Molli{\`e}re et al.\ results, who have been
using the CEA code \citep{Gordon1994,McBride1996}.  Figure~1 in
\citet{Madhusudhan2012} actually shows the same effect, although
Madhusudhan has used a log-scaling of the C/O-axis where deviations of
order 5\% are simply not visible. The reason for this asymmetry is the
formation of the SiO molecule with a dissociation energy of
8.2\,eV. If $\rm C/O\!\la\!1$, the small amount of excess oxygen is
first of all consumed to form SiO molecules, with similar consequences
for the abundances of water and hydro-carbon molecules as used from
the CO-blocking.  We measure numerically that the abundances of H$_2$O
and CH$_4$ intersect at $\rm C/O\!=\!1.00$ when $\epsilon_{\rm Si}\to
0$, at 0.98 when $\epsilon_{\rm Si}\!=\!7.3$, at 0.96 when
$\epsilon_{\rm Si}\!=\!7.51$\,(solar), at 0.92 when $\epsilon_{\rm
  Si}\!=\!7.7$, and at 0.82 when $\epsilon_{\rm Si}\!=\!8$.


Figure~\ref{fig:CzuO_dep} also shows the influence of the CO-blocking
on the condensation. At 1500\,K and 10\,mbar, C[s] {\sl (graphite)}
can condense at $\rm C/O\!\ga\!1.07$, SiC[s] {\sl(silicon carbide)} at
$\rm C/O\!\ga\!0.96$ and TiC[s] {\sl(titanium carbide)} at $\rm
C/O\!\ga\!0.93$. At smaller C/O, the oxygen-containing solids can
form, here in particular Al$_2$O$_3$[s] {\sl(corundum)},
Ca$_2$Al$_2$SiO$_7$[s] {\sl(gehlenite)}, MgAl$_2$O$_4$[s]
{\sl(spinel)}, Ca$_2$MgSi$_2$O$_7$[s] {\sl(akermanite)}, CaTiO$_3$[s]
{\sl(perovskite)} and Mg$_2$SiO$_4$ {\sl(fosterite)}.  Other
condensates like W[s] and ZrO$_2$[s] (not shown) and Fe[s] are not
affected by C/O, because they either do not contain oxygen or carbon,
or have even higher dissociation energies than CO, like ZrO$_2$[s]. 

The feedback of condensation on the molecular abundances for $\rm
C/O\!>\!1$ is more substantial than for $\rm C/O\!<\!1$. The
molecular concentrations of C$_2$H$_2$ and CH$_4$ are found to not
increase any further once carbon dust is present, as any surplus
carbon is simply converted into additional carbon dust, but not into
additional hydro-carbon molecules in phase equilibrium. We have
actually used the total (gas $+$ dust) abundances of oxygen and carbon
to create Fig.~\ref{fig:CzuO_dep}, because the gas-phase C/O only
increases to about 1.1 even if much larger values of the total C/O are
used.

At lower temperatures ($T\!<\!850\,$K at $p\!=\!10\,$mbar), however,
CH$_4$ becomes the dominating carbon gas phase species, whereas CO
becomes a trace species, and so the spell is broken. In a
hydrogen-rich gas, CH$_4$ is thermodynamically more favourable than
any solid carbon compound at low temperatures, which leads to high
water concentrations independent of C/O, and no carbon dust in phase
equilibrium. This confirms our earlier findings in kinetic cloud
formation models for carbon-rich atmospheres \citep[Fig.~6
  in][]{Helling2017}.

\begin{figure*}
\vspace*{-3mm}
\resizebox{189mm}{!}{
\begin{tabular}{ccc}
\hspace*{-6mm}
\includegraphics[height=53mm,trim=10 40 0 10, clip]{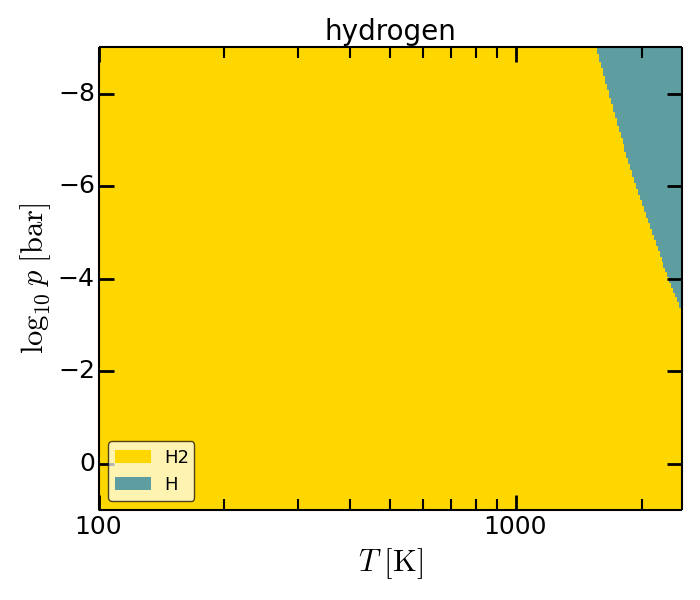} &
\hspace*{-6mm}
\includegraphics[height=53mm,trim=43 40 0 10, clip]{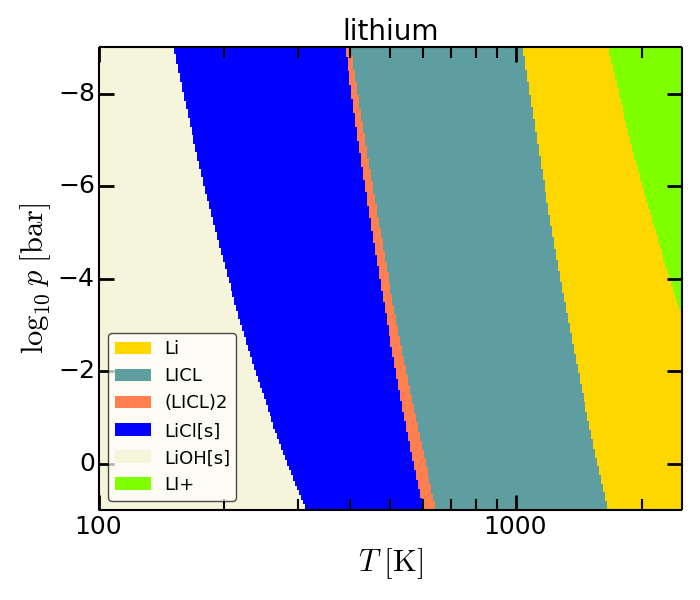}  &
\hspace*{-6mm}
\includegraphics[height=53mm,trim=43 40 0 10, clip]{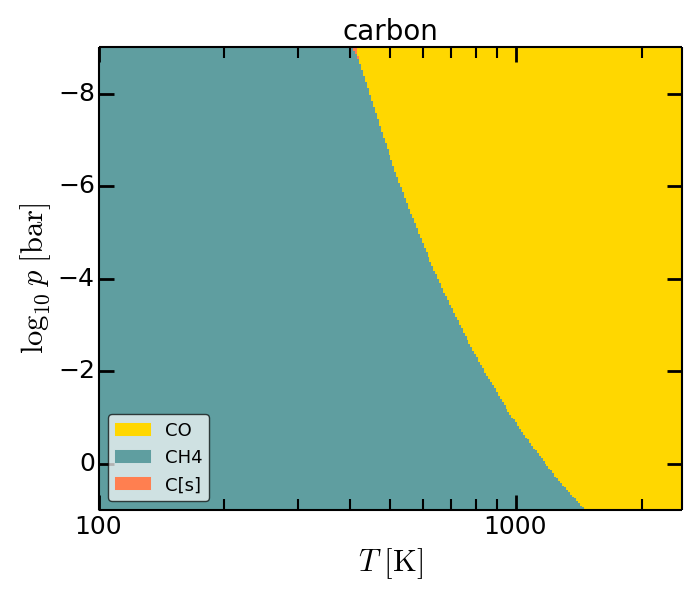} 
\\[-1mm]
\hspace*{-6mm}
\includegraphics[height=53mm,trim=10 40 0 10, clip]{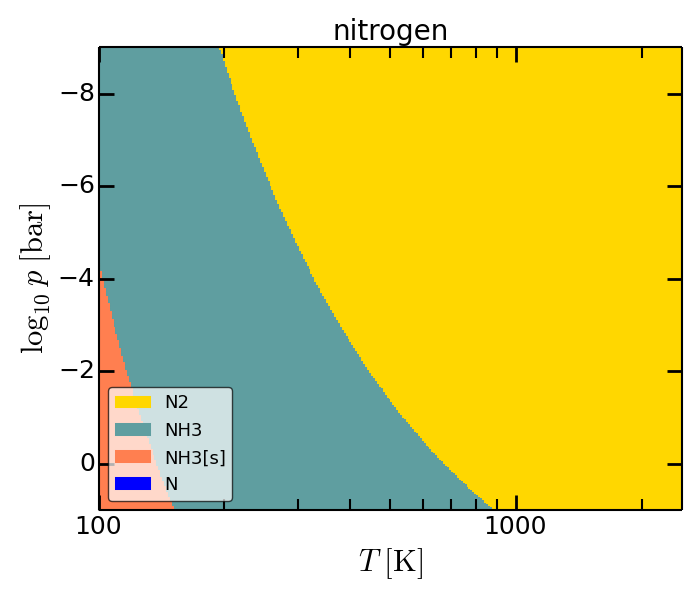} &
\hspace*{-6mm}
\includegraphics[height=53mm,trim=43 40 0 10, clip]{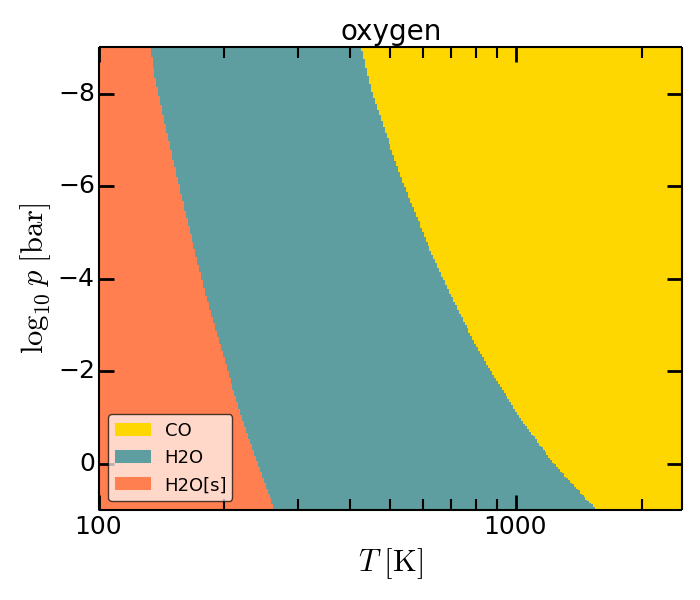}   &
\hspace*{-6mm}
\includegraphics[height=53mm,trim=43 40 0 10, clip]{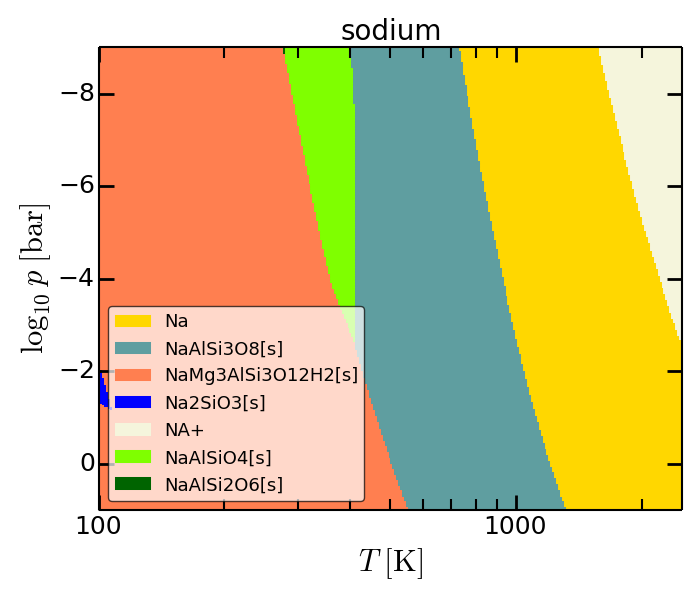}
\\[-1mm]
\hspace*{-6mm}
\includegraphics[height=53mm,trim=10 40 0 10, clip]{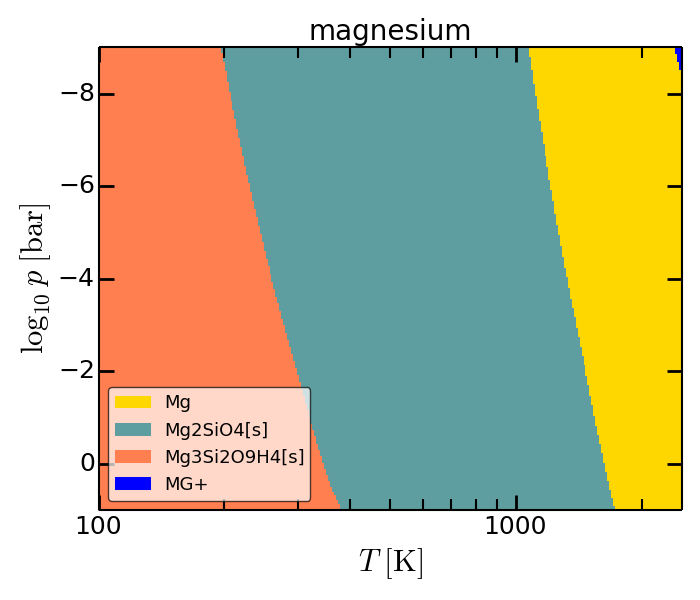} &
\hspace*{-6mm}
\includegraphics[height=53mm,trim=43 40 0 10, clip]{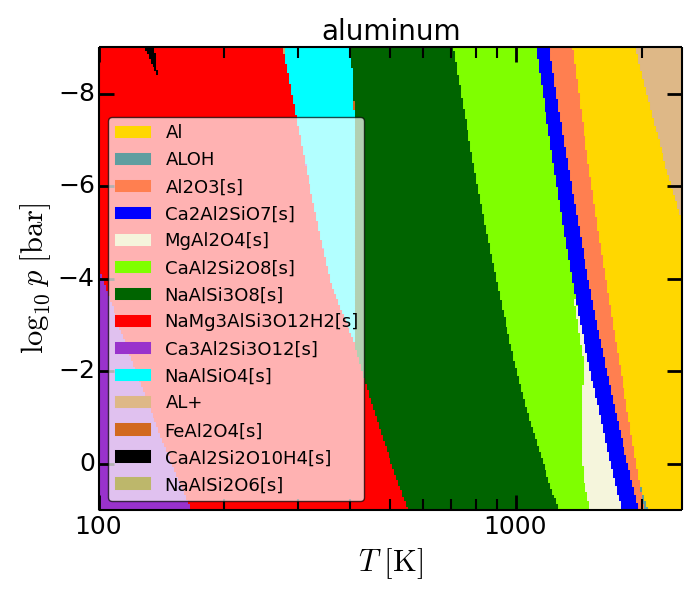} &
\hspace*{-6mm}
\includegraphics[height=53mm,trim=43 40 0 10, clip]{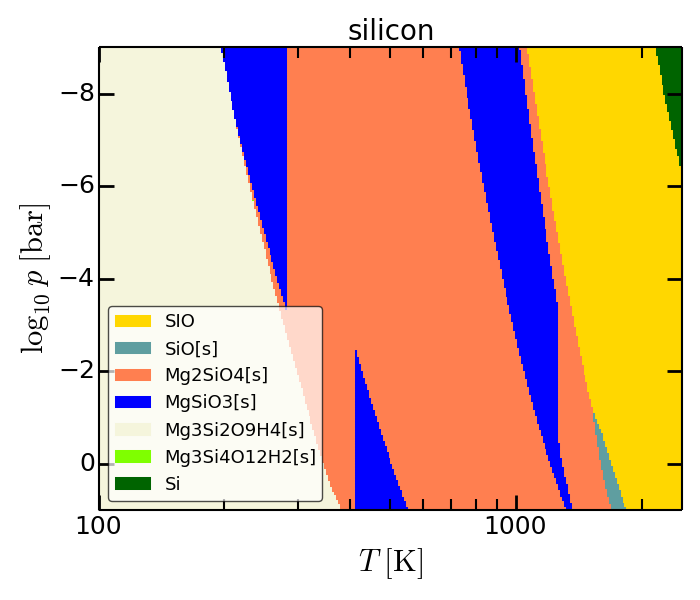} 
\\[-1mm]
\hspace*{-6mm}
\includegraphics[height=55mm,trim=10 17 0 10, clip]{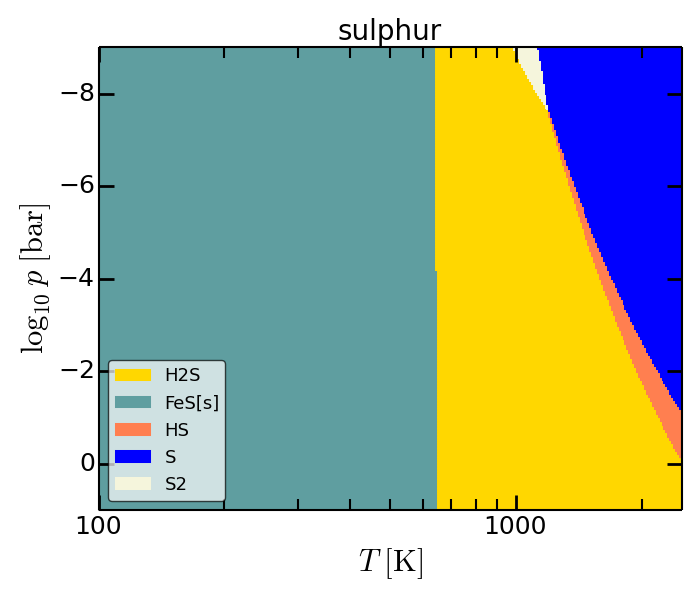} &
\hspace*{-6mm}
\includegraphics[height=55mm,trim=43 17 0 10, clip]{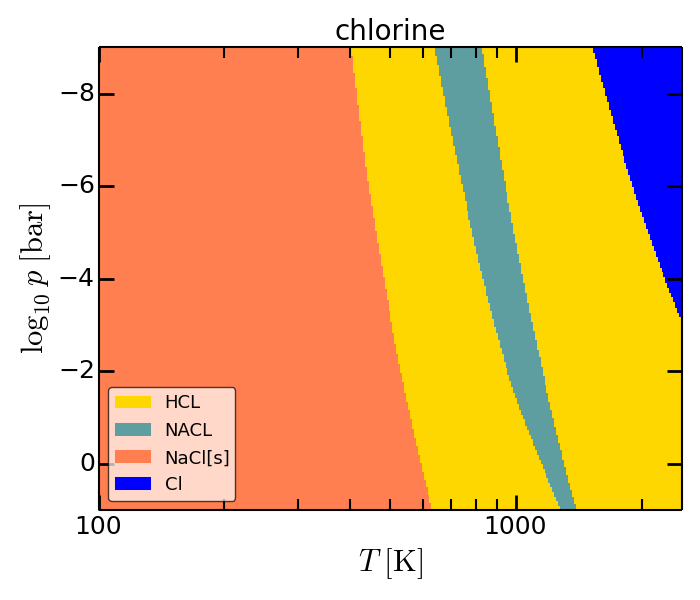} &
\hspace*{-6mm}
\includegraphics[height=55mm,trim=43 17 0 10, clip]{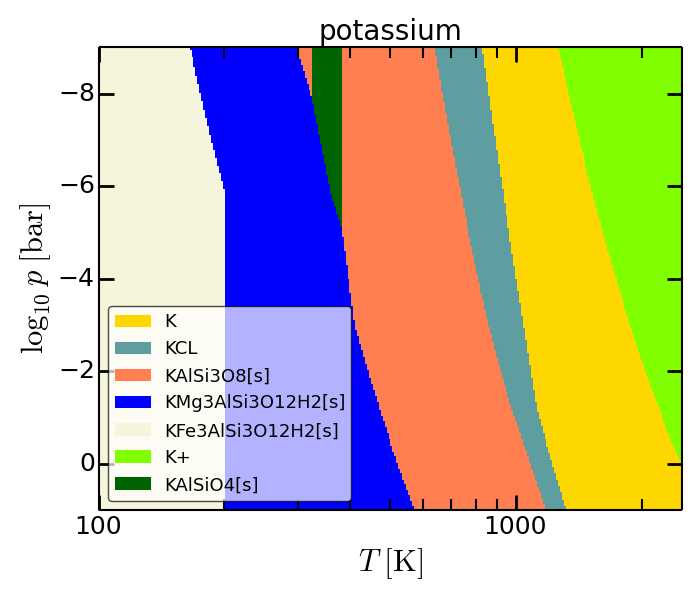} 
\\[-2mm]
\end{tabular}}
\caption{Gas phase/condensed species which contain most of the
  elements in the $(p,T)$-plane. Solid species are
  marked by '[s]' and liquid species by '[l]'.  At low temperatures,
  the elements Mg, Si, Na, Al and K are mostly present in form of
  phyllosilicates.}
\label{fig:phase}
\vspace*{5cm}
\end{figure*}
\addtocounter{figure}{-1}
\begin{figure*}
\vspace*{-3mm}
\resizebox{189mm}{!}{
\begin{tabular}{ccc}
\hspace*{-6mm}
\\[-1mm]
\hspace*{-6mm}
\includegraphics[height=53mm,trim=10 40 0 10, clip]{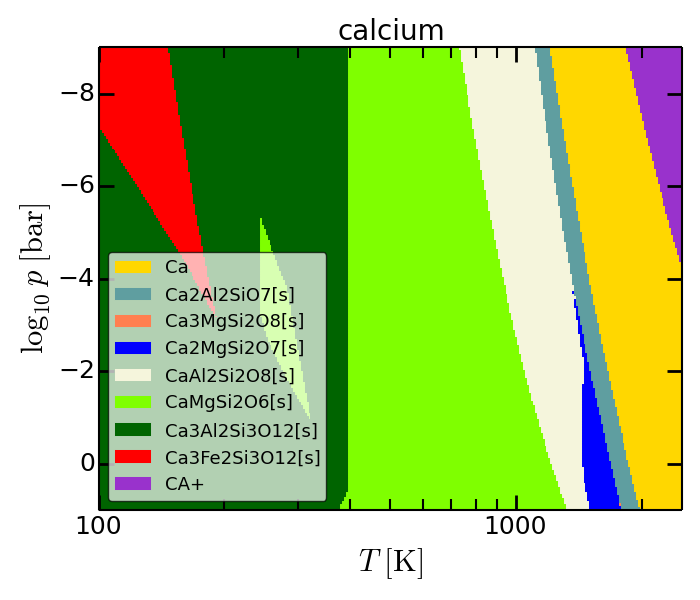} &
\hspace*{-6mm}
\includegraphics[height=53mm,trim=43 40 0 10, clip]{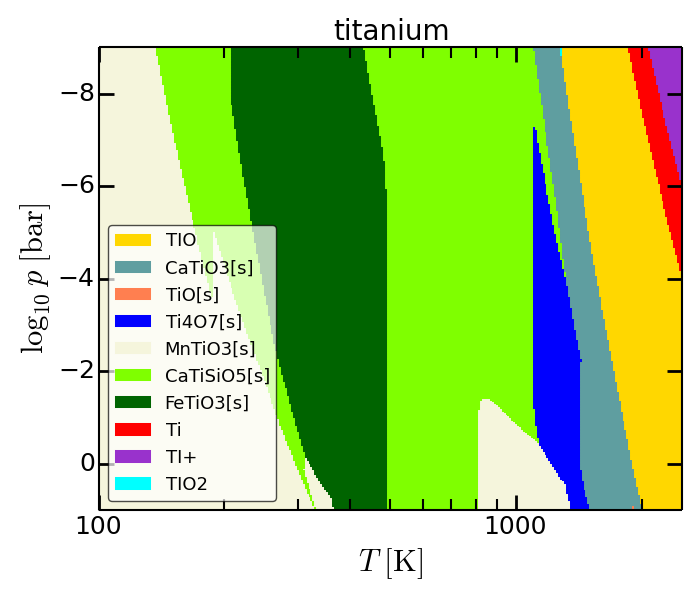} &
\hspace*{-6mm}
\includegraphics[height=53mm,trim=43 40 0 10, clip]{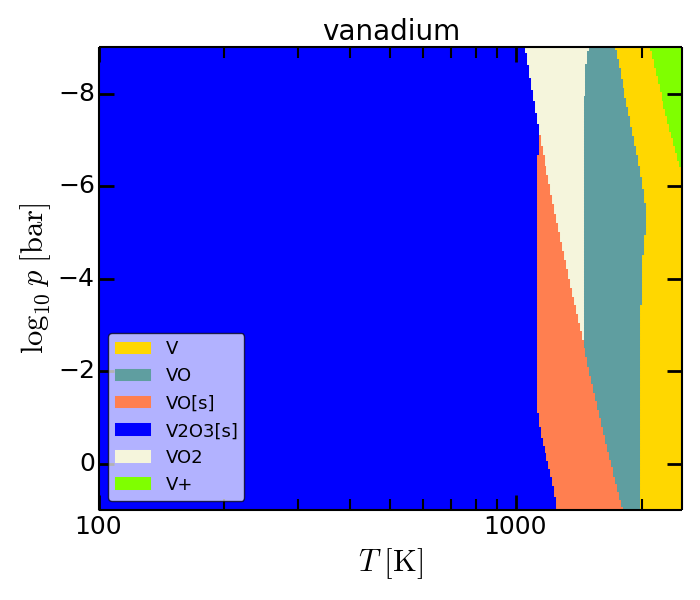} 
\\[-1mm]
\hspace*{-6mm}
\includegraphics[height=53mm,trim=10 40 0 10, clip]{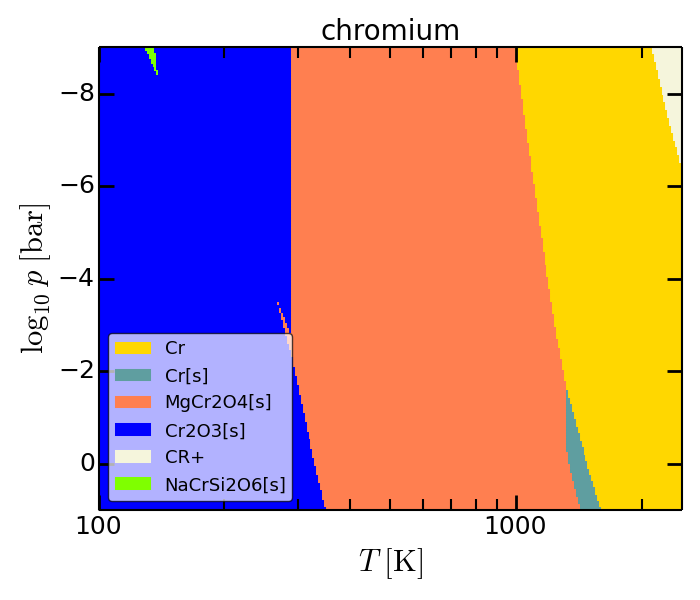} &
\hspace*{-6mm}
\includegraphics[height=53mm,trim=43 40 0 10, clip]{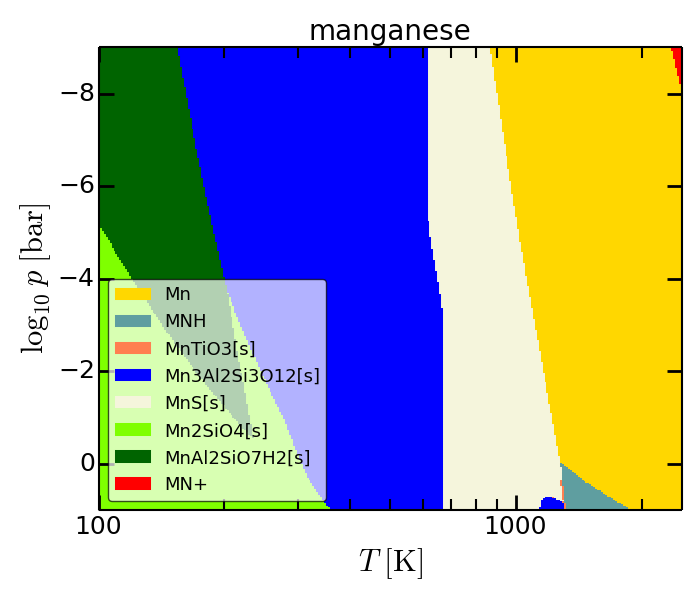} &
\hspace*{-6mm}
\includegraphics[height=53mm,trim=43 40 0 10, clip]{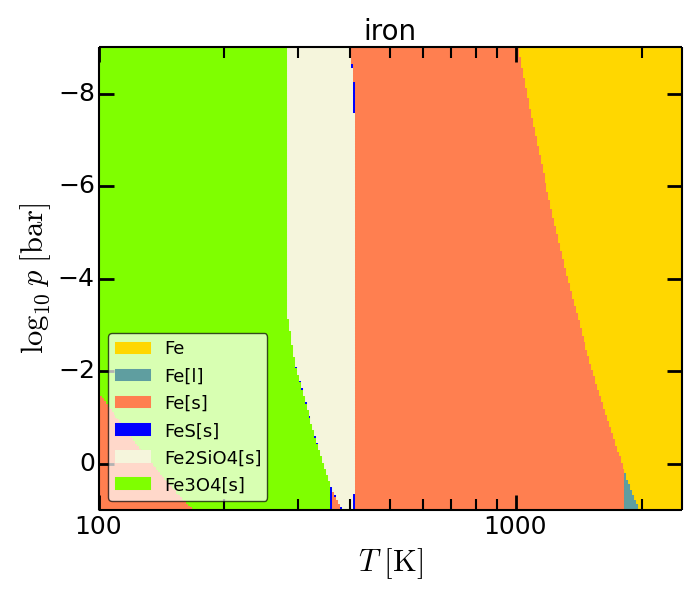} 
\\[-1mm]
\hspace*{-6mm}
\includegraphics[height=55mm,trim=10 17 0 10, clip]{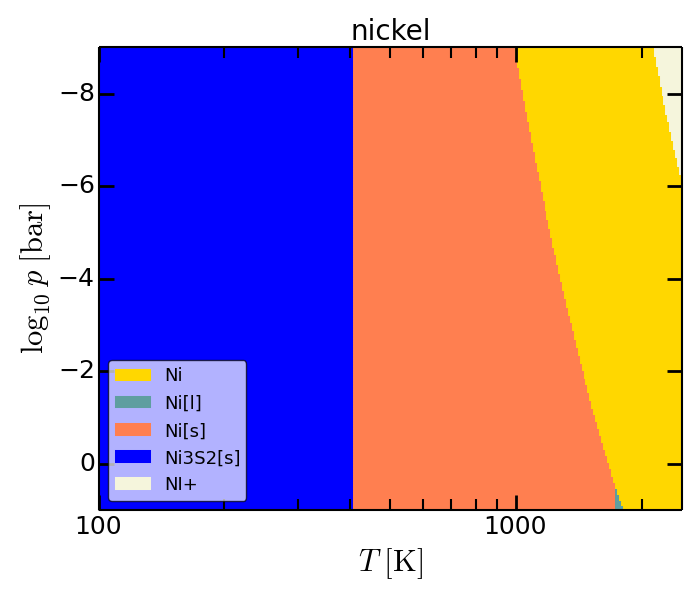} &
\hspace*{-6mm}
\includegraphics[height=55mm,trim=43 17 0 10, clip]{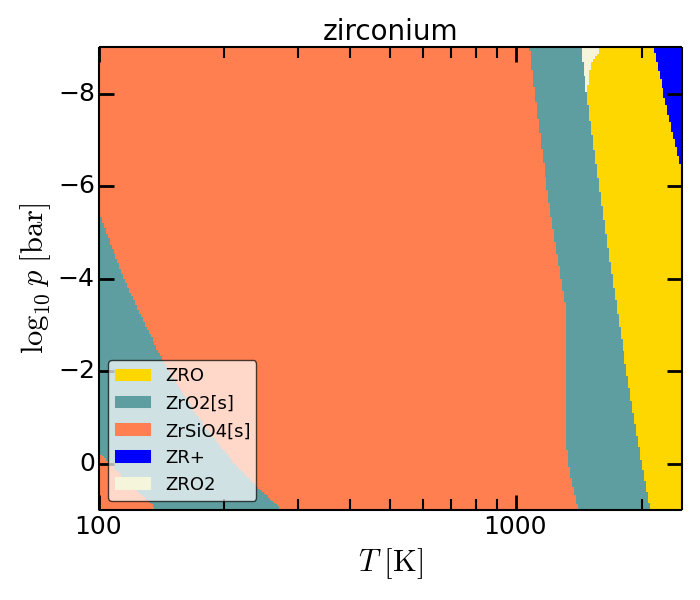} &
\hspace*{-6mm}
\includegraphics[height=55mm,trim=43 17 0 10, clip]{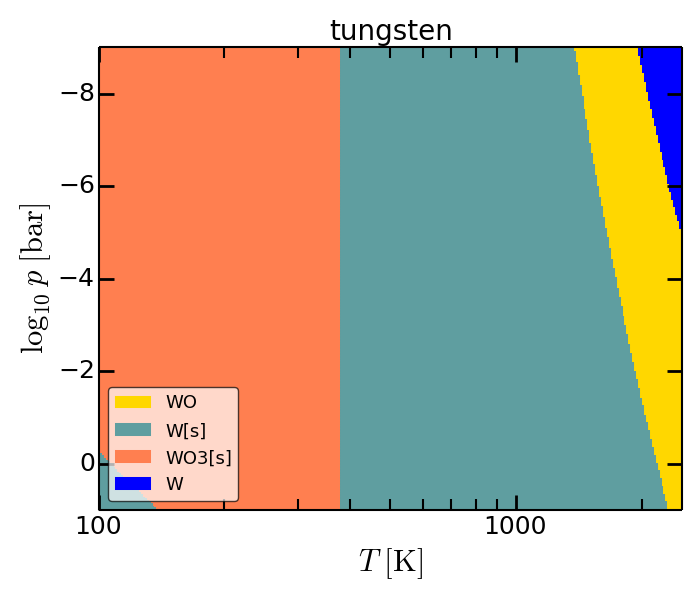} 
\\[-2mm]
\end{tabular}}
\caption{{\bf continued:} The only further element found to form
  phyllosilicates in phase equilibrium is manganese.}
\end{figure*}

\subsection{Phase diagrams of the elements}
\label{sec:phase}

Figure~\ref{fig:phase} shows some results of the equilibrium
condensation model (22 elements, charges, 365 molecules and 190
condensed species, see Sect.~\ref{sec:eqcond}) in the entire
$(p,T)$-plane between $10^{-9}$\,bar and 100\,bar, and between 100\,K and
2500\,K, for solar abundances. The figures indicate the most abundant
species that contain certain elements, after multiplication by their
stoichiometric factor, hence the main reservoirs of the elements as
function of pressure and temperature.

We generally see increasing complexity from high to low temperature,
from ions over atoms to simple molecules, to complex molecules, and
then to condensates with increasingly complex stoichiometry and
phyllosilicates at the lowest temperatures. In the first approximation,
all molecular transitions, as well as the type-1 and type-2 phase
transitions discussed in Sect.~\ref{sec:eqcond} simply occur at lower
temperatures when the pressure is decreased, roughly a factor of 2-4
in temperature for 10 decades of pressure decrease. This leads to
numerous slanted and slightly curved phase transition lines, similar
to phase diagrams in thermodynamics. Most prominent molecular
transitions are $\rm H\to H_2$ (at $\sim$\,2200\,K), $\rm CO\to CH_4$
(at $\sim$\,650\,K) and $\rm N_2\to NH_3$ (at $\sim$\,320\,K), where
the numbers in brackets refer to a pressure of $p\!=\!10^{-4}$\,bar.

The details, however, are in fact much more complicated, in particular
concerning the gas\,$\to$\,solid transitions. The order of certain
phase transitions as described in Sect.~\ref{sec:eqcond} may be
different, and some condensates may be present only in certain parts
of the $(p,T)$-plane. The main silicates can form below about 1700\,K
to 1100\,K, and eventually the elements Mg, Si, Na, Al, K and Mn can form
thermodynamically stable phyllosilicates at temperatures below about 600\,K to
200\,K.

We want to stress, however, that all these results depend on the
chosen element abundances, as demonstrated for the case of varying C/O
in Sect.~\ref{sec:CO-block}. The results do also depend technically on
the choice of elements, so if another element is taken into account in
addition, it will interfere with all others which can potentially
change the results shown in Fig.~\ref{fig:phase}. Therefore, our
``phase diagrams'' are not as general as their name suggests. The
results should not be taken out of context and be applied to e.g. the
atmospheres of rocky planets. The proper way to do this is to run
the {\sc GGchem}-code for the expected pressures,
temperatures and local element abundances in the
atmosphere.

\subsection{The C/O-ratio as affected by metal oxide molecules and condensation}
\label{sec:C/O}

\begin{figure*}
\centering
\begin{tabular}{cc}
\hspace*{6mm}{\sf pure gas-phase model} & 
\hspace*{9mm}{\sf equilibrium condensation model}\\[-1mm]
\includegraphics[width=80mm,trim=0 0 0 0, clip]{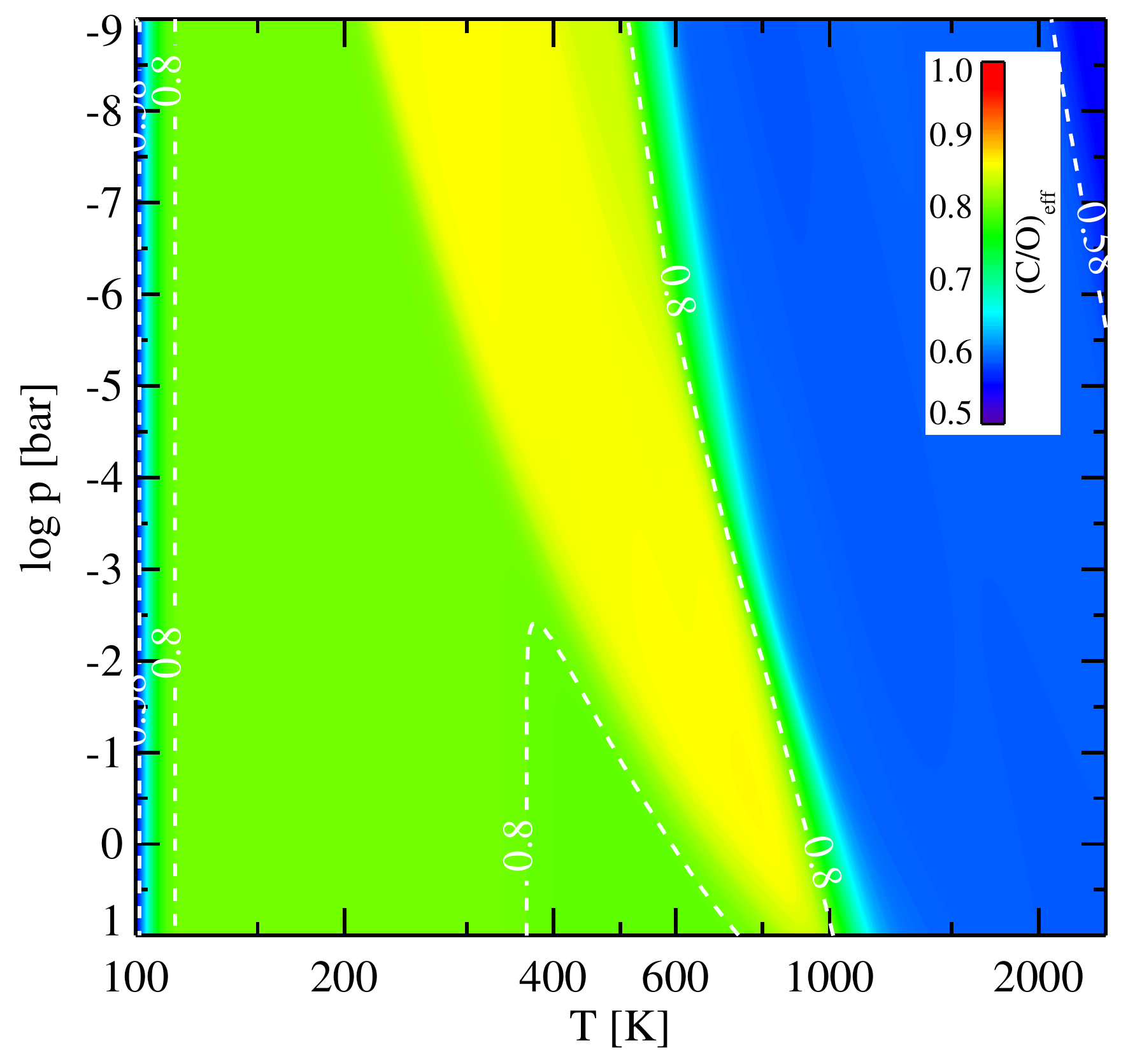} &
\includegraphics[width=80mm,trim=0 0 0 0, clip]{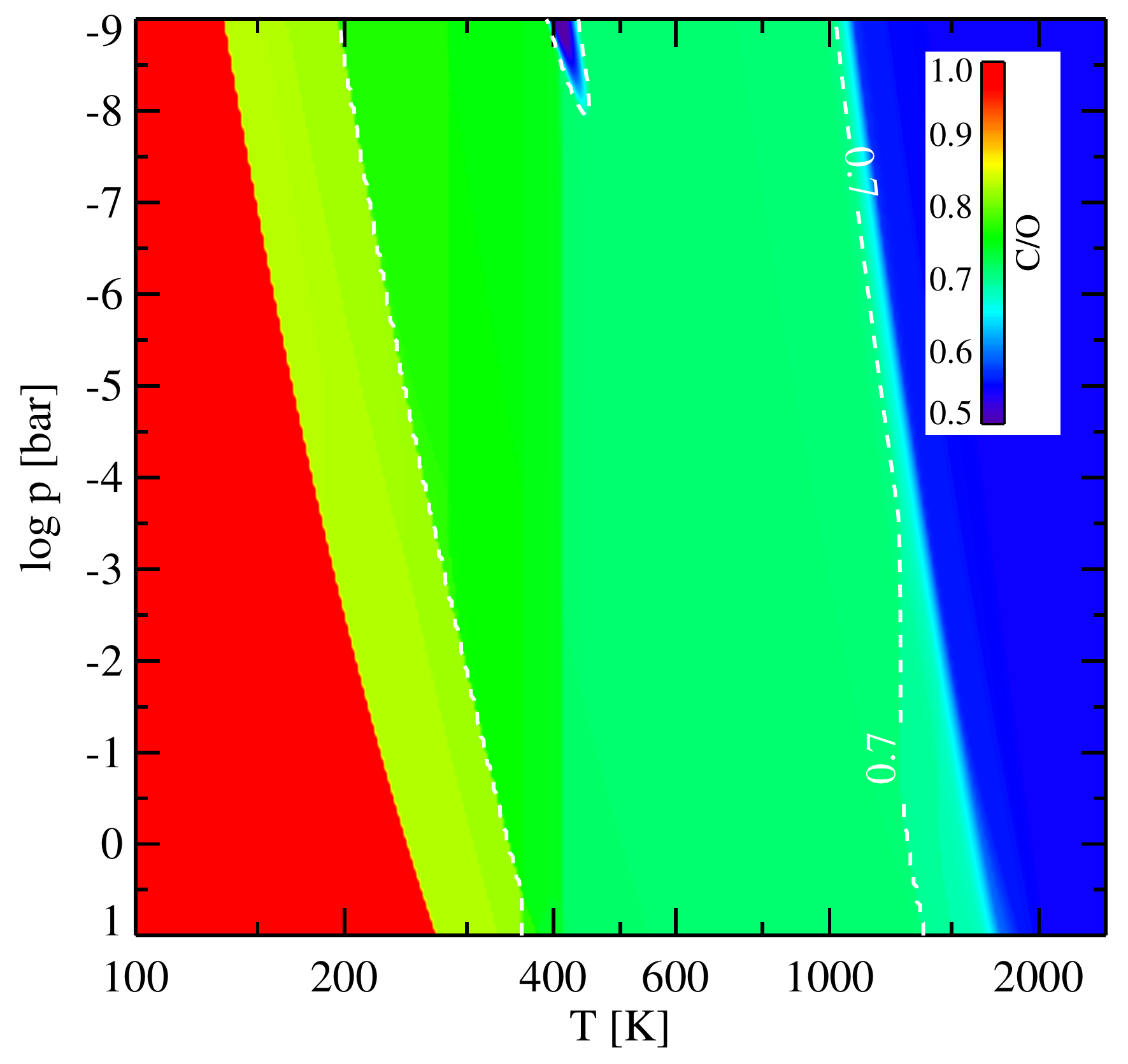}
\end{tabular}\\[-3mm]
\caption{The C/O ratio in the gas phase affected by the formation of
  metal oxide and hydroxide molecules (left) and condensation of
  silicates, phyllosilicates, and water ice (right). On the left side,
  we plot the ``effective'' C/O that is contained in the gas in form
  of molecules made of H-C-N-O only, but do not count metal oxide and
  hydroxide molecules. The contour lines shown are 0.58 (after SiO
  formation) and 0.8 (after Mg(OH)$_2$ and Fe(OH)$_2$ formation). On
  the right side, we plot the resulting gas phase abundances after the
  condensation of solids and liquids. The increase of $\rm
  C/O\!>\!0.7$ is due to the consumption of oxygen by the major
  silicates like MgSiO$_3$[s] and Mg$_2$SiO$_4$[s], and the further
  increase to $>\!0.8$ is due to phyllosilicates. The sharp increase
  at low temperatures (red area) is due to water condensation.}
\label{fig:CzuO}
\vspace*{-0.5mm}
\end{figure*}

Figure~\ref{fig:CzuO} (right side) shows the C/O ratio in the gas
phase as affected by condensation in the $(p,T)$-plane. One of the
most significant results of this paper again shows, namely that the
solar value of about C/O\,$=$\,0.55 only holds prior to the
condensation of the main silicates. The consumption of oxygen by
silicate formation is substantial, and increases the C/O ratio to
about 0.71. Once the silicates become hydrated to form
phyllosilicates, which happens to the most abundant silicates between
about 200\,K to 400\,K in the equilibrium model (depending on
pressure), the C/O ratio jumps to 0.83, and then gets very large once
water condenses\footnote{Such strong changes in C/O, however, were not
  found in kinetic cloud formation models \citep{Helling2016}.}.
The little blue area in the central top of Fig.~\ref{fig:CzuO} is not
a computational error -- it is a small region in the $(p,T)$-plane
where we find solid carbon C[s] {\sl(graphite)} to condense.

\begin{figure*}
\begin{tabular}{ccc}
{\small\sf MgSiO$_3$[s] {\sl(enstatite)}} & 
{\small\sf Mg$_2$SiO$_4$[s] {\sl(fosterite)}} & 
{\small\sf Mg$_3$Si$_2$O$_9$H$_4$[s] {\sl(lizardite)}}\\[-1mm]
\hspace*{-3mm}\includegraphics[width=62.5mm,trim=0 0 0 0, clip]{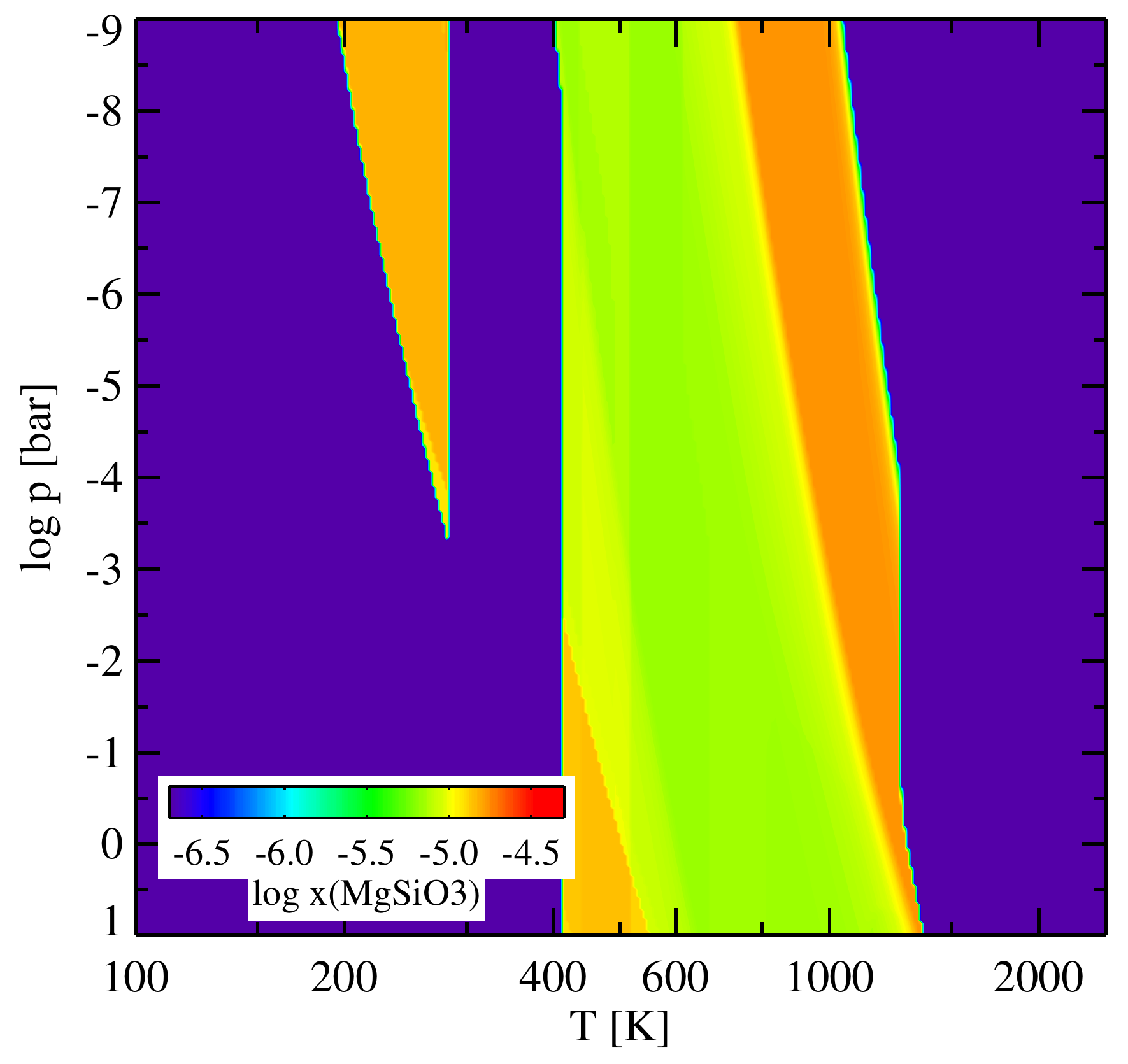} &
\hspace*{-5mm}\includegraphics[width=62.5mm,trim=0 0 0 0, clip]{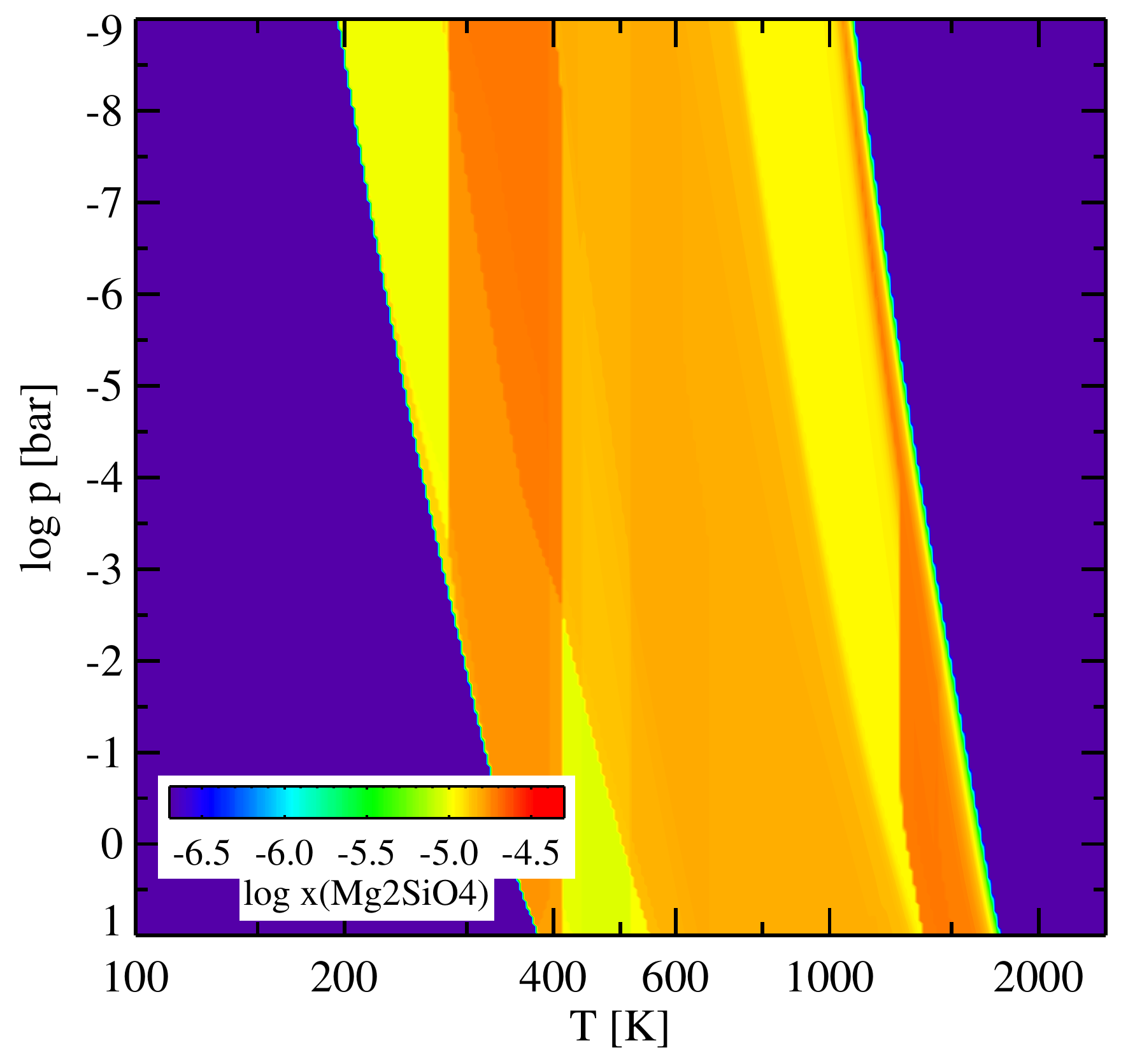} &
\hspace*{-5mm}\includegraphics[width=62.5mm,trim=0 0 0 0, clip]{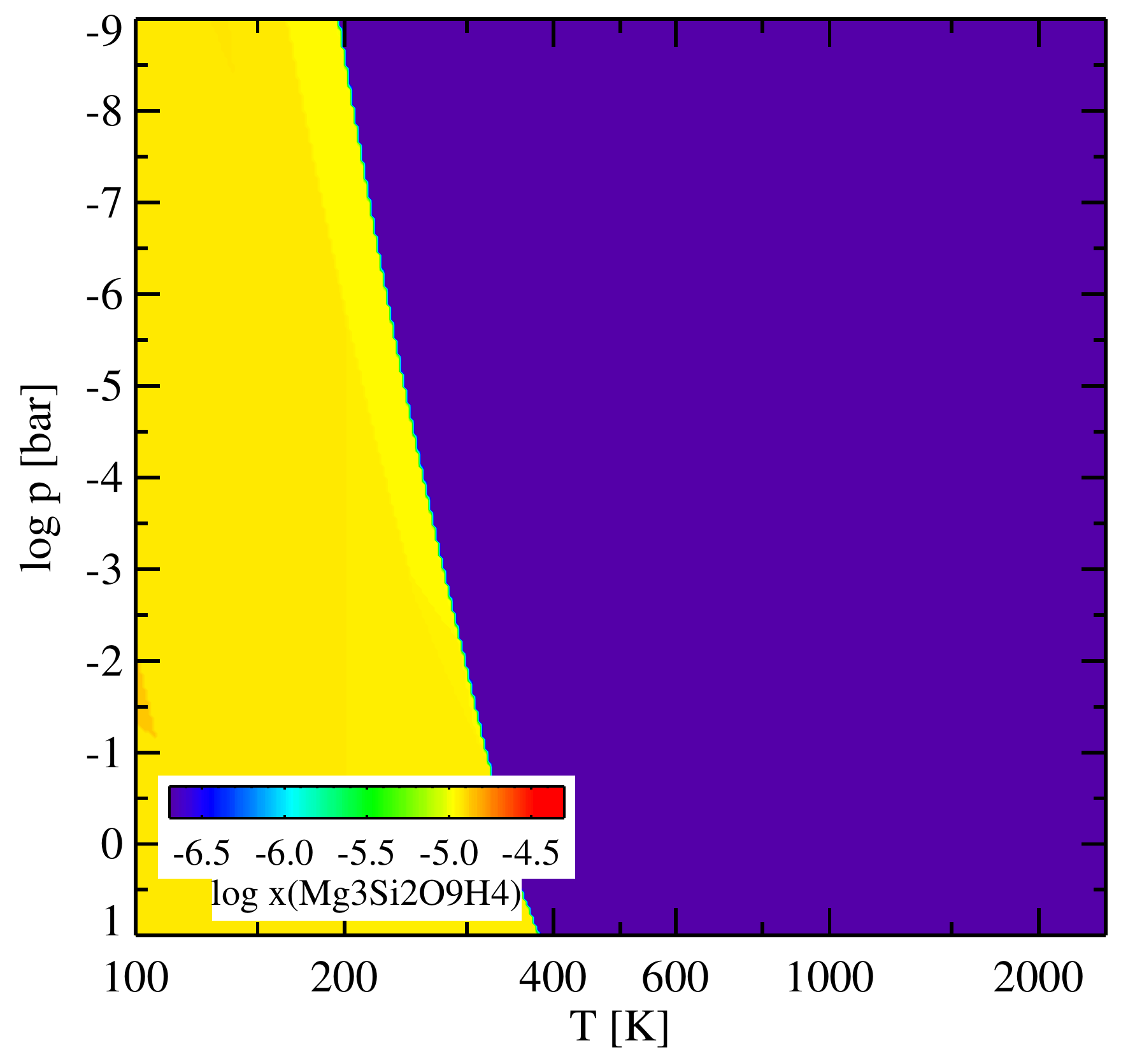}\\[-3mm]
\end{tabular}
\caption{The concentration $(x_j=n_j/\nH)$ of the three major
  magnesium-silicates in the $(p,T)$-plane in the equilibrium
  condensation model.}
\label{fig:silicates}
\vspace*{-2mm}
\end{figure*}

To clarify the effect of condensation on gas phase C/O, we compare our
results to a pure gas phase model (without condensation) where
we compute an ``effective'' C/O as
\begin{eqnarray}
  {\rm (C/O)_{eff}} &\!\!\!=\!\!\!& 
    \tilde{n}_{\langle\rm C\rangle}/\tilde{n}_{\langle\rm O\rangle} 
  \label{eq:C/Oeff}\\
  \tilde{n}_{\langle\rm C\rangle} &\!\!\!=\!\!\!& 
    \sum_i n_i \tilde{s}_{i,\rm C} \quad,\quad
  \tilde{n}_{\langle\rm O\rangle} = \sum_i n_i \tilde{s}_{i,\rm O} \ ,
  \nonumber
\end{eqnarray}   
where only atoms and molecules are counted here if they are made of the
elements H, C, N and O, but we set the stoichiometric coefficient
$\tilde{s}_{i,\rm C}\!=\!0$ \ if a molecule $i$ contains any other
(metal) elements like Mg, Si or Fe.  In a pure gas phase model, as
plotted on the left side of Fig.~\ref{fig:CzuO}, the proper $\rm
C/O\!\approx\!0.55$ is of course constant everywhere in the
$(p,T)$-plane by assumption, but $\rm (C/O)_{eff}$ will be different
because of the oxygen and carbon that is bound in molecules which are
neglected in Eq.\,(\ref{eq:C/Oeff}). We do this because in fast
spectral retrieval models \citep[e.g.][]{Lavie2017}, the influence of
these metals on the concentration of molecules like H$_2$O and CO$_2$
is often neglected, i.e.\ $\rm (C/O)_{eff} \approx C/O$ is assumed.

In Fig.~\ref{fig:CzuO} (left side) we see that $\rm (C/O)_{eff}$ may
become substantially larger than C/O at lower temperatures.  $\rm
(C/O)_{eff}$ increases from the solar value 0.55 to about 0.58 due to
the formation of SiO between about 2000\,K and 3000\,K, and then
increases further to values above 0.8 once Mg(OH)$_2$ and Fe(OH)$_2$
become abundant in the gas phase, which happens at temperatures
between about 600\,K and 1100\,K.

In both discussed cases, the large pure gas phase model and the
equilibrium condensation model, the reason for the increase of the
(effective) C/O ratio is the same -- the consumption of oxygen atoms
by the formation of metal compounds, either oxide and hydroxide
molecules or condensates, which are then missing for the formation of gaseous
molecules like H$_2$O and CO$_2$. The effect is actually similar in
both models, because each Mg, Si and Fe atom, on average, consumes
about $1-2$ oxygen atoms in both cases at low temperatures. The main
difference is that in models including condensation, the establishment
of large C/O ratios already occurs at higher temperatures, between
about 1200\,K to 1600\,K. We discuss this point further in
Sect.~\ref{s:sc}.

Figure~\ref{fig:silicates} shows the abundance of the three most
significant Mg-silicates in the $(p,T)$-plane. As described in
Sect.\,\ref{sec:eqcond}, the two major reservoirs for condensed Mg and
Si at relatively high temperatures are MgSiO$_3$[s] {\sl(enstatite)} and
Mg$_2$SiO$_4$[s] {\sl(fosterite)}. However, some more stable silicates
exist, which contain e.g.\ Ca, Fe, Ti, Mn or Cr. These can form in
smaller amounts only, due to element abundance constraints, so they
``steal'' some Mg and Si from enstatite and fosterite, which affects
the concentrations of enstatite and fosterite in complicated
ways. Fosterite generally becomes stable at slightly higher
temperatures than enstatite, but once formed, enstatite can become the
most abundant condensate in certain parts of the $(p,T)$-plane, along
with Fe[s], Fe$_2$SiO$_4$[s], FeS[s] or Fe$_2$O$_3$[s]. 

\begin{figure*}
\vspace*{-2mm}
\hspace*{-4mm}
\includegraphics[height=63.5mm,trim= 0 7 9 0, clip]{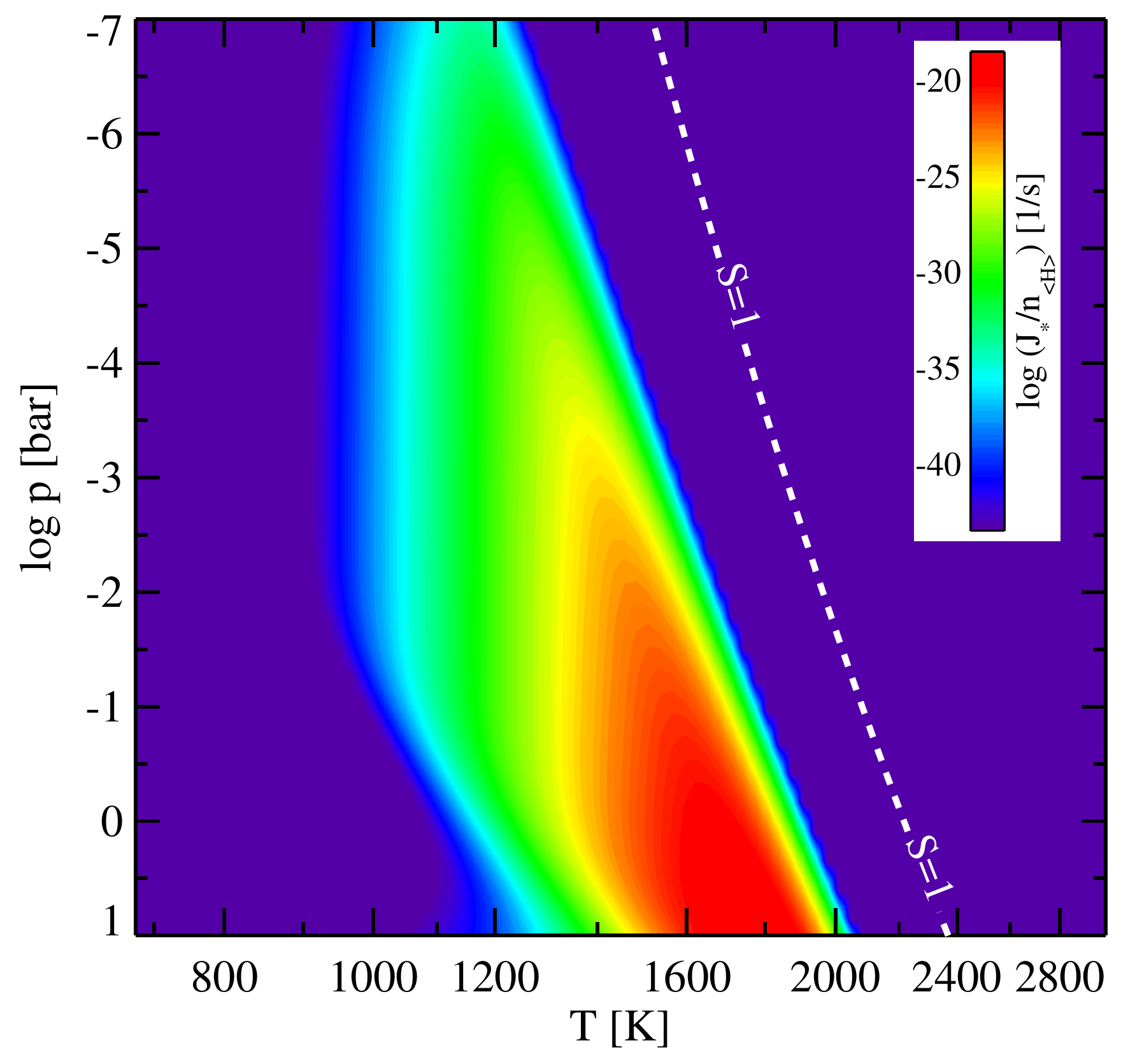}
\hspace*{-1mm}
\includegraphics[height=63.5mm,trim=57 7 9 0, clip]{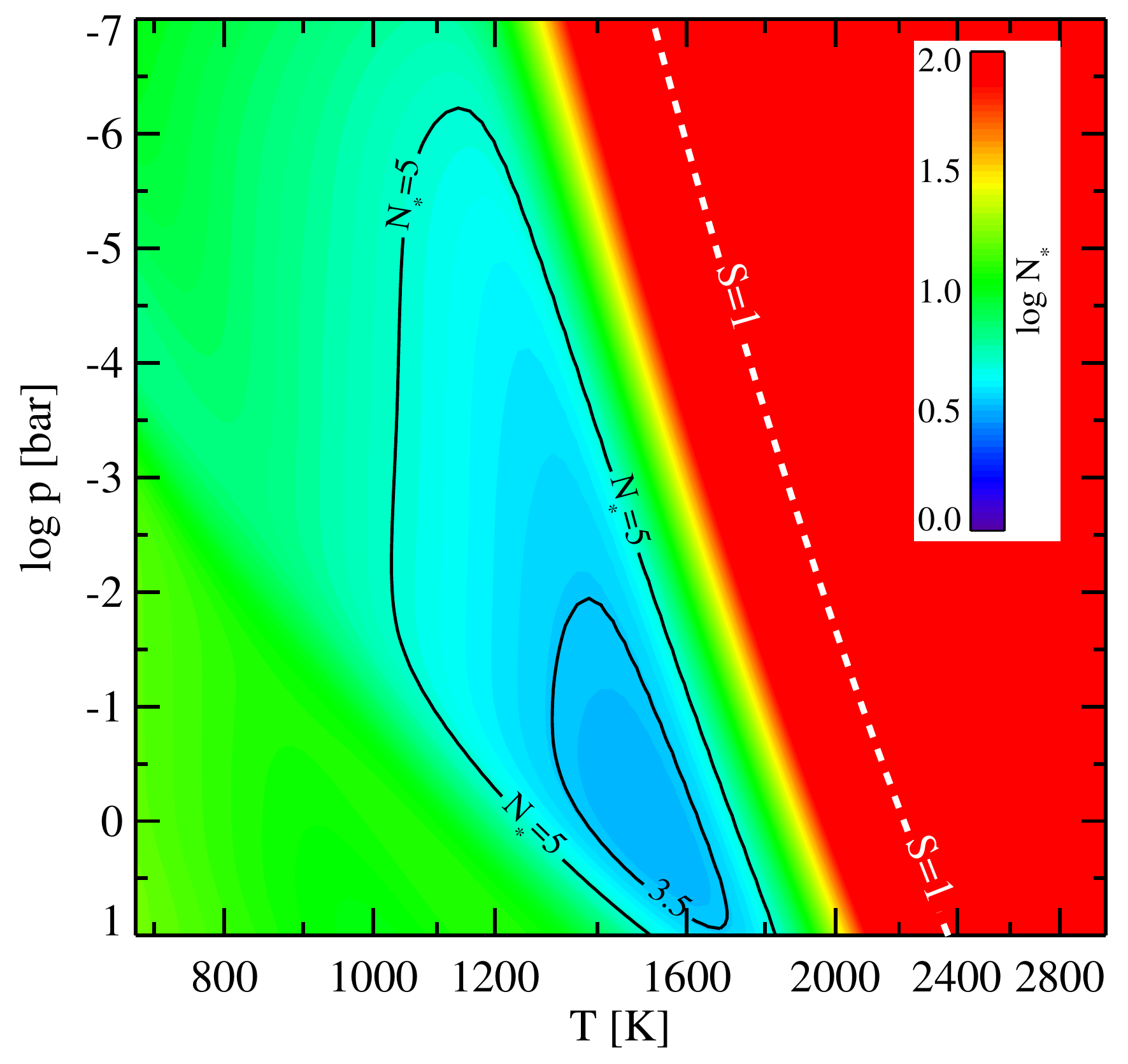}
\hspace*{-1mm}
\includegraphics[height=63.5mm,trim=57 7 9 0, clip]{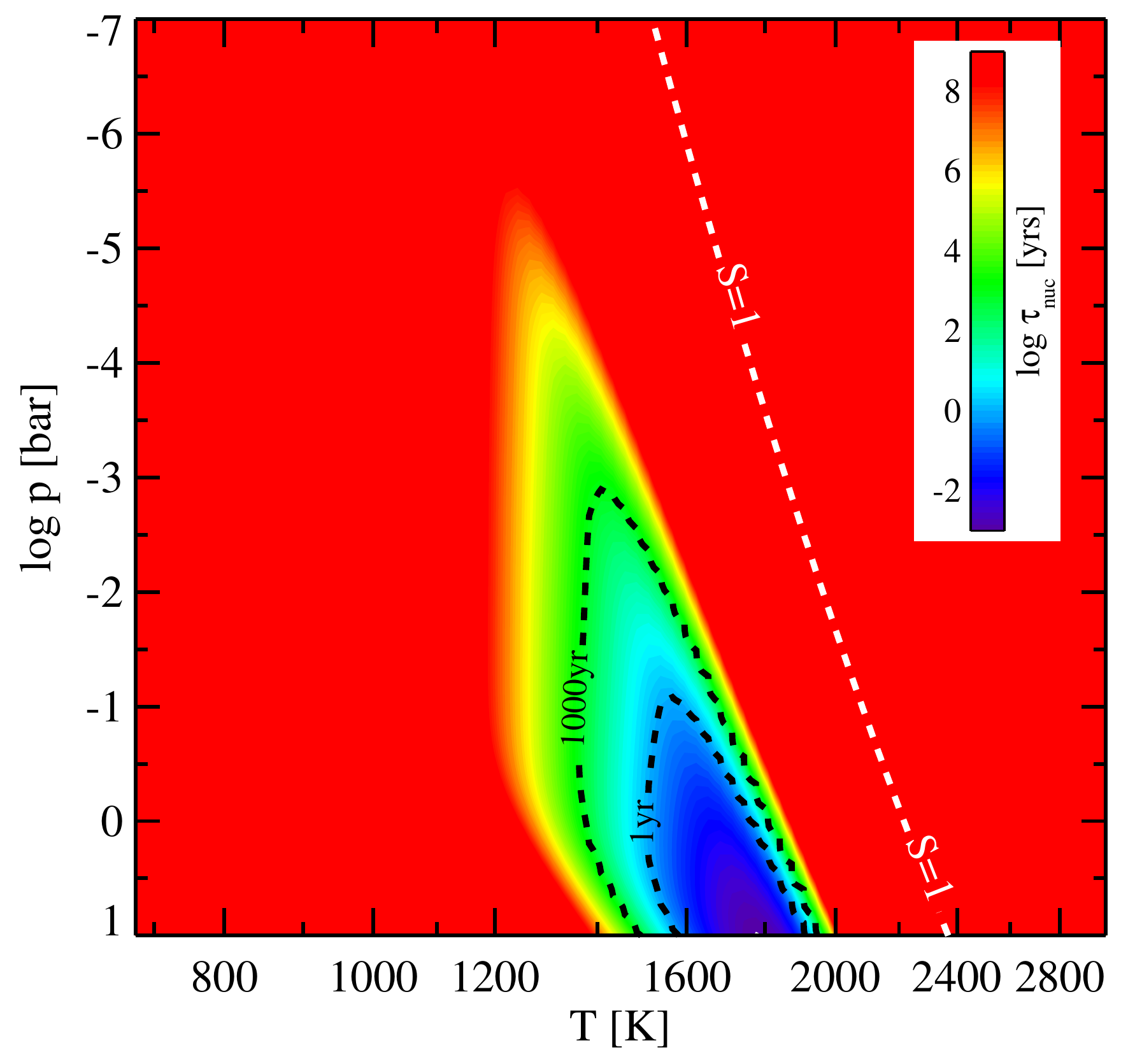}\\[-5mm]
\caption{{\bf Left:} The nucleation rate of tungsten per hydrogen
  nucleus $J_\star/\nH$\,[1/s] as function of gas pressure and
  temperature. {\bf Centre:} The size of the critical cluster
  $N_\star$, {\bf Right:} The nucleation timescale $\tau_{\rm
    nuc}$\,[yrs] according
  to Eq.\,(\ref{eq:taunuc}). The white dashed contour line marks
  saturation $S\!=\!1$, the gas is supersaturated to the left of this
  line.}
\label{fig:Jstar}
\vspace*{-2mm}
\end{figure*}

Below $200-400$\,K (depending on pressure) both Mg-silicates are
combined and hydrated to form the phyllosilicate
Mg$_3$Si$_2$O$_9$H$_4$[s] {\sl(lizardite)}.  Interestingly, there is
no feedback of water condensation on this phyllosilicate, the
concentration of lizardite stays constant towards very low
temperatures, showing that the formation of lizardite is
thermodynamically favoured over the formation of water ice.

We conclude from Fig.~\ref{fig:phase} that
Mg, Si, Na, Al, K and Mn can form thermodynamically stable 
phyllosilicates at low temperatures. Beside the major phyllosilicate
Mg$_3$Si$_2$O$_9$H$_4$[s] {\sl(lizardite)}, we also find 
NaMg$_3$AlSi$_3$O$_{12}$H$_2$[s] {\sl(sodaphlogopite)},
CaAl$_2$Si$_2$O$_{10}$H$_4$[s] {\sl(lawsonite)},
Mg$_3$Si$_4$O$_{12}$H$_2$[s] {\sl(talc)},
KMg$_3$AlSi$_3$O$_{12}$H$_2$[s] {\sl(phlogopite)},
KFe$_3$AlSi$_3$O$_{12}$H$_2$[s] {\sl(annite)},
KMn$_3$AlSi$_3$O$_{12}$H$_2$[s] {\sl(Mn-biotite)} and
MnAl$_2$SiO$_7$H$_2$[s] {\sl(Mn-chloritoid)} is less amounts, due to
element abundance constraints.

\section{The nucleation of tungsten}
\label{s:tung}

An interesting side-result of our models is that solid tungsten W[s]
is found to be the {\sl first condensate}, i.e.\ metallic tungsten is
the first solid material that becomes thermodynamically stable in any
galactic/stellar/atmospheric cooling flow which is initially too hot
to contain any condensed materials. The question arises whether tungsten
could provide the first nucleation seeds required for the growth of
all other, more abundant solid/liquid materials which only become stable
at lower temperatures. See \citep{Gail1998} for a more detailed introduction 
to the search for the first condensate.

\subsection{Is there enough tungsten?}

To estimate the total number of seed particles that can be provided by
tungsten, let us assume that one seed particle consists of $N_l\approx
10-1000$ tungsten atoms.  Using the solar abundance of tungsten
$\epsilon_{\rm W}=7.1\times 10^{-12}$ \citep{Asplund2009}, the
concentration of tungsten seed particles cannot exceed
\begin{equation}
  \frac{n_{\rm seeds}}{\nH} = \frac{\epsilon_{\rm W}}{N_l} 
  \quad\approx\quad 7\times 10^{-15} - 7\times 10^{-13} \ .
\end{equation}
Next, we assume that the various silicates, iron and all other 
abundant condensates will grow on top of these seed particles later,
to form a solid mantle of unique thickness.  The total available 
volume of solid material per H nucleus is about
\begin{equation}
  V_{\rm dust} \approx 
    \big(\epsilon_{\rm Si}+\epsilon_{\rm Mg}+\epsilon_{\rm Fe}\big)\,V_1
  \quad\approx\quad (1.2 - 2.7) \times 10^{-27}\,\rm cm^3 \ ,
\end{equation}
where we have used solar abundances and where $V_1\approx
(1.2-3)\times 10^{-23}\rm\,cm^{-3}$ is the dust volume per heavy
atom (the sum of Si, Mg and Fe atoms in the condensate), estimated
from the mass and solid densities of iron, MgSiO$_3$ and Mg$_2$SiO$_4$.  The
radius of the dust particles forming this way would be given
by\ \ $n_{\rm seeds} \frac{4\pi}{3}\,a^3 = \nH\,V_{\rm dust}$,\ \ i.e.
\begin{equation}
  a = \left(\frac{3}{4\pi}\,V_{\rm dust}\,\frac{\nH}{n_{\rm seeds}}\right)^{1/3}
  \quad\approx\quad 0.075-0.47\,\mu\rm m   \ , 
\end{equation}
which falls surprisingly well into the size range of observed dust
particles in space. This proves nothing of course, but it shows that
there is enough tungsten in a gas of solar abundances to provide the
seed particles needed for astrophysical bulk condensation. In fact,
from the simple estimate presented here we can conclude that the
concentration of seed particles {\em must} not exceed about
$10^{-15}$ to $10^{-13}$ per hydrogen nucleus, otherwise too many, too
small dust particles would form by condensation.

In application to M-type AGB-star winds, \citet{Hofner2016} have
shown that only silicate grains in a certain size range 0.1\,$\mu$m to
1\,$\mu$m can efficiently drive an outflow by radiation pressure, via
their large scattering opacity around $\lambda\!\sim\!1\,\mu$m. In
contrast, if their absorption coefficients dominate at those
wavelengths (as is true when they contain some iron), the grains
rather get heated in the strong IR radiation field and evaporate
\citep{Woitke2006}. In order to obtain these particular particle
sizes, \citet{Hofner2016} needed to assume a certain seed particle
concentration in their time-dependent hydrodynamical and dust
formation models, $n_{\rm seeds}/\nH\approx 3\times10^{-14}$ to
$3\times10^{-15}$.

\subsection{Timescales and supercooling}

The nucleation rate of crystalline tungsten has been calculated
according to classical nucleation theory in the formulation of
\citet{Gail1984}, with surface tension $\sigma\!=\!3340$\,erg/cm$^2$
\citep{Tran2016} and parameter $N_{1/2}\!=\!10$ (the cluster size
where the surface tension reduces to half its bulk value). The
supersaturation ratio $S$ and the particle density of atomic tungsten
$n_{\rm W}$ are calculated by a pure gas-phase {\sc GGchem} model
without condensation.  The resulting nucleation rates per hydrogen
nucleus are shown in Fig.~\ref{fig:Jstar}, from which we estimate that
a supercooling of about $300-400$\,K is needed to start the effective
nucleation of tungsten. The time required to form these seed particles
is
\begin{equation}
  \tau_{\rm nuc} = \frac{n_{\rm seeds}}{J_\star}
  = \frac{n_{\rm seeds}}{\nH}\,\frac{\nH}{J_\star}  
   \quad\approx\quad 10^{-14}\left(\frac{J_\star}{\nH}\right)^{-1}
  \label{eq:taunuc}
\end{equation}
From Fig.\,\ref{fig:Jstar} we read off that the formation timescales
of the tungsten seed particles would be rather long and would require
high gas pressures, for example $\tau_{\rm nuc}\!=\!1\,$yr at
0.1\,bar and 1600\,K. At lower gas pressures, the tungsten 
nucleation timescale increases fast, exceeding 
the age of the universe around $10^{-6}$\,bar. This fast increase is
mainly due to an increase of the critical cluster size $N_\star$
(central part of Fig.~\ref{fig:Jstar}).

\section{Summary and conclusions}
\label{s:sc}

A fast and versatile public code has been developed, called {\sc
  GGchem}, to calculate the chemical composition of astrophysical
gases in chemical equilibrium down to 100\,K, with or without
equilibrium condensation.  We have collected and compared the
thermo-chemical data for molecules and condensates from different
sources, and have presented tables with temperature-fits of molecular
equilibrium constants $k_p(T)$ and Gibbs free energies $\dG(T)$ which
behave robustly towards low temperatures and can safely be applied
down to 100\,K. The data comprises 552 molecules and 257 condensates,
including 38 liquids.

For pure gas-phase applications, {\sc GGchem} has been thoroughly
tested against the TEA-code \citep{Blecic2016}, revealing close to
perfect agreement. Differences arise only when the selection of
molecules differs. In {\sc GGchem}, we are using a number of molecules
with thermo-chemical data from \citet{Barklem2016}, which are not
available in NIST-JANAF, in particular metal hydrides like CaH, TiH
and FeH. {\sc GGchem} is found to converge fast and robustly down to
100\,K under all tested circumstances including very small and very
large selections of elements and molecules, as well as for unusual
element abundances.

Concerning very sparse models with just a few molecules and elements
(for example H, C, N, O to quickly determine the abundances of the
spectroscopically active molecules like H$_2$O and CH$_4$), we want to
emphasise that the combined effect of elements like Mg, Si and Fe is
substantial and might lead to an overestimation of the C/O ratio if
these elements are neglected. For solar abundances ($\rm
C/O\!=\!0.55$) the combined abundance of Si, Mg and Fe alone is about
$10^{-4}$ which is larger than the nitrogen abundance ($\epsilon_{\rm
  N}\!=\!6.8\times 10^{-5}$) and can be compared to the carbon and
oxygen abundances ($\epsilon_{\rm C}\!=\!2.7\times 10^{-4}$,
$\epsilon_{\rm O}\!=\!4.9\times 10^{-4}$). Since molecules like SiO,
$\rm Mg(OH)_2$ and $\rm Fe(OH)_2$ form at low temperatures, every Si,
Mg and Fe atom takes away about $x\approx 1\!-\!2$ oxygen atoms
from the C-N-O system. If we consider the effective C/O ratio
remaining in the C-N-O system, the result for solar element abundances
is
\begin{equation}
  {\rm (C/O)}_{\rm eff} = \frac{\epsilon_{\rm C}}{\epsilon_{\rm O}
         -x\,(\epsilon_{\rm Si}+\epsilon_{\rm Mg}+\epsilon_{\rm Fe})}
  \approx \left\{\begin{array}{cc}
      0.70 & \mbox{for\ } x=1\\
      0.95 & \mbox{for\ } x=2
    \end{array}\right.
\end{equation}
Thus, if C/O ratios are derived from observations based on sparse
chemical models, where elements like Mg, Si and Fe are lacking, the
results must be expected to be too large.

We have discussed the condensation sequence of the elements in phase
equilibrium, generally confirming the results found earlier by
\citet{Sharp1990} and \citet{Lodders2003}. Differences arise mostly
from the different selection/availability of the thermo-chemical data for
condensed species, but we did not find any evidence for systematic
differences due to different numerical techniques.  An important step
here was to create a link to the geophysical database SUPCRTBL
\citep{Zimmer2016}, from which we have extracted the thermo-chemical
data for 121 condensed species, including phyllosilicates.  The
numerical methods rather affect the computational speed and
stability of the codes.

A straightforward result from our models is that the dust/gas mass
ratio in a solar composition gas is expected to be 0.0045 rather than
0.01.  The latter (standard) value of 0.01 is only obtained in our
models once water and ammonia ices condense at low temperatures. The
ice formation increases the mass of the condensates by a factor of
$\sim$\,2.5 and the volume of the condensates by a factor of $\sim$\,6
(from $\rm 2.7\times 10^{-27}\,cm^3$ to $\rm 1.7\times 10^{-26}\,cm^3$
per hydrogen nucleus to be precise).

Our models show that phyllosilicates (``wet silicates'') are the most
abundant condensates at low temperatures in a solar composition gas in
phase equilibrium. The most abundant phyllosilicate is identified to
be Mg$_3$Si$_2$O$_9$H$_4$[s] {\sl(lizardite)} which contains the
majority of Mg and Si below $\rm 200\,K-400\,K$ in phase equilibrium,
depending on pressure. In a similar way, a few other phyllosilicates
are found to consume most of Na, Al, K and Mn, some of which may form
already at 500\,K.  The phyllosilicates have a notable influence on
the dust/gas ratio and will increase the C/O ratio in the gas phase 
to about 0.83 because of the additional intake of oxygen.

Whether or not these phyllosilicates (and all other condensates)
actually form in space is not discussed in this paper, as this is a
paper just about chemical and phase equilibrium. A recent kinetic
approach for the formation of phyllosilicates in protoplanetary discs
has been submitted to A\&A by Thi et al., based on a warm
surface chemistry model. Thi et al.\ find that despite the high
energy barriers involved in chemisorption and diffusion, water and
hydroxyl can efficiently migrate into the bulk lattice within typical
lifetimes of protoplanetary discs if the gas and dust temperatures are of
order 80\,K to 700\,K. The maximum intake of water and hydroxyl
is a parameter in these models.

As a side result, we discussed whether tungsten could be the first
condensate in space and could provide the seed particles for
astrophysical dust formation. We argue that solid tungsten becomes
stable already at very high temperatures of order 2000\,K, that it
could provide just about the right numbers of seed particle per H-atom
to form micron-sized silicate grains, but that a strong super-cooling
of order 300\,K-400\,K is required and that high pressures of the
order of milli-bars are necessary to make tungsten nucleation an
efficient process.

\bigskip\noindent {\bf Acknowledgements:\ } We thank Dr.~Wing-Fai Thi
for pointing us toward the SUPCRTBL database and for his comments on
the manuscript. The research leading to
these results has received funding from the European Union Seventh
Framework Programme FP7-2011 under grant agreement no 284405. 
Jasmina Blecic is supported by NASA trough the NASA
ROSES-2016/Exoplanets Research Program, grant NNX17AC03G.
Christiane Helling highlights the hospitality of the 
Rijksuniversiteit Groningen and the Universiteit van Amsterdam
with travel support from NWO and LKBF.

\bibliography{references}

\appendix

\section{Solving gas phase chemical equilibrium down to 100\,K}
\label{AppC}

Solving the system of non-linear equations (Eq.\,\ref{eq:conserve}),
i.e.\ finding the root vector of atomic partial pressures which
satisfies the element conservation equations after elimination of all
molecular partial pressures in chemical equilibrium according to
Eq.\,(\ref{eq:kp}), seems to be a simple numerical standard
problem. Indeed, a simple Newton-Raphson iteration may lead to quick
success at high temperatures.  However, at low temperatures, the
solution vector becomes more and more degenerate, with some atomic
partial pressures approaching values
$<\!10^{\,-\,\rm{several\ hundred}}$ dyn/cm$^2$, for example carbon,
whereas others like hydrogen stay of order 1. This leads to a number
of numerical problems, in particular
\begin{itemize} 
\item[(i)] the conditional number of the Jacobi matrix becomes 
  large, i.e.\ the resolution of the linearised equation system
  is not possible without substantial losses of precision, and\\[-1.8ex]
\item[(ii)] the Newton-Raphson method only works satisfactory when a
  good initial estimate of the solution vector is provided, which
  becomes increasingly difficult at low temperatures.
\end{itemize}
We have developed a stable algorithm that solves these problems under
all tested circumstances, including unusual element abundances and
pressures, down to 100\,K.  The first necessary step was to switch to
quadruple precision arithmetics for temperatures $T\!<\!1000\,$K,
which solves problem (i).

\begin{figure}
\centering
\vspace*{1mm}
\includegraphics[width=70mm,trim=0 0 0 0, clip]{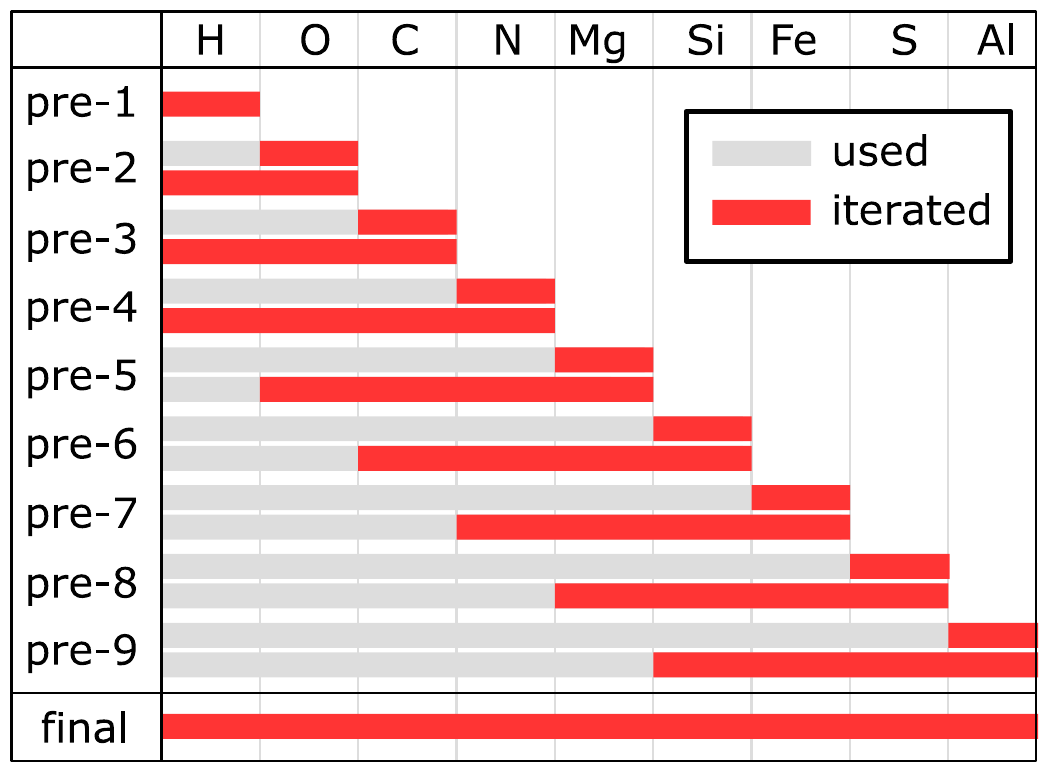}\\[0mm] 
\caption{Algorithm used in {\sc GGchem} to iteratively solve the
  chemical equilibrium problem for total number of elements $K\!=\!9$, 
  and number of elements taken into account during the pre-iterations 
  $N\!=\!4$, see text.}
\label{fig:algorithm}
\vspace*{-2mm}
\end{figure}

Figure~\ref{fig:algorithm} sketches how we solve problem (ii).  The
algorithm starts by sorting the elements according to their
abundances. This hierarchy can change substantially by condensation as
certain elements will be removed almost entirely from the gas phase,
for example Ca and Al, whereas other elements stay in the gas phase
for longer, for example S and Cl. In the example shown in
Fig.~\ref{fig:algorithm}, we consider $K\!=\!9$ elements.  There are
9 pre-iteration steps performed in this case, before a final
Newton-Raphson iteration is carried out to solve the element
conservation equations (Eq.\,\ref{eq:conserve}) for all elements. The
task of the pre-iterations is to provide good initial estimates of the
atomic partial pressures for this final iteration, they are the key to
obtain code stability.

In each pre-iteration, we only consider a subset of elements and
molecules which are composed of these elements. Each pre-iteration
starts with an estimate of the atomic partial pressure of a new
element (short red bar) using the values of all other atomic partial
pressures computed so far (long grey bar). The pre-iteration then
continues by improving the atomic pressures of the last $N$ elements
(long red bars, $N\!=\!4$ in the figure) by taking into account the mutual
feedbacks, however without changing the partial pressures of the other,
even more abundant elements on the left (short grey bars).

For example, pre-iteration 1 (``pre-1'') only considers hydrogen with
species H and H$_2$, from which a first guess of $p_{\rm H}$ is
obtained. Next comes oxygen with additional species O, OH, O$_2$
HO$_2$, H$_2$O and O$_3$.  During the first stage of pre-2, a first
guess of $p_{\rm O}$ is obtained by solving the element conservation
equation for O at given $p_{\rm H}$. During the second stage of pre-2,
the feedback between O on H is taken into account by solving the two
coupled conservation equations for H and O by a Newton-Raphson method
to improve $p_{\rm H},p_{\rm O}$. The next element is carbon with
additional species C, CH, CO, C$_2$, HCO, CH$_2$, H$_2$CO, CH$_3$,
CH$_4$, CO$_2$, C$_2$H, C$_2$H$_2$, C$_2$H$_4$, C$_2$H$_4$O, C$_2$O,
C$_3$, C$_3$O$_2$, C$_4$, C$_5$ from which $p_{\rm C}$ is first
estimated and then determined, including the mutual feedback on H and
O, by solving three coupled non-linear algebraic equations, and so
on. The algorithm continues this way until Mg, where we stop refining
$p_{\rm H}$ in stage 2 of\linebreak pre-5. We are using the
hierarchical order of the elements here, i.e.\ whatever the outcome of
$p_{\rm Mg}$ will be, it can only have a minor feedback on $p_{\rm H}$
since $\epsilon_{\rm Mg}\!\ll\!\epsilon_{\rm H}$. When sulphur enters
the pre-iteration phase at pre-8, the atomic partial
pressures of O, C and N are already frozen.  The algorithm continues
this way, before the full system of element conservation equations
(Eq.\,\ref{eq:conserve}) is solved with the Newton Raphson method, using
the atomic pressures from the pre-iteration phase as initial
estimates.

These pre-iterations in fact work so well, that the final full Newton
Raphson iteration only makes minor corrections $<5\%$ in all our
benchmark tests, at most, which produces no visible changes in
Fig.~\ref{fig:TEA}, for example, only another boost in performance.
Figure~\ref{fig:algorithm} sketches the algorithm for $N\!=\!4$,
however we achieved best overall performance with $N\!=\!5$. Further
improvements of code stability and performance are achieved by
dividing the resulting atomic partial pressures by their initial
guesses after each Newton-Raphson iteration, then store these ratios
into the memory, and use them as correction factors next time to
obtain improved initial guesses. This simple idea was the key to
achieve code convergence down to 100\,K.

At $T\!=\!96\,$K, however, the equilibrium constant $k_p$ of the
molecule $\rm (WO_3)_4$ hits $10^{+4932}$, the maximum number that can
be represented in quadruple precision, and the code crashes. It might be
technically possible to go deeper in temperature by eliminating a few
big molecules, however these are just the molecules which 
dominate at those temperatures, see Fig.~\ref{fig:TEA}.

\section{Solving phase equilibrium}
\label{AppD}

We formulate the element conservation equations in the presence of
condensed species in the following way
\begin{equation}
  \ek^0 = \ek + \sum\limits_{j=1}^N s_{j,k}\;\cj  \ ,
  \label{eq:cons}
\end{equation}
where $\ek^0$ are the total element abundances (or ``initial''
element abundances before condensation), $\ek$ are the element
abundances in the gas phase and $\cj = n_j^{\rm cond}/\nH$ is the
concentration of condensed species $j$ per hydrogen nucleus. $N$ is
the current number of present condensates, and $s_{j,k}$ are the
stoichiometric factors of elements $k$ in condensate $j$.
Equation~(\ref{eq:cons}) is a generalised variant of
(Eq.\,\ref{eq:conserve}).  Note that all equations for the gas phase in
Sect.~\ref{sec:eqgas} remain valid when using $\ek$.

An essential realisation is that Eq.~(\ref{eq:cons}) should
{\it\,never\,} be used in computer codes to ``subtract the dust'', as
this would unavoidably lead to a substantial loss of numerical
precision at low temperatures. Some $\ek$ are reduced by several
hundreds of orders of magnitude at low temperatures, which will soon
exceed the numerical resolution required to perform this subtraction
precisely, even when using quadruple precision arithmetics.

We have developed an iterative algorithm which avoids this problem by 
applying corrections $\delta\ek$ to $\ek$ to eventually solve
\begin{equation}
  \hspace*{17mm}
  \Sj\,(\ek) \;\leq\; 1   \quad\quad\mbox{all condensates}
\end{equation}
at given temperature $T$ and density $\nH$. Provided that we have
selected the present condensates correctly (all non-selected
condensates must have $S\!<\!1$), this equation simplifies to
\begin{equation}
  \hspace*{17mm}
  \Sj\,(\ek) \;=\; 1   \quad\quad\mbox{selected condensates.}
  \label{eq:tosolve}
\end{equation}
The dependencies on $\ek$ are marked here because all gas phase
equilibrium results only depend on $(\nH,T,\ek)$ (where $\nH$ and $T$
are parameters to the problem), and so do the supersaturation ratios
$\Sj$. We choose $N_{\rm ind}$ element abundances as independent
variables of the problem
\begin{equation}
  x_i = \big\{ \delta\epsilon_i\;|\;i=1,...\,,N_{\rm ind} \big\} \ ,
\end{equation}
whereas the dependent variables are
\begin{equation}
  y_j = \sum\limits_i^{N_{\rm ind}} A_{j,i}\,x_i 
      = \left\{\begin{array}{ll}
          \delta \cj             & \mbox{for $j=1,...\,,N$} \\
          \delta \epsilon_{k(j)}  & \mbox{for $j=N+1,...\,,N+D$\ .}
        \end{array}\right. 
\end{equation}
$A_{j,i}$ is the conversion matrix and $D$ the number of dependent 
elements with indices $k(j)$. One iteration step is completed by
\begin{eqnarray}
  \ek &\!\!\to\!\!& \ek + \delta\ek \\
  \cj &\!\!\to\!\!& \cj + \delta\cj 
\end{eqnarray}
The concentrations of the condensed species $\cj$ are simply a
byproduct of these computations, but they are {\it not used} during
the iterations. The element conservation is implemented in the
conversion matrix $A_{j,i}$ as explained below, and we carefully check
that Eq.\,(\ref{eq:cons}) remains valid after each iteration step.

To explain these definitions, let us consider the condensation of
Al$_2$O$_3$, MgSiO$_3$ and Mg$_2$SiO$_4$ as an example. There are 3
condensates is this case ($N\!=\!3$) and 4 affected elements ($N_{\rm
  ind}\!+\!D=4$).  In order to proceed, we must select three of them
($N_{\rm ind}\!=\!3, D\!=\!1$) to obtain 3 equations (Eq.\,\ref{eq:tosolve})
for 3 unknowns. As default rule, we pick the least abundant elements
(here Al, Si and Mg) as independent variables, i.e. $i\!=\!1$ is Al,
$i\!=\!2$ is Si and $i=3$ is Mg. The other elements (here O,
$D\!=\!1$) become dependent elements, $k(4)$ is O. To find the
conversion matrix $A_{j,i}$ we write down the conservation equations
for all elements affected by condensation as
\begin{equation}
  \begin{array}{crrll}
        2\,\delta c_{\rm Al_2O_3} & & 
   &    +\;\delta\epsilon_{\rm Al} &= 0\\
   &       \delta c_{\rm MgSiO_3} 
   &    +\;\delta c_{\rm Mg_2SiO_4} 
   &    +\;\delta\epsilon_{\rm Si} &= 0\\
   &       \delta c_{\rm MgSiO_3} 
   & +\;2\,\delta c_{\rm Mg_2SiO_4} 
   &    +\;\delta\epsilon_{\rm Mg} &= 0\\
        3\,\delta c_{\rm Al_2O_3} 
   & +\;3\,\delta c_{\rm MgSiO_3} 
   & +\;4\,\delta c_{\rm Mg_2SiO_4} 
   &    +\;\delta\epsilon_{\rm O} &= 0 \ .\\
  \end{array}\nonumber
\end{equation}
By a stepwise elimination scheme, we find the conversion matrix
$A$ in this case to be
\begin{equation}
  \;\left(\!\!\begin{array}{c} \delta c_{\rm Al_2O_3}\\ 
                         \delta c_{\rm MgSiO_3}\\ 
                         \delta c_{\rm Mg_2SiO_4}\\
                         \delta\epsilon_{\rm O}
  \end{array}\!\!\!\right)    
  \;=\; {A}\cdot\left(\!\!\begin{array}{c} 
                \delta\epsilon_{\rm Al}\\ 
                \delta\epsilon_{\rm Si}\\ 
                \delta\epsilon_{\rm Mg}\end{array}\!\!\right)
  \;=\; \left(\begin{array}{ccc}
          -\frac{1}{2} & 0 & 0\\
          0 & -2 & 1 \\
          0 & 1 & -1\\
          \frac{3}{2} & 2 & 1
          \end{array}\right)\cdot\left(\!\!\begin{array}{c} 
                \delta\epsilon_{\rm Al}\\ 
                \delta\epsilon_{\rm Si}\\ 
                \delta\epsilon_{\rm Mg}\end{array}\!\!\right)
  \nonumber
\end{equation}
The conversion matrix $A$ is a $N\times(N\!+\!D)$ matrix which
describes how the dependent variables change when the independent
element abundances are changed.  This transformation allows us to
solve Eq.\,(\ref{eq:tosolve}) as system of $N\!=\!3$ coupled
non-linear algebraic equations for the $N\!=\!3$ unknowns
$(\delta\epsilon_{\rm Al}, \delta\epsilon_{\rm Si},
\delta\epsilon_{\rm Mg})$, which succeeds easily by a Newton-Raphson
iteration with several internal calls of the equilibrium chemistry to
compute the Jacobi matrix at every iteration step.  We have to make
sure, however, to limit the Newton-Raphson step $\delta x$ to 
prevent negative $\cj$ or negative $\ek$. This method is found to
generally converge very quickly at all temperatures, within less than
about 10 iteration steps, depending on the quality of the initial
guesses of $\ek$ (here $\epsilon_{\rm Al}, \epsilon_{\rm Si},
\epsilon_{\rm Mg})$.

To solve the problem of bad initial guesses for $\ek$, we store
all successfully computed results ($\ek, \cj$) as function of
($\nH,T$) into a database and take the initial guesses from the
closest $(\nH,T)$ database point next time. This way, the
program automatically becomes more stable and faster over time. But
initially, we must run {\sc GGchem} with small temperature steps
from warm to cold.

Some practical problems may still arise, because of the somewhat
unclear selection of condensed species and independent elements.  The
selection of condensed species is guided by monitoring the
supersaturation ratios of all condensates during the iterations. If a
new condensate becomes supersaturated, it will be added to the set of
selected condensates, or will replace one of them.  We have
implemented a number and hand-crafted criteria here. For example, one
condensate cannot be present simultaneously as liquid and
solid. Application of the algorithm explained above would fail
  already in the first step, as there is only one affected element
but two condensates -- only one can be present. Another example is
CaTiO$_3$, MgTiO$_3$ and CaMgSi$_2$O$_6$, where the latter is a linear
combination of the former two, in which case the elimination scheme to
obtain the conversion matrix fails -- only two of them can be present
under any circumstances. This leads us to some quite interesting
general insights into the nature of phase equilibrium:
\begin{itemize}
\item[a)] $N=N_{\rm ind}$ must hold, i.e.\ the number of
  simultaneously present condensates must equal the number of
  independent elements.  $N_{\rm ind}$ is limited by the number of
  elements affected by condensation $K$ ($K=N_{\rm ind}+D$ with $D\ge
  0$). For any given set of condensates, this means that we can simply
  count the number of elements from which they are made, $K$, and the
  number of present condensates in phase equilibrium $N$ must not
  exceed that number: $N\le K$.\\[-1.8ex]
\item[b)] Stoichiometric linear combinations of condensates must not
  be present simultaneously in phase equilibrium. This applies in
  particular to the solid/liquid phase transitions.\\[-1.8ex]
\item[c)] The number of dependent elements $D$ is given by the number
  of elements which cannot be completely converted into condensates, 
  due to element conservation constraints, at temperature $T$. Toward
  low $T$, $N$ increases and $D$ decreases
  monotonically. Eventually, $D$ becomes very small, only counting elements
  like O, N, C and H which are affected by condensation but need 
  very low temperatures to completely condense\footnote{Although
    eventually, O and N will condense anyway, in form of 
    water and ammonia.}. All other elements eventually vanish almost
  entirely from the gas phase at low temperatures.
\end{itemize}
For example, $N$ increases by one around 600\,K, as soon as FeS[s]
becomes stable. Since $\epsilon^0_{\rm Fe}>\epsilon^0_{\rm S}$, the
stoichiometry of that condensate makes it possible to completely
remove sulphur from the gas phase. Without such new opportunities to
lock away elements, some sudden transformations might occur between
several condensates (type-2 phase transitions), but the total number
of them will remain constant.  Practically, we observe that condition
(b) is also valid for linear combinations of condensates which
conserve all elements but oxygen, the abundance of which is only
marginally affected anyway by condensation.

\section{The molecular $k_p$ data}
\label{AppA}

During the preparation phase for this paper, we have collected
molecular $k_p$ data from \citep{Tsuji1973}, \citep{Sharp1990},
\citep{Luttke2002}, \citep{Stock2008}, and \citep{Barklem2016} and
compared these to our own previous data collection going back to
\cite{Gail1986}. The names of 1155 molecules have been homogenised and
the different functional fits of $k_p(T)$ have been overplotted, similar to
Fig.~\ref{fig:kps}, and characterised to assess the level of agreement.

\begin{figure}
\hspace*{-2mm}\includegraphics[height=78mm,trim=48 0 0 0,clip]{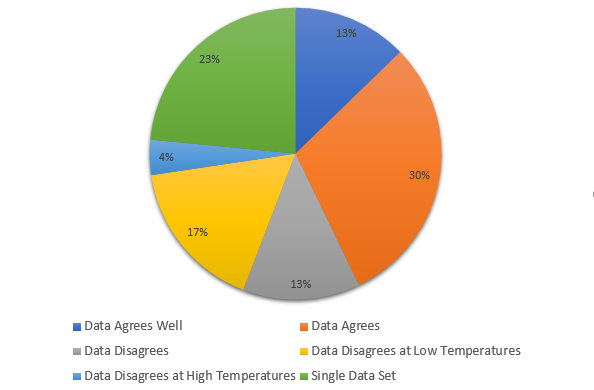}\\[1mm]
\resizebox{88mm}{!}{
\begin{tabular}{l|c|c}
\hline
&&\\[-2.1ex]
& $\sigma[\log k_p(200\,{\rm K})]$ & $\sigma[\log k_p(3000\,{\rm K})]$\\
&&\\[-2.1ex]
\hline
&&\\[-2.1ex]
data agrees well           & $<0.1$ & $<0.05$\\
data agrees                & $<0.4$ & $<0.1$\\
data disagrees at low $T$  & $>0.4$ & $<0.1$\\
data disagrees at high $T$ & $<0.4$ & $>0.1$\\
data disagrees             & $>0.4$ & $>0.1$\\
\hline
\end{tabular}}
\caption{Statistics of molecular $k_p(T)$ agreement between the
  functional fits provided by \citep{Sharp1990}, \citep{Luttke2002},
  \citep{Stock2008}, and \citep{Barklem2016}.}
\label{fig:pie}
\vspace*{2mm}
\end{figure}

Figure~\ref{fig:pie} shows the results of the complete comparison
study \citep{Worters2017}.  Most of the molecules fall into the
category ``data agrees'' (30\%) with deviations better than 0.4\,dex
at 200\,K and better than 0.1\,dex at 3000\,K, disregarding the Tsuji
data. However, for the majority of the molecules, the disagreement is
worse, with some cases of quite obvious disagreements.

\def\Aplu{\rm A^{\!\scriptstyle+}}
\def\Amin{\rm A^{\!\scriptstyle-}} 
\def\ABplu{\rm AB^{\!\scriptstyle+}} 
\def\ABmin{\rm AB^{\!\scriptstyle-}}

Similar conclusions have recently been published by 
\citet{Barklem2016}, who have reviewed the thermo-chemical data of 291
diatomic and charged diatomic molecules. These deviations are mostly
due to uncertainties in the dissociation energies. \citet{Barklem2016}
published their data in terms of partition functions and dissociation
energies 
\begin{equation}
  k_{\rm Bark} =
  \frac{p_{\rm A}p_{\rm B}}{p_{\rm AB}} 
  = \left(\frac{2\pi\,m_{\rm red}\,kT}{h^2}\right)^{3/2}\!\!\!\!kT\;
    \frac{Q_{\rm A}Q_{\rm B}}{Q_{\rm AB}}\,
    \exp\left(-\frac{D^{\,0}_{\rm AB}}{kT}\right) \,,
  \label{eq:kBark}
\end{equation}
where A and B are atoms, AB is a diatomic molecule, $m_{\rm
  red}=m_{\rm A} m_{\rm B}/(m_{\rm A}+m_{\rm B})$ is the reduced mass,
$m_{\rm el}$ the electron mass, $Q_{\rm A}$, $Q_{\rm B}$ and $Q_{\rm
  AB}$ are the partition functions, and $D^{\,0}_{\rm AB}$ is the
dissociation energy. To compute the equilibrium constants $k_p(T)$
according to our definition from this data, also for charged
molecules, we need to combine Eq.\,(\ref{eq:kBark}) with Saha
functions
\begin{eqnarray}
  S_{\!+} &\!\!\!=\!\!& \frac{p_{\Aplu}\,p_{\rm el}}{p_{\rm A}} 
  ~=~  \left(\frac{2\pi\,m_{\rm el}\,kT}{h^2}\right)^{3/2}\!\!\!kT\;
    \frac{2 Q_{\Aplu}}{Q_{\rm A}}\,
    \exp\left(-\frac{\;\chi_{\rm A}}{kT}\right) \\
  S_{\!-} &\!\!\!=\!\!& \frac{p_{\rm A}\,p_{\rm el}}{p_{\Amin}} 
  ~=~  \left(\frac{2\pi\,m_{\rm el}\,kT}{h^2}\right)^{3/2}\!\!\!kT\;
    \frac{2 Q_{\rm A}}{Q_{\Amin}}\,
    \exp\left(-\frac{\;\chi_{\Amin}}{kT}\right) 
\end{eqnarray}
to obtain
\begin{eqnarray}
  k_p^{\rm AB} &\!\!=\!\!& \frac{p_{\rm AB}}{p_{\rm A}p_{\rm B}} 
          ~=~ \Big(k_{\rm Bark}({\rm AB})\Big)^{-1} \\
  k_p^{\ABplu} &\!\!=\!\!& \frac{p_{\ABplu}\,p_{\rm el}}{p_{\rm A}p_{\rm B}} 
          ~=~ \Big(k_{\rm Bark}({\ABplu})\Big)^{-1}\;S_{\!+} \\
  k_p^{\ABmin} &\!\!=\!\!& \frac{p_{\ABmin}}{p_{\rm A}p_{\rm B}p_{\rm el}} 
          ~=~ \Big(k_{\rm Bark}({\ABmin})\Big)^{-1}\;\Big(S_{\!-}\Big)^{-1}  \ .
\end{eqnarray}
$\chi_{\rm A}$ is the ionisation potential of the neutral atom
and $\chi_{\Amin}$ is the electron affinity of the negative ion.
Tables for $D^0_{\rm AB}$, $\chi_{\rm A}$, $Q_{\rm A}(T)$, $Q_{\Aplu}(T)$,
$Q_{\Amin}(T)$, $Q_{\rm AB}(T)$, $Q_{\ABplu}(T)$ and $Q_{\ABmin}(T)$ are given
in the electronic appendices of \citet{Barklem2016}, on 42 temperatures
points between $10^{-5}$\,K and $10^4$\,K. The electron affinities are
not included, however, so we adopt the following values from
\url{https://en.wikipedia.org/wiki/Electron_affinity_(data_page)}
\begin{equation}
  \begin{array}{r@{\hspace*{1mm}}c@{\hspace*{1mm}}lc
                r@{\hspace*{1mm}}c@{\hspace*{1mm}}l}
  \chi_{\rm H^{\scriptstyle-}}  &=& 0.7542\,\text{eV} & , &
  \chi_{\rm C^{\scriptstyle-}}  &=& 1.2621\,\text{eV} \\
  \chi_{\rm O^{\scriptstyle-}}  &=& 1.4611\,\text{eV} & , &
  \chi_{\rm F^{\scriptstyle-}}  &=& 3.4012\,\text{eV} \\
  \chi_{\rm Si^{\scriptstyle-}} &=& 1.3895\,\text{eV} & , & 
  \chi_{\rm S^{\scriptstyle-}}  &=& 2.0771\,\text{eV} \\
  \chi_{\rm Cl^{\scriptstyle-}} &=& 3.6127\,\text{eV} \ .
  \end{array}
\end{equation}
We have fitted these data between 50\,K and 6000\,K with a Stock-function
(Eq.~\ref{eq:Stock}).

Our final choice of $k_p(T)$ data for {\sc GGchem} is shown in
Table~\ref{tab:kp}.  We have given preference to the NIST-JANAF data
fitted by \citep{Stock2008}. For molecules available in
\citep{Barklem2016}, but not available in NIST-JANAF, we use the
Barklem\,\&\,Collet data. The TEA benchmark test revealed a few
mismatches between the \citep{Stock2008} data and the TEA data. For
each of those molecules, we have re-fitted the NIST-JANAF data
ourselves. We do not use any of the \citep{Tsuji1973} or
\citep{Sharp1990} data directly, although some of the
\citep{Stock2008} data we are using are based on the Tsuji-data.

\section{The condensed phase data}
\label{AppB}

\begin{figure}
\vspace*{1mm}\hspace*{-2mm}
\includegraphics[width=91mm,height=65mm,trim=5 0 15 15, clip]
                {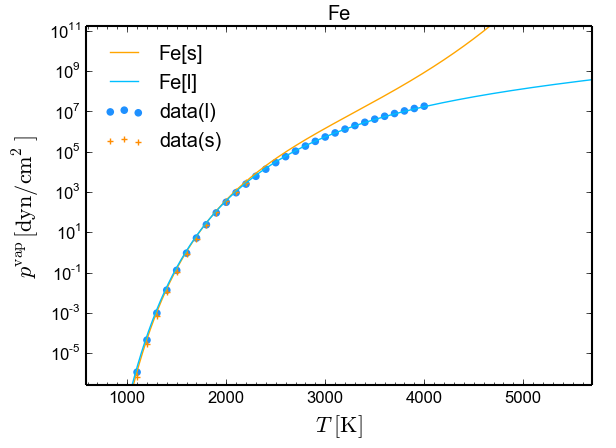}\\[-7mm] 
\caption{Fitting the vapour pressures of solid and liquid iron from
  the NIST-JANAF database. The two fitted $\pvap(T)$ curves intersect
  at 1824\,K which is close to the known melting point of 1809\,K
  according to NIST-JANAF. The model will automatically give
  preference to the phase with the lower vapour pressure. Solid data
  exist between 100\,K and 2200\,K, and liquid data between 298\,K and
  4000\,K. The solid curve shoots upward for $T>2500\,$K, but this
  does not matter as iron condenses as a liquid at those
  temperatures. When designing our solid Fe vapour pressure
  $\pvap(T)$-fit, we just have to make sure that is does not produce a
  second, spurious intersection point with the liquid curve.}
\label{fig:pvapfit}
\vspace*{-1mm}
\end{figure}

We have extracted condensed phase Gibbs free energy data from two
sources. First, the NIST-JANAF database
\citep{Chase1982,Chase1986}\footnote{\url{http://kinetics.nist.gov/janaf/}}
provides $\dGNIST(T)$-data [kJ/mol] at $\pst\!=\!1\,$bar in the
7$^{\rm th}$ column of their data-files. In NIST-JANAF, these are the
Gibbs free energies of formation from the respective NIST reference 
states of the elements, which are temperature-dependent
\begin{eqnarray}
  &&\dGNIST = \GNIST(\mol[{\rm cond}],T) \\
  && - a\,\GNIST({\rm Aref},T) 
     - b\,\GNIST({\rm Bref},T) 
     - c\,\GNIST({\rm Cref},T) \nonumber\ .
\end{eqnarray}
To obtain the Gibbs free energy of formation of a condensate from free
gas atoms, our $\dG$ as input for Eq.\,(\ref{eq:dGsolid}), we need to
subtract as
\begin{eqnarray}
  &&\dG = \dGNIST(\mol[{\rm cond}],T) \\
  && - a\,\dGNIST({\rm A},T) 
     - b\,\dGNIST({\rm B},T) 
     - c\,\dGNIST({\rm C},T) \nonumber
\end{eqnarray}
The required input for the vapour pressure (Eq.\,\ref{eq:pvap}) is
obtained in a similar way by subtracting $\dGNIST(\mol,T)$
instead.

Second, the geophysical SUPCRTBL database
\citep{Zimmer2016,Johnson1992}\footnote{\url{http://www.indiana.edu/~hydrogeo/supcrtbl.html}}
provides condensed phase $\dGSU(T,p)$ data. In this database, the
Gibbs free energies of formation are defined in a different way,
namely with respect to the reference state of the elements at
reference temperature $\Tref\!=\!298.15$\,K
\begin{eqnarray}
  &&\dGSU = \GSU(\mol[{\rm cond}],T) \\
  && - a\,\GSU({\rm Aref},\Tref) 
     - b\,\GSU({\rm Bref},\Tref) 
     - c\,\GSU({\rm Cref},\Tref) \nonumber \ .
\end{eqnarray}
The database actually provides $\dGSU(p,T)$ according to a term $\int
V^\st_i\,dP$ in Eq.\,(2) of \citep{Zimmer2016}. However, $V^\st$ is of
order of a few J/bar, and can safely be neglected at 
$p\!<$\,a few bar as assumed in this paper. In order to arrive at the input
for our Eq.\,(\ref{eq:dGsolid}), we need to subtract corrections as
\begin{equation}
  \dG = \dGSU - a\G_{\rm\!corr}({\rm A}) - b\G_{\rm\!corr}({\rm B}) 
              - c\G_{\rm\!corr}({\rm C}) \ ,
  \label{eq:SUdata}
\end{equation}
namely the Gibbs free energy differences between the free atoms at
temperature $T$ and the respective elements in their standard state at
$\Tref$. We could not figure out a way how to consistently perform
these conversions from the SUPCRTBL data, as most of the atoms are
lacking. However, assuming that the standard states of the elements at
$\Tref$ are identical in both NIST and SUPCRTBL, we can utilise
NIST-JANAF as
\begin{eqnarray}
  \G_{\rm\!corr}({\rm A}) &\!\!\!=\!\!\!& \GNIST({\rm A},T)-\GNIST({\rm Aref},\Tref) \\
   &\!\!\!=\!\!\!& \dGNIST({\rm A},T) \nonumber\\
   &\!\!\!+\!\!\!& \big[H^\st_{\rm NIST}({\rm Aref},T)-H^\st_{\rm NIST}({\rm Aref},\Tref)\big] 
  \nonumber\\
   &\!\!\!-\!\!\!& T\,S^\st_{\rm NIST}({\rm Aref},T) + \Tref\,S^\st_{\rm NIST}({\rm Aref},\Tref)
  \ . \label{eq:AtomCorr}
\end{eqnarray}
where $\dGNIST$, [$H^\st_{\rm NIST}(T)\!-\!H^\st_{\rm NIST}(\Tref)$] and
$S^\st_{\rm NIST}$ are given in the 7$^{\rm th}$, 5$^{\rm th}$ and
3$^{\rm rd}$ columns of the NIST-JANAF data-files.  We have computed
high-precision functional fits of these conversions and list the
fit coefficients and functional form in Table~\ref{tab:Acorr}.

By applying these atomic corrections to the SUPCRTBL $\dGSU$ data according to
Eq.\,(\ref{eq:SUdata}), we have converted that data to our reference states of
neutral atoms needed in Eq.\,(\ref{eq:dGsolid}). We have extracted
$\dG$ data this way for a large number of
minerals known on Earth listed in Table~\ref{tab:SUdata}. The table
lists functional fits for $\dG$ for 121 solid species, interestingly
including some phyllosilicates which host OH or H$_2$O in their lattice
structure. We have ordered the minerals in Table~\ref{tab:SUdata},
somewhat arbitrarily, by $\dGSU(\Tref)/N$, where $N$ is the sum of
stoichiometric factors, which gives a first expression of what could
be the most thermodynamically stable solid materials (note that
SUPCRTBL does not have tungsten). We did not include ``aqueous''
species (i.e.\ species solved in liquid water), and applied the
following additional selection criteria:
\begin{itemize}
\item in case of multiple minerals with the same stoichiometric
  factors we only included the most stable compound at standard
  temperature,
\item minerals with broken stoichiometric factors, or with any 
  stoichiometric factor $>\!16$ are ignored,
\item we did not include arsenic (As) or gallium (Ga) compounds. 
\end{itemize}

\begin{table*}[!p]
\caption{Temperature-fits of $k_p$ data for 552 molecules and ions
  made from H, He, Li, C, N, O, F, Na, Mg, Al, Si,
  P, S, Cl, K, Ca, Ti, V, Cr, Mn, Fe, Ni, Zr and W (24 elements).}
\label{tab:kp}
\vspace*{-2mm}
\centering
\resizebox{14cm}{!}{\begin{tabular}{l|l|c|c|rrrrr|l}
\hline
&&&&&&&&&\\[-2ex]
 & molecule & source & fit & $a_0$ & $a_1$ & $a_2$ & $a_3$ & $a_4$ & $\sigma$[kJ/mol]\\
&&&&&&&&&\\[-2ex]
\hline
&&&&&&&&&\\[-2ex]
   1 & H2           & (2) & 4 &  5.19096E+04 & -1.80117E+00 &  8.72246E-02 &  2.56139E-04 & -5.35403E-09 & \\
   2 & LI2          & (2) & 4 &  1.23866E+04 &  1.56495E-02 & -9.78117E+00 & -3.29003E-04 &  2.77588E-08 & \\
   3 & C2           & (2) & 4 &  7.13486E+04 & -7.53302E-01 & -8.82633E+00 &  8.40914E-05 & -8.34907E-10 & \\
   4 & N2           & (2) & 4 &  1.13210E+05 & -1.79949E+00 & -2.03355E+00 &  4.31303E-04 & -2.48433E-08 & \\
   5 & O2           & (2) & 4 &  5.95336E+04 & -1.61517E+00 & -3.60976E+00 &  4.06274E-04 & -1.86946E-08 & \\
   6 & F2           & (2) & 4 &  1.87418E+04 & -1.49389E+00 & -4.22069E+00 &  4.84494E-04 & -3.34918E-08 & \\
   7 & NA2          & (2) & 4 &  8.85584E+03 &  3.60393E-01 & -1.15966E+01 & -5.20312E-04 &  3.52831E-08 & \\
   8 & MG2          & (2) & 4 &  2.95153E+02 & -1.54214E+00 &  3.81836E+00 & -7.93259E-04 &  9.60535E-08 & \\
   9 & AL2          & (2) & 4 &  2.08161E+04 &  1.98351E-01 & -1.28542E+01 & -1.49552E-04 &  1.17923E-08 & \\
  10 & SI2          & (2) & 4 &  3.73518E+04 & -2.45866E-01 & -1.15745E+01 &  1.50992E-04 & -1.59304E-08 & \\
  11 & P2           & (2) & 4 &  5.85956E+04 & -9.74048E-01 & -6.75364E+00 &  2.83200E-04 & -3.16725E-08 & \\
  12 & S2           & (2) & 4 &  5.08019E+04 & -1.51882E+00 & -3.18934E+00 &  4.61697E-04 & -2.60901E-08 & \\
  13 & CL2          & (2) & 4 &  2.87966E+04 & -1.37354E+00 & -3.87637E+00 &  2.85709E-04 & -1.07670E-08 & \\
  14 & K2           & (2) & 4 &  6.57112E+03 &  1.69191E-01 & -9.50003E+00 & -4.24701E-04 & -9.32751E-09 & \\
  15 & H2+          & (2) & 4 & -1.27090E+05 &  6.76863E-01 & -1.27744E+01 &  4.97800E-04 & -2.04564E-08 & \\
  16 & HE2+         & (1) & 4 & -2.56952E+05 &  8.42165E-01 & -1.31376E+01 &  4.52726E-04 & -1.79413E-08 & $\pm$0.04\\
  17 & C2+          & (1) & 4 & -5.86384E+04 &  7.92144E-01 & -1.67128E+01 &  5.48351E-04 & -3.85179E-08 & $\pm$0.20\\
  18 & N2+          & (2) & 4 & -6.76237E+04 &  6.57659E-01 & -1.52323E+01 &  4.51807E-04 & -1.70741E-08 & \\
  19 & O2+          & (2) & 4 & -8.05914E+04 &  7.70714E-01 & -1.69329E+01 &  4.25784E-04 & -2.32586E-08 & \\
  20 & P2+          & (1) & 4 & -6.42167E+04 &  1.41044E+00 & -1.81067E+01 &  4.70925E-04 & -4.58882E-08 & $\pm$0.26\\
  21 & S2+          & (1) & 4 & -6.05511E+04 &  7.79066E-01 & -1.59703E+01 &  6.14185E-04 & -4.51795E-08 & $\pm$0.33\\
  22 & H2-          & (2) & 4 &  2.30070E+04 & -3.98549E+00 &  1.35395E+01 &  4.01297E-04 & -1.74259E-08 & \\
  23 & C2-          & (2) & 4 &  1.17887E+05 & -4.09781E+00 &  1.02566E+01 &  3.78388E-04 & -1.57139E-08 & \\
  24 & LIH          & (2) & 4 &  2.81110E+04 & -1.41584E+00 & -7.86666E-01 &  4.18528E-04 & -2.55524E-08 & \\
  25 & CH           & (2) & 4 &  4.03796E+04 & -2.06522E+00 &  2.73563E+00 &  6.03534E-04 & -2.96038E-08 & \\
  26 & NH           & (2) & 4 &  3.72483E+04 & -1.85316E+00 &  1.80469E+00 &  3.61042E-04 & -1.42478E-08 & \\
  27 & OH           & (2) & 4 &  5.09586E+04 & -1.86657E+00 &  1.28296E+00 &  3.27612E-04 & -1.27895E-08 & \\
  28 & HF           & (2) & 4 &  6.79757E+04 & -2.00046E+00 &  1.23667E+00 &  3.35407E-04 & -1.20171E-08 & \\
  29 & NAH          & (2) & 4 &  2.38678E+04 & -1.26079E+00 & -1.49175E+00 &  3.72601E-04 & -2.18874E-08 & \\
  30 & MGH          & (2) & 4 &  2.32206E+04 & -1.42911E+00 &  8.11874E-01 &  3.89732E-04 & -2.05585E-08 & \\
  31 & ALH          & (2) & 4 &  3.42887E+04 & -1.51855E+00 & -1.14794E+00 &  3.83819E-04 & -1.87632E-08 & \\
  32 & SIH          & (2) & 4 &  3.46400E+04 & -1.54955E+00 & -8.92506E-02 &  3.23215E-04 & -1.63317E-08 & \\
  33 & PH           & (3) & 4 &  3.33009E+04 & -1.75432E+00 &  1.65582E+00 &  4.35721E-04 & -2.40582E-08 & $\pm$0.06\\
  34 & HS           & (3) & 4 &  4.23352E+04 & -1.58869E+00 & -1.54130E-02 &  2.52514E-04 & -9.19992E-09 & $\pm$0.13\\
  35 & HCL          & (2) & 4 &  5.13028E+04 & -2.13173E+00 &  2.87709E+00 &  4.45605E-04 & -1.92144E-08 & \\
  36 & KH           & (2) & 4 &  2.18271E+04 & -1.21500E+00 & -1.44207E+00 &  4.13969E-04 & -3.15761E-08 & \\
  37 & CAH          & (1) & 4 &  2.64607E+04 & -1.68708E+00 &  2.79074E+00 &  6.21435E-04 & -4.76821E-08 & $\pm$0.03\\
  38 & TIH          & (1) & 4 &  2.41513E+04 & -1.87758E+00 &  2.14663E+00 &  1.12481E-03 & -8.96218E-08 & $\pm$0.46\\
  39 & CRH          & (1) & 4 &  2.23915E+04 & -1.63632E+00 &  1.31500E+00 &  4.48117E-04 & -3.10873E-08 & $\pm$0.17\\
  40 & MNH          & (1) & 4 &  3.01855E+04 & -1.68515E+00 &  1.97207E+00 &  5.43045E-04 & -3.20198E-08 & $\pm$0.03\\
  41 & FEH          & (1) & 4 &  1.73894E+04 & -1.45574E+00 & -9.26441E-01 &  8.18399E-04 & -6.07297E-08 & $\pm$0.22\\
  42 & NIH          & (1) & 4 &  2.84479E+04 & -1.86698E+00 &  1.56057E+00 &  4.66092E-04 & -2.32895E-08 & $\pm$0.15\\
  43 & HEH+         & (1) & 4 & -1.36386E+05 &  8.66741E-01 & -1.41095E+01 &  2.66518E-04 & -2.54550E-10 & $\pm$0.23\\
  44 & CH+          & (2) & 4 & -8.30023E+04 &  6.96425E-01 & -1.45227E+01 &  3.80894E-04 & -1.77873E-08 & \\
  45 & NH+          & (1) & 4 & -1.16665E+05 &  1.26447E+00 & -1.55925E+01 &  5.59511E-05 &  6.26295E-09 & $\pm$0.18\\
  46 & OH+          & (2) & 4 & -1.02017E+05 &  6.07635E-01 & -1.29560E+01 &  3.89881E-04 & -1.60349E-08 & \\
  47 & HF+          & (1) & 4 & -1.18079E+05 &  7.25450E-01 & -1.26387E+01 &  2.19710E-04 & -1.57439E-09 & $\pm$0.11\\
  48 & MGH+         & (1) & 4 & -6.62374E+04 &  8.32839E-01 & -1.25711E+01 &  4.57171E-04 & -2.33883E-08 & $\pm$0.03\\
  49 & ALH+         & (1) & 4 & -5.05309E+04 &  5.24296E-01 & -1.17158E+01 &  7.03377E-04 & -4.16652E-08 & $\pm$0.21\\
  50 & SIH+         & (2) & 4 & -5.73762E+04 &  8.56881E-01 & -1.50535E+01 &  3.49121E-04 & -1.81757E-08 & \\
  51 & PH+          & (1) & 4 & -8.24772E+04 &  8.49004E-01 & -1.29265E+01 &  3.42109E-04 & -2.13700E-08 & $\pm$0.07\\
  52 & SH+          & (1) & 4 & -7.88343E+04 &  5.93436E-01 & -1.25291E+01 &  3.78752E-04 & -1.67501E-08 & $\pm$0.11\\
  53 & HCL+         & (1) & 4 & -9.64393E+04 &  8.22741E-01 & -1.28902E+01 &  1.66628E-04 &  4.90184E-09 & $\pm$0.21\\
  54 & CH-          & (1) & 4 &  5.46372E+04 & -4.15997E+00 &  1.40591E+01 &  2.79170E-04 & -5.39916E-09 & $\pm$0.10\\
  55 & OH-          & (2) & 4 &  7.21674E+04 & -4.40770E+00 &  1.44059E+01 &  3.38619E-04 & -1.35099E-08 & \\
  56 & SIH-         & (1) & 4 &  4.96635E+04 & -4.73103E+00 &  1.83213E+01 &  7.93958E-04 & -4.73984E-08 & $\pm$0.29\\
  57 & HS-          & (1) & 4 &  6.85217E+04 & -4.39583E+00 &  1.51490E+01 &  3.32732E-04 & -9.75243E-09 & $\pm$0.13\\
  58 & CN           & (2) & 4 &  9.02324E+04 & -1.75861E+00 & -1.56088E+00 &  4.20862E-04 & -1.59741E-08 & \\
  59 & CO           & (2) & 4 &  1.28998E+05 & -1.75498E+00 & -3.16258E+00 &  4.13362E-04 & -2.35800E-08 & \\
  60 & CF           & (2) & 4 &  6.47022E+04 & -1.43350E+00 & -3.23877E+00 &  3.54145E-04 & -2.03239E-08 & \\
  61 & SIC          & (2) & 4 &  5.38781E+04 &  4.16289E-01 & -1.62625E+01 & -2.85599E-04 &  1.44412E-08 & \\
  62 & CP           & (2) & 4 &  6.98314E+04 & -1.27244E+00 & -4.42830E+00 &  3.42881E-04 & -2.86410E-08 & \\
  63 & CS           & (2) & 4 &  8.54423E+04 & -1.49342E+00 & -4.26643E+00 &  3.66935E-04 & -2.34112E-08 & \\
  64 & CCL          & (2) & 4 &  4.00861E+04 & -1.21265E+00 & -3.98298E+00 &  2.56087E-04 & -1.37754E-08 & \\
  65 & CN+          & (2) & 4 & -7.34125E+04 &  1.01025E+00 & -1.63642E+01 &  4.25271E-04 & -2.35543E-08 & \\
  66 & CO+          & (1) & 4 & -3.35805E+04 &  5.48177E-01 & -1.53832E+01 &  5.17815E-04 & -2.97321E-08 & $\pm$0.14\\
  67 & CN-          & (2) & 4 &  1.34538E+05 & -4.23936E+00 &  1.17219E+01 &  4.10824E-04 & -2.43610E-08 & \\
  68 & CS-          & (1) & 4 &  8.88849E+04 & -4.44100E+00 &  1.43923E+01 &  6.13513E-04 & -4.02613E-08 & $\pm$0.17\\
\hline
\end{tabular}}\\[1mm]
\begin{minipage}{18cm}
\footnotesize All $k_p$ data for standard pressure $\pst\!=\!1\,$bar
in cgs-units [$\rm(dyn/cm^2)^{1-n}$], where $n$ is the number of atoms
and electrons in the molecule.  Data sources: (1) $=$
\citet{Barklem2016} converted and fitted between 50\,K and
6000\,K. (2) $=$ \citet{Stock2008} based on NIST-JANAF
\citep{Chase1986}.  (3) $=$ fits made by the authors to NIST-JANAF
data for the full available temperature range, usually 100\,K to
6000\,K. The last column is the standard deviation of the fit from the
data points [kJ/mol]. (4) $=$ Andreas Gauger (priv.comm). (5) $=$
\citep{Tsuji1973} data re-fitted by \citet{Stock2008}. The following
formula apply, where $\theta\!=\!5040/T$:\\[1.5mm]
\begin{tabular}{crp{1mm}l}
$\rm fit=4$: & $  \ln k_p(T)$    &=& $(1-n)\ln(1{\rm\,bar}) +\left(a_0/T + a_1\,\ln T + a_2 + a_3\,T + a_4\,T^2\right)$\\[0.5mm]
$\rm fit=5$: & $\log_{10} k_p(T)$ &=& $-a_0 -a_1\theta -a_2\log_{10}\theta -a_3(\log_{\rm 10}\theta)^2 -a_4(\log_{\rm 10}\theta)^3$
\end{tabular}
\end{minipage}
\end{table*}
\addtocounter{table}{-1}
\begin{table*}[!p]
\caption{{\bf continued}}
\vspace*{-2mm}
\centering
\resizebox{14cm}{!}{\begin{tabular}{l|l|c|c|rrrrr|l}
\hline
&&&&&&&&&\\[-2ex]
 & molecule & source & fit & $a_0$ & $a_1$ & $a_2$ & $a_3$ & $a_4$ & $\sigma$[kJ/mol]\\
&&&&&&&&&\\[-2ex]
\hline
&&&&&&&&&\\[-2ex]
  69 & NO           & (2) & 4 &  7.55243E+04 & -1.66485E+00 & -1.61409E+00 &  4.13077E-04 & -2.37492E-08 & \\
  70 & FN           & (2) & 4 &  3.61331E+04 & -1.34903E+00 & -2.94868E+00 &  3.47267E-04 & -1.96760E-08 & \\
  71 & ALN          & (2) & 4 &  3.33804E+04 & -9.12019E-01 & -4.84637E+00 &  1.97911E-04 & -1.11046E-08 & \\
  72 & SIN          & (2) & 4 &  6.57987E+04 & -1.53565E+00 & -2.64502E+00 &  5.01934E-04 & -2.62642E-08 & \\
\hline
\end{tabular}}\\[1mm]
\end{table*}
\addtocounter{table}{-1}
\begin{table*}[!p]
\caption{{\bf continued}}
\vspace*{-2mm}
\centering
\resizebox{14cm}{!}{\begin{tabular}{l|l|c|c|rrrrr|l}
\hline
&&&&&&&&&\\[-2ex]
 & molecule & source & fit & $a_0$ & $a_1$ & $a_2$ & $a_3$ & $a_4$ & $\sigma$[kJ/mol]\\
&&&&&&&&&\\[-2ex]
\hline
&&&&&&&&&\\[-2ex]
 152 & CFCLO        & (2) & 4 &  1.90938E+05 & -3.22705E+00 & -2.35717E+01 &  1.11849E-03 & -6.39039E-08 & \\
 153 & CCLF3        & (2) & 4 &  2.13906E+05 & -3.54875E+00 & -3.96988E+01 &  1.43940E-03 & -8.20706E-08 & \\
 154 & CCLN         & (2) & 4 &  1.40543E+05 & -2.11454E+00 & -1.53492E+01 &  7.62648E-04 & -4.01053E-08 & \\
 155 & CCLO         & (2) & 4 &  1.37740E+05 & -2.24974E+00 & -1.16913E+01 &  5.79713E-04 & -3.34800E-08 & \\
\hline
\end{tabular}}\\[1mm]
\end{table*}
\addtocounter{table}{-1}
\begin{table*}[!p]
\caption{{\bf continued}}
\vspace*{-2mm}
\centering
\resizebox{14cm}{!}{\begin{tabular}{l|l|c|c|rrrrr|l}
\hline
&&&&&&&&&\\[-2ex]
 & molecule & source & fit & $a_0$ & $a_1$ & $a_2$ & $a_3$ & $a_4$ & $\sigma$[kJ/mol]\\
&&&&&&&&&\\[-2ex]
\hline
&&&&&&&&&\\[-2ex]
 235 & CLF5         & (2) & 4 &  9.06733E+04 & -2.37569E+00 & -6.35600E+01 &  1.21159E-03 & -6.79411E-08 & \\
 236 & CLF5S        & (2) & 4 &  2.20114E+05 & -3.43455E+00 & -7.62559E+01 &  1.78547E-03 & -1.02728E-07 & \\
 237 & CHCLF2       & (2) & 4 &  2.02940E+05 & -4.98328E+00 & -2.54851E+01 &  1.80645E-03 & -9.85202E-08 & \\
 238 & CHCL2F       & (2) & 4 &  1.84147E+05 & -4.76213E+00 & -2.60846E+01 &  1.70179E-03 & -9.19490E-08 & \\
\hline
\end{tabular}}\\[1mm]
\end{table*}
\addtocounter{table}{-1}
\begin{table*}[!p]
\caption{{\bf continued}}
\vspace*{-2mm}
\centering
\resizebox{14cm}{!}{\begin{tabular}{l|l|c|c|rrrrr|l}
\hline
&&&&&&&&&\\[-2ex]
 & molecule & source & fit & $a_0$ & $a_1$ & $a_2$ & $a_3$ & $a_4$ & $\sigma$[kJ/mol]\\
&&&&&&&&&\\[-2ex]
\hline
&&&&&&&&&\\[-2ex]
 319 & NF3          & (2) & 4 &  1.00865E+05 & -2.71701E+00 & -2.73662E+01 &  1.00793E-03 & -5.86664E-08 & \\
 320 & F3NO         & (2) & 4 &  1.34498E+05 & -3.24725E+00 & -4.14236E+01 &  1.36643E-03 & -7.89717E-08 & \\
 321 & POF3         & (2) & 4 &  2.46952E+05 & -3.19634E+00 & -4.21379E+01 &  1.38688E-03 & -8.68175E-08 & \\
 322 & PF3          & (2) & 4 &  1.81552E+05 & -2.27768E+00 & -2.98605E+01 &  8.79252E-04 & -5.78887E-08 & \\
\hline
\end{tabular}}\\[1mm]
\end{table*}
\addtocounter{table}{-1}
\begin{table*}[!p]
\caption{{\bf continued}}
\vspace*{-2mm}
\centering
\resizebox{14cm}{!}{\begin{tabular}{l|l|c|c|rrrrr|l}
\hline
&&&&&&&&&\\[-2ex]
 & molecule & source & fit & $a_0$ & $a_1$ & $a_2$ & $a_3$ & $a_4$ & $\sigma$[kJ/mol]\\
&&&&&&&&&\\[-2ex]
\hline
&&&&&&&&&\\[-2ex]
 404 & ALF2O-       & (2) & 4 &  2.45136E+05 & -4.40725E+00 & -1.69695E+01 &  7.08766E-04 & -4.12110E-08 & \\
 405 & ALF4-        & (2) & 4 &  3.16357E+05 & -4.40893E+00 & -3.48436E+01 &  8.87754E-04 & -5.06459E-08 & \\
 406 & ALOH+        & (2) & 4 &  3.04243E+04 & -6.37579E-01 & -1.95500E+01 &  9.55290E-04 & -4.98767E-08 & \\
 407 & ALOH-        & (2) & 4 &  1.22012E+05 & -5.51846E+00 &  9.67964E+00 &  7.33455E-04 & -3.74812E-08 & \\
\hline
\end{tabular}}\\[1mm]
\end{table*}
\addtocounter{table}{-1}
\begin{table*}[!p]
\caption{{\bf continued}}
\vspace*{-2mm}
\centering
\resizebox{14cm}{!}{\begin{tabular}{l|l|c|c|rrrrr|l}
\hline
&&&&&&&&&\\[-2ex]
 & molecule & source & fit & $a_0$ & $a_1$ & $a_2$ & $a_3$ & $a_4$ & $\sigma$[kJ/mol]\\
&&&&&&&&&\\[-2ex]
\hline
&&&&&&&&&\\[-2ex]
 487 & LIF4AL       & (2) & 4 &  3.19697E+05 & -1.96460E+00 & -6.18249E+01 &  1.14317E-03 & -7.07626E-08 & \\
 488 & CLFLI2       & (2) & 4 &  1.52729E+05 & -1.98975E+00 & -2.75831E+01 &  7.99301E-04 & -5.56238E-08 & \\
 489 & LIOCL        & (2) & 4 &  6.49261E+04 & -2.13332E+00 & -1.14020E+01 &  6.20542E-04 & -4.05228E-08 & \\
 490 & WCL          & (3) & 4 &  5.00275E+04 & -1.10972E+00 & -1.72470E+00 & -2.74247E-04 &  1.37409E-08 & $\pm$0.50\\
\hline
\end{tabular}}
\end{table*}

\begin{table*}[!p]
\caption{Condensed phase data fitted and collected from different sources.}
\label{tab:SOLID}
\vspace*{-2mm}\hspace*{-1mm}\resizebox{18.5cm}{!}{
\begin{tabular}{r|ll|c|c|c|ccccc|c}
\hline
&&&&&&&&&&&\\[-2ex]
no & formula & name  & \!$\rho$\,[g/cm$^3$]\!\! & fit &\!restriction\!& $c_0$ & $c_1$ & $c_2$ & $c_3$ & $c_4$ & \!$\sigma$\,[kJ/mol]\!\!\\
&&&&&&&&&&&\\[-2ex]
\hline
&&&&&&&&&&&\\[-2.2ex]
  1 & Al2O3[s]             & CORUNDUM($\alpha$)        &  3.97 &  5 &          &  3.68836E+05 & -7.75003E+00 & -4.32225E+01 &  7.53364E-03 & -7.64141E-07 & $\pm$0.54\\
  2 & Al2O3[l]             &                           &  2.78 &  5 &          &  3.64246E+05 & -1.63572E+00 & -8.13311E+01 &  2.51029E-03 &  3.18884E-08 & $\pm$1.80\\
  3 & MgAl2O4[s]           & SPINEL                    &  3.58 &  5 &          &  4.90917E+05 & -1.01783E+01 & -6.09577E+01 &  1.00839E-02 & -9.21697E-07 & $\pm$0.57\\
  4 & MgAl2O4[l]           &                           &  2.75 &  5 &          &  4.70478E+05 & -1.35143E+00 & -1.08539E+02 &  2.76930E-03 & -7.78980E-08 & $\pm$0.86\\
\hline                                                                                                                                         
\end{tabular}}\\[1mm]                                                                                                                           
\footnotesize 
Solid compounds are marked with '[s]' and liquids with '[l]', whereas
'[s/l]' denote condensates treated as a single combined species.  
The standard pressure is $\pst\!=\!1\,$bar unless otherwise noted. 
$T_C\!=\!T-273.15$ is the temperature in $^\circ$C.
Restriction: this fit should not be applied beyond the indicated temperature as
it will create a secondary spurious intersection point with the other
phase. In {\sc GGchem}, we deal with this by artificially increasing $\pvap$ 
above/below this temperature for solid/liquid species, respectively,
which is far away from the melting point anyway. $\sigma$ is 
the standard deviation of the fitted values from the data points [kJ/mol].\\*[-4mm]
\begin{center}
\begin{tabular}{lrp{1mm}ll}
$\rm fit=1$: & $\dG[{\rm cal/mol}]$     &=& $c_0/T + c_1 + c_2\,T + c_3\,T^2 + c_4\,T^3$,  & data from \citet{Sharp1990}, note that $\pst\!=\!1\,$atm\\[0.5mm]
$\rm fit=2$: & $\dG[{\rm J/mol}]$       &=& $c_0/T + c_1 + c_2\,T + c_3\,T^2 + c_4\,T^3$,  & fitted to NIST-JANAF data \citep{Chase1986} by the authors\\[0.5mm]
$\rm fit=3$: & $\ln\pvap[\rm dyn/cm^2]$ &=& $c_0/T + c_1 + c_2\,T + c_3\,T^2 + c_4\,T^3$   & fitted to NIST-JANAF data \citep{Chase1986} by the authors\\[0.5mm]
$\rm fit=4$: & $\ln\pvap[\rm dyn/cm^2]$ &=& $c_0 + c_1/(T + c_2)$                          & fitted to NIST-JANAF data \citep{Chase1986} by the authors\\[0.5mm]
$\rm fit=5$: & $-\dG/RT$                &=& $c_0/T + c_1\ln(T) + c_2 + c_3\,T + c_4\,T^2$  & fitted to NIST-JANAF data \citep{Chase1986} by the authors\\[0.5mm]
$\rm fit=6$: & $\log_{10}\pvap[\rm mmHg]$&=& $c_0 + c_1/T + c_2\,\log_{10}T + c_3\,T + c_4\,T^2$ & data from \citet{Yaws1999}\\[0.5mm]
$\rm fit=7$: & $\pvap[\rm dyn/cm^2]$    &=& $c_0 \exp[(c_1\,T_C + T_C^2/c_2)/(T_C + c_3)]$ & data from \citet{Ackerman2001}\\[0.5mm]
$\rm fit=8$: & $\ln\pvap[\rm bar]$      &=& $c_0 + c_1/T + c_2/T^2$                       & data from \citet{Weast1971}\\[0.5mm]
$\rm fit=9$: & $\log_{10}\pvap[\rm bar]$&=& $c_0 + c_1/(T + c_2)$                          & data from \citet{Prydz1972}\\[0.5mm]
$\rm fit=10$:& $\ln\pvap[\rm dyn/cm^2]$ &=& $c_0/T + c_1$                                  & data from \citet{Gail2013}
\end{tabular}
\end{center}
\end{table*}
\addtocounter{table}{-1}
\begin{table*}[!p]
\caption{{\bf continued}}
\vspace*{-2mm}\hspace*{-1mm}\resizebox{18.5cm}{!}{
\begin{tabular}{r|ll|c|c|c|ccccc|c}
\hline
&&&&&&&&&&&\\[-2ex]
no & formula & name  & \!$\rho$\,[g/cm$^3$]\!\! & fit &\!restriction\!& $c_0$ & $c_1$ & $c_2$ & $c_3$ & $c_4$ & \!$\sigma$\,[kJ/mol]\!\!\\
&&&&&&&&&&&\\[-2ex]
\hline
&&&&&&&&&&&\\[-2.2ex]
 62 & AlCl3[s]             & AL-TRICHLORIDE            &  2.48 &  2 & $<$1998K &  4.17188E+05 & -1.40425E+06 &  5.68472E+02 & -1.79150E-02 & -4.81449E-06 & $\pm$0.08\\
 63 & AlCl3[l]             &                           &  2.00 &  2 &          & -1.51097E+06 & -1.36116E+06 &  4.88316E+02 & -2.96109E-02 &  3.95859E-06 & $\pm$0.01\\
 64 & CaO[s]               & LIME                      &  3.35 &  2 &          & -4.94275E+04 & -1.06229E+06 &  2.81700E+02 & -7.00317E-03 &  4.60182E-07 & $\pm$0.20\\
 65 & CaO[l]               &                           &  2.70 &  2 &          &  7.88379E+05 & -9.87984E+05 &  2.61388E+02 & -8.44845E-03 &  6.19197E-07 & $\pm$0.15\\
 66 & CaCl2[s]             & CA-DICHLORIDE             &  2.15 &  3 &          & -3.91597E+04 &  3.71795E+01 & -1.93815E-03 &  1.18087E-07 &  0.00000E+00 & $\pm$0.09\\
 67 & CaCl2[l]             &                           &  1.70 &  3 &          & -3.68100E+04 &  3.61822E+01 & -3.53647E-03 &  5.03350E-07 & -3.05198E-11 & $\pm$0.03\\
 68 & LiH[s]               & LI-HYDRIDE                &  0.78 &  3 &          & -2.76405E+04 &  3.07600E+01 &  3.06028E-03 & -3.26876E-06 &  7.70757E-10 & $\pm$0.04\\
 69 & LiH[l]               &                           &  0.62 &  3 &          & -2.63155E+04 &  3.30329E+01 & -3.77190E-03 &  6.41033E-07 &  0.00000E+00 & $\pm$0.04\\
 70 & MgTiO3[s]            & GEIKIELITE                &  3.88 &  2 &          &  8.18358E+05 & -2.94809E+06 &  7.57824E+02 & -1.56613E-02 &  1.01510E-06 & $\pm$0.20\\
 71 & MgTiO3[l]            &                           &  3.10 &  2 &          &  2.95447E+06 & -2.88536E+06 &  7.40614E+02 & -2.52836E-02 &  1.86167E-06 & $\pm$0.19\\
 72 & K2SiO3[s]            & K-SILICATE                &  2.40 &  2 &          &  1.08754E+06 & -2.93466E+06 &  8.61167E+02 & -3.66511E-02 &  3.06399E-06 & $\pm$0.09\\
 73 & K2SiO3[l]            &                           &  1.90 &  2 &          & -2.09568E+06 & -2.87130E+06 &  8.04917E+02 & -3.03325E-02 &  2.75492E-06 & $\pm$0.34\\
 74 & Ti[s]                & TITANIUM(beta)            &  4.51 &  2 & $<$4302K &  1.31154E+05 & -4.67808E+05 &  1.45369E+02 & -3.49413E-03 &  0.00000E+00 & $\pm$0.02\\
 75 & Ti[l]                &                           &  4.14 &  2 &          &  1.72969E+06 & -4.68384E+05 &  1.54389E+02 & -9.64953E-03 &  7.21551E-07 & $\pm$0.18\\
 76 & TiO[s]               & TI-MONOXIDE($\beta$)      &  4.95 &  3 &          & -7.14013E+04 &  3.80934E+01 & -1.49104E-03 &  6.06190E-08 &  0.00000E+00 & $\pm$0.01\\
 77 & TiO[l]               &                           &  4.00 &  3 &          & -6.61504E+04 &  3.58743E+01 & -1.97510E-03 &  2.33017E-07 & -1.21467E-11 & $\pm$0.01\\
 78 & LiOH[s]              & LI-HYDROXIDE              &  1.46 &  3 &          & -3.01062E+04 &  3.37718E+01 &  9.66853E-04 & -2.02320E-06 &  5.34064E-10 & $\pm$0.04\\
 79 & LiOH[l]              &                           &  1.20 &  3 &          & -2.87677E+04 &  3.42848E+01 & -3.81580E-03 &  5.39401E-07 &  0.00000E+00 & $\pm$0.07\\
 80 & VO[s]                & V-MONOXIDE                &  5.76 &  3 & $<$3404K & -6.74603E+04 &  3.82717E+01 & -2.78551E-03 &  5.72078E-07 & -7.41840E-11 & $\pm$0.02\\
 81 & VO[l]                &                           &  5.20 &  3 &          & -6.01063E+04 &  3.45413E+01 & -2.64854E-03 &  4.70724E-07 & -3.85068E-11 & $\pm$0.01\\
 82 & V2O3[s]              & KARELIANITE               &  4.95 &  5 &          &  3.59474E+05 & -4.09857E+00 & -6.48556E+01 &  4.73778E-03 & -3.36133E-07 & $\pm$0.32\\
 83 & V2O4[s]              & PARAMONTROSEITE\!\!       &  4.09 &  5 &          &  4.14455E+05 & -2.22575E+00 & -9.36637E+01 &  3.73856E-03 & -2.64367E-07 & $\pm$0.53\\
 84 & V2O5[s]              & SHCHERBINAITE             &  3.37 &  5 &          &  4.58651E+05 & -6.83064E+00 & -8.22840E+01 &  6.02078E-03 &  0.00000E+00 & $\pm$0.51\\
 85 & CaS[s]               & CALCIUM-SULFIDE           &  2.59 &  2 &          & -8.11371E+04 & -9.28021E+05 &  2.69568E+02 & -7.11550E-03 &  5.21453E-07 & $\pm$0.13\\
 86 & FeS2[s]              & PYRITE                    &  4.90 &  2 &          &  5.30142E+05 & -1.14524E+06 &  4.72929E+02 & -4.33681E-03 & -1.49521E-06 & $\pm$0.05\\
 87 & Na2S[s]              & NA-SULFIDE                &  1.86 &  2 &          &  1.99053E+06 & -8.70027E+05 &  4.06721E+02 & -2.78298E-02 &  0.00000E+00 & $\pm$0.30\\
 88 & Mn[s]                & MANGANESE($\alpha-\delta$)&  7.43 &  3 &          & -3.41316E+04 &  3.12732E+01 & -5.59461E-04 & -4.63498E-07 &  1.01677E-10 & $\pm$0.01\\
 89 & Mn[l]                &                           &  5.95 &  3 &          & -3.22207E+04 &  3.01880E+01 & -1.14796E-03 & -3.02939E-08 &  2.15830E-11 & $\pm$0.00\\
 90 & MnS[s]               & ALABANDITE                &  4.08 &  1 &          &  1.12482E+05 & -1.81938E+05 &  5.87107E+01 &  8.89360E-05 & -4.20876E-09 &          \\
 91 & Ni[s]                & NICKEL                    &  8.91 &  3 & $<$2915K & -5.17726E+04 &  3.25355E+01 & -8.54276E-04 &  1.93715E-07 & -3.72769E-11 & $\pm$0.03\\
 92 & Ni[l]                &                           &  7.81 &  3 &          & -4.96579E+04 &  3.09204E+01 & -2.80251E-04 & -1.00993E-07 &  1.63913E-11 & $\pm$0.00\\
 93 & Cr[s]                & CHROMIUM                  &  7.19 &  3 & $<$2981K & -4.78455E+04 &  3.22423E+01 & -5.28710E-04 & -6.17347E-08 &  2.88469E-12 & $\pm$0.00\\
 94 & Cr[l]                &                           &  6.30 &  3 &          & -4.47712E+04 &  3.09753E+01 & -7.84094E-04 & -5.92580E-10 &  1.25866E-11 & $\pm$0.01\\
 95 & CrN[s]               & CARLSBERGITE              &  6.09 &  3 &          & -7.51246E+04 &  3.88434E+01 & -3.48286E-03 &  1.12997E-06 & -1.66901E-10 & $\pm$0.06\\
 96 & CaSiO3[s]            & WOLLASTONITE              &  2.91 &  5 &          &  3.60350E+05 & -6.69046E+00 & -4.57724E+01 &  7.14526E-03 & -8.10884E-07 & $\pm$0.26\\
 97 & CaTiO3[s]            & PEROVSKITE                &  3.98 &  1 &          &  1.19107E+04 & -7.30327E+05 &  1.75930E+02 & -2.84630E-03 &  1.10392E-07 &          \\
 98 & NiS[s]               & MILLERITE                 &  5.37 &  1 &          & -4.32421E+04 & -1.89653E+05 &  7.16749E+01 & -2.50684E-03 &  0.00000E+00 &          \\
 99 & NiS2[s]              & VAESITE                   &  4.45 &  1 &          &  0.00000E+00 & -2.66982E+05 &  1.08176E+02 & -1.96812E-03 &  0.00000E+00 &          \\
100 & Ni3S2[s]             & HEAZLEWOODITE             &  5.87 &  1 &          &  3.12877E+05 & -5.01379E+05 &  2.14522E+02 & -3.99302E-02 &  7.88559E-06 &          \\
\hline                                                                                                                                         
\end{tabular}}
\end{table*}

\begin{table}
\caption{Overview of pairs of solid/liquid data and melting points.}
\label{tab:liquid}
\vspace*{-2mm}
\resizebox{!}{58mm}{
\begin{tabular}{l|c|c|c}
\hline
condensate & intersection point [K] & melting point [K] & ref.\\
\hline
&&&\\[-2.2ex]
W                 &  3702  & 3680 & (1)\\
TiC               &  3292  & 3290 & (1)\\
CaO               &  3200  & 3200 & (1)\\
MgO               &  3105  & 3105 & (1)\\
ZrO$_2$           &  2952  & 2950 & (1)\\
MgAl$_2$O$_4$     &  2408  & 2408 & (1)\\
Al$_2$O$_3$       &  2327  & 2327 & (1)\\
Mg$_2$SiO$_4$     &  2171  & 2171 & (1)\\
TiO$_2$           &  2129  & 2130 & (1)\\
Cr                &  2137  & 2130 & (1)\\
VO                &  2063  & 2063 & (1)\\
TiO               &  2025  & 2023 & (1)\\
MgTi$_2$O$_5$     &  1960  & 1963 & (1)\\
MgTiO$_3$         &  1949  & 1953 & (1)\\
Ti$_4$O$_7$       &  1947  & 1950 & (1)\\
Ti                &  1919  & 1939 & (1)\\
MgSiO$_3$         &  1852  & 1850 & (1)\\
Fe                &  1822  & 1809 & (1)\\
Ni                &  1725  & 1728 & (1)\\
FeS               &  1490  & 1463 & (1)\\
Mn                &  1516  & 1519 & (1)\\
SiO$_2$           &  1698  & 1696 & (1)\\
Zr                &  2116  & 2125 & (1)\\
WO$_3$            &  1748  & 1745 & (1)\\
FeO               &  1651  & 1650 & (1)\\
SiS$_2$           &  1364  & 1363 & (1)\\
Na$_2$SiO$_3$     &  1362  & 1362 & (1)\\
K$_2$SiO$_3$      &  1245  & 1249 & (1)\\
NaCl              &  1078  & 1074 & (1)\\
KCl               &  1054  & 1044 & (1)\\
CaCl$_2$          &  1035  & 1045 & (1)\\
LiH               &   966  &  962 & (1)\\
LiCl              &   877  &  883 & (1)\\
LiOH              &   733  &  744 & (1)\\
AlCl$_3$          &   466  &  466 & (1)\\
S                 &   385  &  388 & (1)\\
Na                &   368  &  371 & (1)\\
H$_2$O            &   273  &  273 & (2)\\
\hline
\end{tabular}}\\[1mm]
\footnotesize
(1) data from NIST-JANAF \citep{Chase1986}, fitted by the authors\\
(2) fits from \citep{Ackerman2001} and \citep{Yaws1999}
\end{table}

\begin{table*}[!p]
\caption{Condensed phase Gibbs free energy data extracted from the SUPCRTBL database \citep{Zimmer2016}.}
\label{tab:SUdata}
\vspace*{-2mm}\hspace*{-1mm}\resizebox{18.5cm}{!}{
\begin{tabular}{r|ll|c|ccccc|c}
\hline
&&&&&&&&&\\[-2ex]
no & formula & name &\!\!$\dGSU(\Tref)/N$\!\!& $b_0$ & $b_1$ & $b_2$ & $b_3$ & $b_4$ & \!$\sigma$\,[kJ/mol]\!\! \\
&&&&&&&&&\\[-2ex]
\hline
&&&&&&&&&\\[-2.2ex]
   1 &            ZrO2 &          BADDELEYITE & -347.6 &  2.648723e+05 & -3.340550e+00 & -3.414969e+01 &  3.635220e-03 & -4.584324e-07 & $\pm$0.14\\ 
   2 &          ZrSiO4 &               ZIRCON & -320.1 &  4.900611e+05 & -8.787114e+00 & -5.527139e+01 &  8.613074e-03 & -1.027249e-06 & $\pm$0.26\\ 
   3 &           Al2O3 &             CORUNDUM & -316.4 &  3.686315e+05 & -8.641215e+00 & -3.798714e+01 &  8.874690e-03 & -1.024209e-06 & $\pm$0.42\\ 
   4 &      Ca2Al2SiO7 &            GEHLENITE & -315.9 &  8.625975e+05 & -1.533457e+01 & -1.143058e+02 &  1.632128e-02 & -1.836678e-06 & $\pm$0.53\\ 
   5 &    Ca3Al2Si3O12 &            GROSSULAR & -314.1 &  1.457200e+06 & -3.049843e+01 & -1.691883e+02 &  3.164365e-02 & -3.657780e-06 & $\pm$1.28\\ 
   6 &         Ca2SiO4 &              LARNITE & -313.3 &  4.925717e+05 & -7.561371e+00 & -7.350215e+01 &  8.818589e-03 & -9.618190e-07 & $\pm$0.23\\ 
   7 &       CaAl2SiO6 &        Ca-TSCHERMAKS & -313.3 &  7.292797e+05 & -1.488941e+01 & -8.659854e+01 &  1.582366e-02 & -1.810913e-06 & $\pm$0.57\\ 
   8 &        Ca3Si2O7 &            RANKINITE & -312.1 &  8.533219e+05 & -1.389306e+01 & -1.213161e+02 &  1.564223e-02 & -1.814195e-06 & $\pm$0.60\\ 
   9 &         MgAl2O4 &               SPINEL & -311.0 &  4.912197e+05 & -1.043835e+01 & -6.070647e+01 &  1.107213e-02 & -1.213724e-06 & $\pm$0.42\\ 
  10 &          CaSiO3 &         WOLLASTONITE & -309.8 &  3.603498e+05 & -6.690462e+00 & -4.577239e+01 &  7.145265e-03 & -8.108841e-07 & $\pm$0.26\\ 
  11 &       Ca5P3O12F &         FLUORAPATITE & -309.1 &  1.411299e+06 & -2.348482e+01 & -2.134680e+02 &  2.699044e-02 & -3.118360e-06 & $\pm$0.84\\ 
  12 &      Ca3MgSi2O8 &            MERWINITE & -308.5 &  9.728542e+05 & -1.618741e+01 & -1.392443e+02 &  1.804333e-02 & -2.020680e-06 & $\pm$0.68\\ 
  13 &      CaAl2Si2O8 &            ANORTHITE & -308.3 &  9.529318e+05 & -2.009489e+01 & -1.046593e+02 &  2.000063e-02 & -2.277610e-06 & $\pm$0.84\\ 
  14 &        CaTiSiO5 &               SPHENE & -307.4 &  5.922714e+05 & -1.238208e+01 & -6.590011e+01 &  1.323755e-02 & -1.584733e-06 & $\pm$0.61\\ 
  15 &      Ca2MgSi2O7 &           AKERMANITE & -305.8 &  8.405318e+05 & -1.269676e+01 & -1.275788e+02 &  1.371402e-02 & -1.450183e-06 & $\pm$0.37\\ 
  16 &         Al2SiO5 &              KYANITE & -305.4 &  5.917326e+05 & -1.433362e+01 & -5.773181e+01 &  1.438186e-02 & -1.670007e-06 & $\pm$0.55\\ 
  17 &        CaMgSiO4 &         MONTICELLITE & -304.9 &  4.818448e+05 & -8.787430e+00 & -6.715236e+01 &  9.912023e-03 & -1.160460e-06 & $\pm$0.34\\ 
  18 &       CaMgSi2O6 &             DIOPSIDE & -302.8 &  7.087080e+05 & -1.484631e+01 & -8.419222e+01 &  1.555722e-02 & -1.793114e-06 & $\pm$0.62\\ 
  19 &       MgAl2SiO6 &        Mg-TSCHERMAKS & -302.0 &  7.118491e+05 & -1.538535e+01 & -8.324974e+01 &  1.591182e-02 & -1.724839e-06 & $\pm$0.46\\ 
  20 &             CaO &                 LIME & -301.6 &  1.273800e+05 & -1.455968e+00 & -2.479564e+01 &  2.333874e-03 & -2.664018e-07 & $\pm$0.11\\ 
  21 &  KMg3AlSi3O10F2 &      FLUORPHLOGOPITE & -300.9 &  1.342665e+06 & -2.662439e+01 & -1.797165e+02 &  2.659617e-02 & -2.835390e-06 & $\pm$1.12\\ 
  22 &    Mg3Al2Si3O12 &               PYROPE & -296.7 &  1.402656e+06 & -3.072607e+01 & -1.635848e+02 &  3.136489e-02 & -3.582534e-06 & $\pm$1.23\\ 
  23 &            TiO2 &               RUTILE & -296.4 &  2.296961e+05 & -3.392573e+00 & -3.338416e+01 &  3.426975e-03 & -3.572185e-07 & $\pm$0.07\\ 
  24 &          MgTiO3 &           GEIKIELITE & -296.1 &  3.517143e+05 & -6.698669e+00 & -4.773530e+01 &  7.802853e-03 & -9.366533e-07 & $\pm$0.34\\ 
  25 &   Ca2Al3Si3O13H &         CLINOZOISITE & -295.6 &  1.560747e+06 & -3.510366e+01 & -1.680363e+02 &  3.352398e-02 & -3.806883e-06 & $\pm$1.37\\ 
  26 &         CaSi2O5 &        CaSi-TITANITE & -293.7 &  5.759983e+05 & -1.513590e+01 & -4.985255e+01 &  1.548385e-02 & -1.893205e-06 & $\pm$0.83\\ 
  27 &         Mg2SiO4 &           FORSTERITE & -293.4 &  4.685652e+05 & -9.176566e+00 & -6.544643e+01 &  9.627992e-03 & -1.038920e-06 & $\pm$0.30\\ 
  28 &          MgSiO3 &            ENSTATITE & -291.6 &  3.458775e+05 & -7.257035e+00 & -4.345839e+01 &  7.516612e-03 & -8.200336e-07 & $\pm$0.22\\ 
  29 &      Ca5Si2CO11 &             SPURRITE & -291.4 &  1.329424e+06 & -2.132758e+01 & -1.950763e+02 &  2.422856e-02 & -2.735895e-06 & $\pm$0.83\\ 
  30 &        KAlSi3O8 &           MICROCLINE & -288.5 &  9.258438e+05 & -2.002663e+01 & -1.044972e+02 &  1.928080e-02 & -2.158177e-06 & $\pm$0.60\\ 
  31 &       Ca5P3O13H &       HYDROXYAPATITE & -288.1 &  1.440061e+06 & -2.948136e+01 & -1.882099e+02 &  3.109138e-02 & -3.217938e-06 & $\pm$1.61\\ 
  32 &         KAlSiO4 &            KALSILITE & -286.9 &  4.774370e+05 & -9.688874e+00 & -6.070594e+01 &  1.032885e-02 & -1.168394e-06 & $\pm$0.35\\ 
  33 &        KAlSi2O6 &              LEUCITE & -286.7 &  6.994399e+05 & -1.408596e+01 & -8.441743e+01 &  1.342159e-02 & -1.489176e-06 & $\pm$0.33\\ 
  34 &       NaAlSi3O8 &               ALBITE & -285.6 &  9.232283e+05 & -2.012507e+01 & -1.039484e+02 &  1.953868e-02 & -2.217366e-06 & $\pm$0.63\\ 
  35 &            SiO2 &               QUARTZ & -285.5 &  2.222506e+05 & -5.478967e+00 & -1.969170e+01 &  4.907829e-03 & -5.612476e-07 & $\pm$0.18\\ 
  36 &       NaAlSi2O6 &              JADEITE & -284.7 &  7.011353e+05 & -1.413739e+01 & -9.120226e+01 &  1.397186e-02 & -1.513554e-06 & $\pm$0.45\\ 
  37 &             MgO &            PERICLASE & -284.6 &  1.195366e+05 & -2.104910e+00 & -2.122681e+01 &  2.564436e-03 & -2.957752e-07 & $\pm$0.11\\ 
  38 &        NaAlSiO4 &            NEPHELINE & -282.9 &  4.759180e+05 & -1.098517e+01 & -5.306336e+01 &  1.100831e-02 & -1.241451e-06 & $\pm$0.32\\ 
  39 & Ca2MnAl2Si3O13H &  PIEMONTITE(ORDERED) & -280.0 &  1.513754e+06 & -3.077556e+01 & -1.907176e+02 &  2.793700e-02 & -3.063564e-06 & $\pm$1.09\\ 
  40 &   CaAl4Si2O12H2 &            MARGARITE & -279.0 &  1.443099e+06 & -3.373819e+01 & -1.571716e+02 &  3.303015e-02 & -3.705003e-06 & $\pm$1.07\\ 
  41 &  Ca2Al2Si3O12H2 &             PREHNITE & -277.4 &  1.434914e+06 & -3.196401e+01 & -1.641458e+02 &  3.145388e-02 & -3.528334e-06 & $\pm$1.01\\ 
  42 & Ca2FeAl2Si3O13H &     EPIDOTE(ORDERED) & -276.6 &  1.521106e+06 & -3.241005e+01 & -1.857802e+02 &  3.245728e-02 & -3.668356e-06 & $\pm$1.36\\ 
  43 &     Ca5Si2C2O13 &            TILLEYITE & -273.3 &  1.537055e+06 & -2.652043e+01 & -2.135833e+02 &  2.876369e-02 & -3.172582e-06 & $\pm$0.87\\ 
  44 &    Ca3Fe2Si3O12 &            ANDRADITE & -271.4 &  1.373546e+06 & -2.766753e+01 & -1.840431e+02 &  3.011350e-02 & -3.455637e-06 & $\pm$1.16\\ 
  45 & KMg2Al3Si2O12H2 &            EASTONITE & -270.2 &  1.439124e+06 & -3.232163e+01 & -1.763878e+02 &  3.221755e-02 & -3.756024e-06 & $\pm$1.09\\ 
  46 &    Mn3Al2Si3O12 &          SPESSARTINE & -268.3 &  1.381612e+06 & -2.857798e+01 & -1.787062e+02 &  3.043611e-02 & -3.453339e-06 & $\pm$1.15\\ 
  47 &       CaFeSi2O6 &         HEDENBERGITE & -268.0 &  6.981350e+05 & -1.348390e+01 & -9.302132e+01 &  1.450960e-02 & -1.588910e-06 & $\pm$0.45\\ 
  48 &    Mg3Cr2Si3O12 &          KNORRINGITE & -267.8 &  1.348934e+06 & -2.497408e+01 & -1.966352e+02 &  2.701188e-02 & -3.090497e-06 & $\pm$1.02\\ 
  49 &         K2Si4O9 &           Si-WADEITE & -267.5 &  1.016675e+06 & -2.009772e+01 & -1.385171e+02 &  1.940090e-02 & -2.058947e-06 & $\pm$0.50\\ 
  50 &  Mg2Al2Si3O12H2 &       TSCHERMAK-TALC & -267.2 &  1.400979e+06 & -3.625955e+01 & -1.386779e+02 &  3.417640e-02 & -3.826802e-06 & $\pm$1.51\\ 
  51 &    KAl3Si3O12H2 &            MUSCOVITE & -266.5 &  1.414776e+06 & -3.449122e+01 & -1.503871e+02 &  3.394598e-02 & -3.828482e-06 & $\pm$1.10\\ 
  52 &  KMg3AlSi3O12H2 &           PHLOGOPITE & -265.2 &  1.417930e+06 & -3.144794e+01 & -1.792887e+02 &  3.133804e-02 & -3.671162e-06 & $\pm$1.07\\ 
  53 &   NaAl3Si3O12H2 &           PARAGONITE & -264.9 &  1.413363e+06 & -3.295879e+01 & -1.610931e+02 &  3.236865e-02 & -3.608740e-06 & $\pm$0.90\\ 
  54 &        AlSi2O6H &         PYROPHYLLITE & -263.4 &  6.892108e+05 & -1.659504e+01 & -7.266395e+01 &  1.542338e-02 & -1.735233e-06 & $\pm$0.40\\ 
  55 & NaMg3AlSi3O12H2 &       SODAPHLOGOPITE & -263.3 &  1.414930e+06 & -3.155196e+01 & -1.788363e+02 &  3.159871e-02 & -3.730675e-06 & $\pm$1.11\\ 
  56 &         FeAl2O4 &            HERCYNITE & -262.9 &  4.816286e+05 & -1.048218e+01 & -6.061770e+01 &  1.156339e-02 & -1.306461e-06 & $\pm$0.48\\ 
  57 &     Mg3Si4O12H2 &                 TALC & -262.8 &  1.382604e+06 & -3.420699e+01 & -1.502025e+02 &  3.251454e-02 & -3.772182e-06 & $\pm$1.35\\ 
  58 &   KMgAlSi4O12H2 &           CELADONITE & -259.8 &  1.390437e+06 & -3.361618e+01 & -1.544995e+02 &  3.306604e-02 & -3.743581e-06 & $\pm$1.09\\ 
  59 &       NaCrSi2O6 &           KOSMOCHLOR & -257.5 &  6.765279e+05 & -1.125471e+01 & -1.086861e+02 &  1.179478e-02 & -1.267462e-06 & $\pm$0.34\\ 
  60 & Ca2FeAlSi3O12H2 &       FERRI-PREHNITE & -257.1 &  1.393161e+06 & -3.104842e+01 & -1.689573e+02 &  3.137888e-02 & -3.535514e-06 & $\pm$1.00\\ 
  61 &          MnTiO3 &          PYROPHANITE & -256.7 &  3.435012e+05 & -5.536905e+00 & -5.436775e+01 &  7.170529e-03 & -8.519687e-07 & $\pm$0.32\\ 
  62 & Ca2Fe2AlSi3O13H &           Fe-EPIDOTE & -256.5 &  1.478087e+06 & -3.103944e+01 & -1.946633e+02 &  3.139057e-02 & -3.529827e-06 & $\pm$1.35\\ 
  63 &     MgAl2SiO7H2 &        Mg-CHLORITOID & -254.5 &  8.346090e+05 & -2.320869e+01 & -7.849664e+01 &  2.128252e-02 & -2.422756e-06 & $\pm$0.80\\ 
  64 &          MnSiO3 &         PYROXMANGITE & -249.2 &  3.355832e+05 & -6.945405e+00 & -4.433766e+01 &  7.426667e-03 & -8.613014e-07 & $\pm$0.34\\ 
  65 &   CaAl2Si4O14H4 &            WAIRAKITE & -248.8 &  1.633745e+06 & -3.834118e+01 & -1.809888e+02 &  3.572679e-02 & -4.035846e-06 & $\pm$1.15\\ 
  66 &      KAlSi3O9H2 &            K-CYMRITE & -247.4 &  1.038223e+06 & -2.399976e+01 & -1.183881e+02 &  2.236149e-02 & -2.482468e-06 & $\pm$0.68\\ 
  67 &    Fe3Al2Si3O12 &            ALMANDINE & -246.9 &  1.376944e+06 & -3.016241e+01 & -1.707578e+02 &  3.187560e-02 & -3.563550e-06 & $\pm$1.11\\ 
  68 &       Al2SiO6H2 &        HYDROXY-TOPAZ & -244.2 &  7.100468e+05 & -1.892466e+01 & -7.372280e+01 &  1.753940e-02 & -1.941487e-06 & $\pm$0.55\\ 
  69 &   KFeAlSi4O12H2 &      FERROCELADONITE & -243.0 &  1.378812e+06 & -3.315759e+01 & -1.566005e+02 &  3.278410e-02 & -3.675750e-06 & $\pm$1.02\\ 
  70 &       NaFeSi2O6 &               ACMITE & -241.7 &  6.583105e+05 & -1.441096e+01 & -8.729880e+01 &  1.476475e-02 & -1.542019e-06 & $\pm$0.48\\ 
  71 &     MnAl2SiO7H2 &        Mn-CHLORITOID & -238.6 &  8.259787e+05 & -2.111654e+01 & -9.285036e+01 &  2.038330e-02 & -2.293290e-06 & $\pm$0.66\\ 
  72 &     NaAlSi2O7H2 &             ANALCITE & -237.6 &  8.176590e+05 & -1.519133e+01 & -1.214758e+02 &  1.562853e-02 & -1.269436e-06 & $\pm$0.38\\ 
  73 &   CaAl2Si2O10H4 &            LAWSONITE & -237.6 &  1.191594e+06 & -3.176561e+01 & -1.226891e+02 &  2.966229e-02 & -3.109709e-06 & $\pm$0.77\\ 
  74 &         MgCr2O4 &        PICROCHROMITE & -236.0 &  4.431007e+05 & -8.938663e+00 & -6.835976e+01 &  1.066679e-02 & -1.296484e-06 & $\pm$0.52\\ 
  75 &         Mn2SiO4 &            TEPHROITE & -233.4 &  4.489460e+05 & -7.633874e+00 & -7.402296e+01 &  8.919088e-03 & -1.011800e-06 & $\pm$0.33\\ 
  76 &  KMn3AlSi3O12H2 &           Mn-BIOTITE & -233.1 &  1.379102e+06 & -2.873641e+01 & -1.929652e+02 &  3.004383e-02 & -3.637965e-06 & $\pm$1.01\\ 
\hline
\end{tabular}}\\[1mm]
\footnotesize The condensates are sorted according to $\dGSU(\Tref)/N$,
i.e.\ SUPCRTBL's Gibbs free energy of formation [kJ/mol] from the elements
at reference pressure $\pst\!=\!1\,$bar and reference temperature $\Tref\!=\!298.15\,$K,
devided by $N$, the sum of stoichiometric factors. $\dG$ are the Gibbs free
energies of formation of the condensates from free atoms at
$\pst\!=\!1\,$bar and temperature $T$, needed for
Eq.\,(\ref{eq:dGsolid}), which are fitted according to
$-\dG/(RT) = b_0/T + b_1\ln(T) + b_2 + b_3\,T + b_4\,T^2$. The
last column is the standard deviation between the generated SUPCRTBL
data points and the functional fit between 100\,K and 2500\,K in kJ/mol.
\end{table*}
\addtocounter{table}{-1}
\begin{table*}[!p]
\caption{{\bf continued}}
\vspace*{-2mm}\hspace*{-1mm}\resizebox{18.5cm}{!}{
\begin{tabular}{r|ll|c|ccccc|c}
\hline
&&&&&&&&&\\[-2ex]
no & formula & name &\!\!$\dGSU(\Tref)/N$\!\!& $b_0$ & $b_1$ & $b_2$ & $b_3$ & $b_4$ & \!$\sigma$\,[kJ/mol]\!\! \\
&&&&&&&&&\\[-2ex]
\hline
&&&&&&&&&\\[-2.2ex]
  77 &   MgAl2Si2O10H4 & MAGNESIOCARPHOLITE\!\!&-232.4 &  1.176370e+06 & -3.070094e+01 & -1.290393e+02 &  2.805212e-02 & -3.037731e-06 & $\pm$0.76\\ 
  78 &          FeTiO3 &             ILMENITE & -230.9 &  3.435788e+05 & -5.480155e+00 & -5.476246e+01 &  6.285519e-03 & -7.119639e-07 & $\pm$0.24\\ 
  79 &           AlO2H &             DIASPORE & -230.3 &  2.443624e+05 & -7.279776e+00 & -2.289286e+01 &  6.441950e-03 & -5.931350e-07 & $\pm$0.12\\ 
  80 &     FeAl2SiO7H2 &        Fe-CHLORITOID & -228.9 &  8.258670e+05 & -2.327161e+01 & -7.938651e+01 &  2.116909e-02 & -2.404658e-06 & $\pm$0.79\\ 
  81 &     Mg7Si2O14H6 &               PHASEA & -228.0 &  1.656847e+06 & -4.230253e+01 & -2.158169e+02 &  4.247591e-02 & -4.797719e-06 & $\pm$1.52\\ 
  82 &           CaCO3 &              CALCITE & -226.0 &  3.411928e+05 & -6.591620e+00 & -4.368617e+01 &  6.352508e-03 & -6.931829e-07 & $\pm$0.23\\ 
  83 &      Mg3Si2O9H4 &            LIZARDITE & -224.4 &  1.054418e+06 & -2.910595e+01 & -1.166171e+02 &  2.654429e-02 & -3.029452e-06 & $\pm$0.88\\ 
  84 &      Al2Si2O9H4 &            KAOLINITE & -223.7 &  1.050606e+06 & -2.819000e+01 & -1.056585e+02 &  1.966983e-02 & -2.504474e-06 & $\pm$0.88\\ 
  85 &          FeSiO3 &          FERROSILITE & -223.5 &  3.361566e+05 & -6.749696e+00 & -4.728394e+01 &  7.277447e-03 & -7.573112e-07 & $\pm$0.14\\ 
  86 &           CaSO4 &            ANHYDRITE & -220.4 &  3.449608e+05 & -9.113763e+00 & -4.690825e+01 &  8.600984e-03 & -6.567852e-07 & $\pm$0.24\\ 
  87 &  KFe3AlSi3O12H2 &               ANNITE & -218.0 &  1.386502e+06 & -3.005579e+01 & -1.891851e+02 &  3.068660e-02 & -3.467367e-06 & $\pm$0.85\\ 
  88 &        CaMgC2O6 &             DOLOMITE & -216.4 &  6.676592e+05 & -1.445077e+01 & -8.264391e+01 &  1.429423e-02 & -1.546931e-06 & $\pm$0.40\\ 
  89 &   CaAl2Si4O16H8 &           LAUMONTITE & -216.3 &  1.868803e+06 & -4.628609e+01 & -2.147229e+02 &  4.189556e-02 & -4.685634e-06 & $\pm$1.32\\ 
  90 &   FeAl2Si2O10H4 &      FERROCARPHOLITE & -214.1 &  1.165522e+06 & -3.037176e+01 & -1.316984e+02 &  2.831269e-02 & -3.060630e-06 & $\pm$0.76\\ 
  91 &     Fe3Si4O12H2 &         MINNESOTAITE & -213.0 &  1.349960e+06 & -3.399941e+01 & -1.519206e+02 &  3.265094e-02 & -3.538871e-06 & $\pm$1.27\\ 
  92 &           Cr2O3 &            ESKOLAITE & -211.3 &  3.216934e+05 & -2.887880e+00 & -7.289829e+01 &  4.520548e-03 & -5.320269e-07 & $\pm$0.21\\ 
  93 &           MgCO3 &            MAGNESITE & -205.5 &  3.255978e+05 & -7.798617e+00 & -3.888094e+01 &  7.661015e-03 & -8.339379e-07 & $\pm$0.22\\ 
  94 &             KCl &              SYLVITE & -204.5 &  7.801979e+04 &  4.147547e-01 & -3.263640e+01 &  1.037051e-03 &  6.931110e-09 & $\pm$0.00\\ 
  95 &         Fe2TiO4 &           ULVOSPINEL & -199.7 &  4.536963e+05 & -9.821801e+00 & -5.985428e+01 &  1.017629e-02 & -6.533094e-07 & $\pm$0.73\\ 
  96 &         Fe2SiO4 &             FAYALITE & -197.0 &  4.495737e+05 & -9.271411e+00 & -6.615000e+01 &  1.041250e-02 & -1.092031e-06 & $\pm$0.38\\ 
  97 &            NaCl &               HALITE & -192.3 &  7.717022e+04 &  3.607723e-01 & -3.269074e+01 &  9.594929e-04 &  1.834797e-08 & $\pm$0.00\\ 
  98 &         MgFe2O4 &      MAGNESIOFERRITE & -190.1 &  4.092716e+05 & -8.152176e+00 & -7.527872e+01 &  1.095039e-02 & -1.255442e-06 & $\pm$0.37\\ 
  99 &        CaFeC2O6 &             ANKERITE & -182.0 &  6.574137e+05 & -1.363162e+01 & -8.757014e+01 &  1.322217e-02 & -1.409341e-06 & $\pm$0.35\\ 
 100 &             MnO &          MANGANOSITE & -181.7 &  1.103123e+05 & -6.866758e-01 & -2.930136e+01 &  1.564511e-03 & -1.440846e-07 & $\pm$0.04\\ 
 101 &       NaAlCO5H2 &            DAWSONITE & -178.3 &  5.687719e+05 & -2.621938e+01 &  1.013648e+01 &  7.713633e-04 & -7.242316e-08 & $\pm$0.01\\ 
 102 &           Mn2O3 &             BIXBYITE & -176.5 &  2.728569e+05 & -2.903180e+00 & -6.914515e+01 &  4.215259e-03 & -2.567819e-07 & $\pm$0.03\\ 
 103 &          MgO2H2 &              BRUCITE & -167.0 &  2.396506e+05 & -7.315243e+00 & -3.115064e+01 &  7.048891e-03 & -8.256913e-07 & $\pm$0.28\\ 
 104 &      Fe3Si2O9H4 &           GREENALITE & -166.8 &  1.022771e+06 & -2.804547e+01 & -1.240261e+02 &  2.772103e-02 & -3.109114e-06 & $\pm$1.01\\ 
 105 &          AlO3H3 &             GIBBSITE & -165.0 &  3.602374e+05 & -1.611001e+01 & -1.217224e+01 &  1.567462e-02 & -4.410396e-07 & $\pm$0.10\\ 
 106 &           MnCO3 &        RHODOCHROSITE & -163.8 &  3.159405e+05 & -6.661948e+00 & -4.500950e+01 &  6.514539e-03 & -6.894651e-07 & $\pm$0.18\\ 
 107 &           Fe2O3 &             HEMATITE & -148.8 &  2.876529e+05 & -6.810027e+00 & -4.976293e+01 &  8.724692e-03 & -1.038569e-06 & $\pm$0.40\\ 
 108 &           Fe3O4 &            MAGNETITE & -144.6 &  4.019320e+05 & -8.960632e+00 & -7.101785e+01 &  1.191424e-02 & -1.404312e-06 & $\pm$0.46\\ 
 109 &           FeCO3 &             SIDERITE & -137.8 &  3.161250e+05 & -6.980649e+00 & -4.434616e+01 &  6.585898e-03 & -6.958439e-07 & $\pm$0.17\\ 
 110 &             FeO &       FERROPERICLASE & -125.8 &  1.121860e+05 & -2.152325e+00 & -2.083896e+01 &  2.956682e-03 & -3.223296e-07 & $\pm$0.15\\ 
 111 &           FeO2H &             GOETHITE & -122.8 &  2.026749e+05 & -4.655123e+00 & -3.861734e+01 &  5.315705e-03 & -5.712016e-07 & $\pm$0.15\\ 
 112 &             NiO &               NICKEL & -105.7 &  1.103349e+05 & -8.363217e-01 & -3.178273e+01 &  1.348578e-03 & -1.056011e-07 & $\pm$0.06\\ 
 113 &             CuO &             TENORITE & -64.2  &  8.900429e+04 & -1.587221e+00 & -2.477537e+01 &  2.420378e-03 & -2.394143e-07 & $\pm$0.14\\ 
 114 &            FeS2 &               PYRITE & -53.5  &  1.364603e+05 & -3.370114e+00 & -3.495254e+01 &  3.621202e-03 & -2.916809e-07 & $\pm$0.19\\ 
 115 &             FeS &             TROILITE & -50.7  &  9.450871e+04 & -2.495865e+00 & -1.829773e+01 &  3.174381e-03 & -3.069607e-07 & $\pm$0.13\\ 
 116 &            Cu2O &              CUPRITE & -49.4  &  1.317125e+05 & -6.268504e-01 & -4.560941e+01 &  2.920620e-03 & -3.213970e-07 & $\pm$0.08\\ 
 117 &              Cu &               COPPER &  -0.1  &  4.065093e+04 &  3.278646e-02 & -1.652274e+01 &  5.549892e-04 & -2.618359e-08 & $\pm$0.03\\ 
 118 &               S &              SULPHUR &  -0.1  &  3.327240e+04 & -4.494524e-01 & -1.393588e+01 &  1.047391e-03 & -9.902057e-08 & $\pm$0.02\\ 
 119 &              Ni &               NICKEL &  -0.1  &  5.178431e+04 & -5.284788e-02 & -1.837439e+01 &  6.292585e-04 & -3.003321e-08 & $\pm$0.03\\ 
 120 &              Fe &                 IRON &  -0.0  &  4.989867e+04 & -6.889399e-01 & -1.461748e+01 &  1.161476e-03 & -5.259868e-08 & $\pm$0.04\\ 
 121 &               C &             GRAPHITE &  -0.0  &  8.563846e+04 & -2.345056e+00 & -3.634772e+00 &  1.828533e-03 & -1.986659e-07 & $\pm$0.03\\ 
\hline
\end{tabular}}
\end{table*}

\begin{table*}[!p] 
\caption{Fit coefficients for the atomic corrections
  $\G_{\rm\!corr}(T)$ to convert the SUPCRTBL data to our reference
  states of neutral atoms, see Eq.\,(\ref{eq:AtomCorr}), derived from
  the NIST-JANAF database.}
\label{tab:Acorr}
\vspace*{-2mm}
\resizebox{12cm}{!}{
\begin{tabular}{c|ccccc|l}
\hline
atom & $a_0$ & $a_1$ & $a_2$ & $a_3$ & $a_4$ & error\,[kJ/mol] \\
\hline
&&&&&&\\[-2.2ex]
  H  & -2.78172E+4 & 2.49885 &-2.93973E+0 & 1.14321E-6 &-1.34658E-10 & $\pm$0.001\\
  Li & -1.94521E+4 & 2.52691 &-2.27778E-1 &-2.14827E-5 & 2.96822E-09 & $\pm$0.039\\
  C  & -8.56514E+4 & 2.52074 & 2.14377E+0 &-2.28858E-5 & 3.90851E-09 & $\pm$0.002\\
  N  & -5.95408E+4 & 2.50073 & 1.69028E+0 &-2.07541E-6 & 4.79627E-10 & $\pm$0.004\\
  O  & -3.28428E+4 & 2.71152 & 1.26593E+0 &-1.35177E-4 & 1.26727E-08 & $\pm$0.027\\
  F  & -1.23701E+4 & 2.76996 & 6.19482E-1 &-1.28288E-4 & 9.91217E-09 & $\pm$0.013\\
  Na & -1.40060E+4 & 2.48726 & 1.80092E+0 & 2.38611E-5 &-3.17527E-09 & $\pm$0.058\\
  Mg & -1.81146E+4 & 2.51860 & 1.00708E+0 &-8.33362E-6 & 3.74599E-10 & $\pm$0.033\\
  Al & -3.98785E+4 & 2.65608 & 2.03015E+0 &-9.87400E-5 & 8.59574E-09 & $\pm$0.032\\
  Si & -5.39603E+4 & 2.82697 & 1.34719E+0 &-2.23157E-4 & 2.56696E-08 & $\pm$0.062\\
  P  & -3.87736E+4 & 2.54304 & 2.61689E+0 &-4.90659E-5 & 8.96721E-09 & $\pm$0.045\\
  S  & -3.36266E+4 & 2.82008 & 1.33086E+0 &-7.74986E-5 & 0.00000E+00 & $\pm$0.17 \\
  Cl & -1.78181E+4 & 2.58828 & 2.50684E+0 & 9.72935E-5 &-1.51727E-08 & $\pm$0.073\\
  K  & -1.22688E+4 & 2.55355 & 2.20391E+0 &-5.36966E-5 & 8.24159E-09 & $\pm$0.050\\
  Ca & -2.21218E+4 & 2.55750 & 1.53540E+0 &-7.09263E-5 & 1.21389E-08 & $\pm$0.067\\
  Ti & -5.71391E+4 & 3.20557 & 4.48870E-1 &-5.18014E-4 & 7.00646E-08 & $\pm$0.10 \\
  Cr & -4.78986E+4 & 2.56936 & 3.82072E+0 &-1.27308E-4 & 3.48156E-08 & $\pm$0.041\\
  Mn & -3.44654E+4 & 2.52906 & 3.96202E+0 &-2.51476E-5 & 3.59496E-09 & $\pm$0.046\\
  Fe & -5.00408E+4 & 3.17358 & 6.57213E-1 &-2.96787E-4 & 3.12523E-08 & $\pm$0.29 \\
  Ni & -5.19591E+4 & 2.80126 & 3.06725E+0 & 1.48777E-4 &-2.21945E-08 & $\pm$0.084\\
  Cu & -4.10420E+4 & 2.54198 & 3.01655E+0 &-5.98838E-5 & 1.27558E-08 & $\pm$0.062\\
  Zr & -7.39272E+4 & 2.79313 & 3.25202E+0 & 1.26147E-4 & 1.15173E-08 & $\pm$0.072\\
\hline
\end{tabular}}\\
\hspace*{3mm}$-\G_{\rm\!corr}/(RT) = a_0/T + a_1\ln(T) + a_2 + a_3\,T + a_4\,T^2$. 
\end{table*}

\end{document}